\documentclass[12pt,letterpaper]{article}
\usepackage[]{epsfig,graphics}
\usepackage{natbib}
\newcommand  \gtsim  {\lower.5ex\hbox{$\; \buildrel > \over \sim \;$}}
\newcommand  \ltsim  {\lower.5ex\hbox{$\; \buildrel < \over \sim \;$}}
\newcommand{\twobytwo}[4]
{\left(\begin{array}{cc}#1&#2\\#3&#4\end{array}\right)}

\newcommand	\nsectionstar[1]	{%
	\vspace{-0.2ex}		
	\section*{\large #1}%
	\vspace{-0.3ex}}	

\def\3he{$^3{\rm He}$}

\def\beqa{\begin{eqnarray}}
\def\eeqa{\end{eqnarray}}

\def\beq{\begin{equation}}
\def\eeq{\end{equation}}
\begin{document}

\title{CMB Telescopes and Optical Systems \\
                To appear in: \\
 Planets, Stars and Stellar Systems (PSSS) \\
 Volume 1: Telescopes and Instrumentation}

\author{Shaul Hanany  {\small (hanany@umn.edu)}\\
{\small University of Minnesota, School of Physics and Astronomy, Minneapolis, MN,  USA, }\\
Michael Niemack  {\small (niemack@nist.gov)} \\
{\small National Institute of Standards and Technology and University of Colorado, Boulder, CO, USA, }\\
~and~Lyman Page  {\small (page@princeton.edu)} \\
{\small Princeton University, Department of Physics, Princeton NJ, USA.}}

\date{March 26, 2012}

\maketitle

\nsectionstar{Abstract}
The cosmic microwave background radiation (CMB) is now firmly established as a fundamental and essential probe of
the geometry, constituents, and birth of the Universe. The CMB is a potent observable because it can be measured with
precision and accuracy. Just as importantly, theoretical models of the Universe can predict the characteristics of the CMB to high accuracy,
and those predictions can be directly compared to observations. There are multiple aspects associated
with making a precise measurement. In this review, we focus on optical components for the instrumentation used to measure the
CMB polarization and temperature anisotropy. We begin with an overview of general considerations for CMB observations
and discuss common concepts used in the community.
We next consider a variety of alternatives available for a designer of a CMB telescope. Our discussion is
guided by the ground and balloon-based instruments that have been implemented over the years.  In the same vein,
we compare the arc-minute resolution Atacama Cosmology Telescope (ACT) and the South Pole Telescope (SPT).
CMB interferometers are presented briefly. We conclude with a comparison of the four CMB satellites, Relikt, COBE, WMAP, and Planck,
to demonstrate a remarkable evolution in design, sensitivity, resolution, and complexity over the past  thirty years.

\nsectionstar{Index terms}
CMB, Gregorian telescope, Satellites, WMAP, Relikt, COBE, Planck, Detector noise, modes, throughput, DLFOV, Polarization anisotropy, Temperature anisotropy, SPT, ACT, Balloons, Stokes parameters, Interferometers

\nsectionstar{Keywords}
CMB, telescopes.

\section{Introduction}
\label{sec:introduction}

The tremendous scientific payoff from studies of the cosmic microwave background radiation (CMB)
has driven researchers to develop new detectors and new detection techniques.
For the most part, CMB measurements have been made with dedicated instruments in which the optical elements are designed specifically to
mate with the detectors (rather than in facility-type telescopes).  The instruments run the gamut of radio and mm-wave detection
techniques: heterodyne receivers, direct power receivers,
correlation receivers, interferometers, Fourier transform spectrometers, and single and multi-mode bolometric receivers.  The quest for ever more
sensitive measurements of the CMB, including its polarization,  has led to the development of arrays  of hundreds to thousands of detectors,
some of which are polarization sensitive. These arrays are coupled to unique, large-throughput optical systems.
In this article we will focus primarily on optical systems for instruments that are used to measure the temperature anisotropy 
and polarization of the CMB.
In other words, instruments that are designed to measure only the temperature difference or polarization as a 
function of angle on the sky.

Before explicitly discussing the optical systems, we introduce in this section the celestial emission spectrum 
at CMB frequencies, discuss how the
instrument resolution is determined, and present the angular power spectrum. We then introduce the concepts of throughput and modes and
end with a discussion of the limits imposed by system noise because it is one of  the driving considerations for any optical design.
In Section~\ref{sec:groundballoon} we review the various choices available for a CMB optics designer, and the main optical systems
that have been used to date.
We also discuss more recent developments with the introduction of large focal plane arrays and the efforts to characterize
the polarization of the CMB. The ACT and SPT instruments are the highest resolution telescopes dedicated to
CMB measurements to date. They are also good examples for the state-of-the-art in CMB optical design at the time of their design, mid-decade 2000.
They are described and compared in Section~\ref{sec:largeground}. CMB Interferometers are briefly
presented in Section~\ref{sec:interferometers},
and the optical systems of the four CMB satellites to date are reviewed in Section~\ref{sec:satellites}.

\subsection{Celestial emission at CMB frequencies}
\label{sec:celestialemission}

\begin{figure}[htb]
\begin{center}
\vspace{-0.18in}
\includegraphics[width=4.5in]{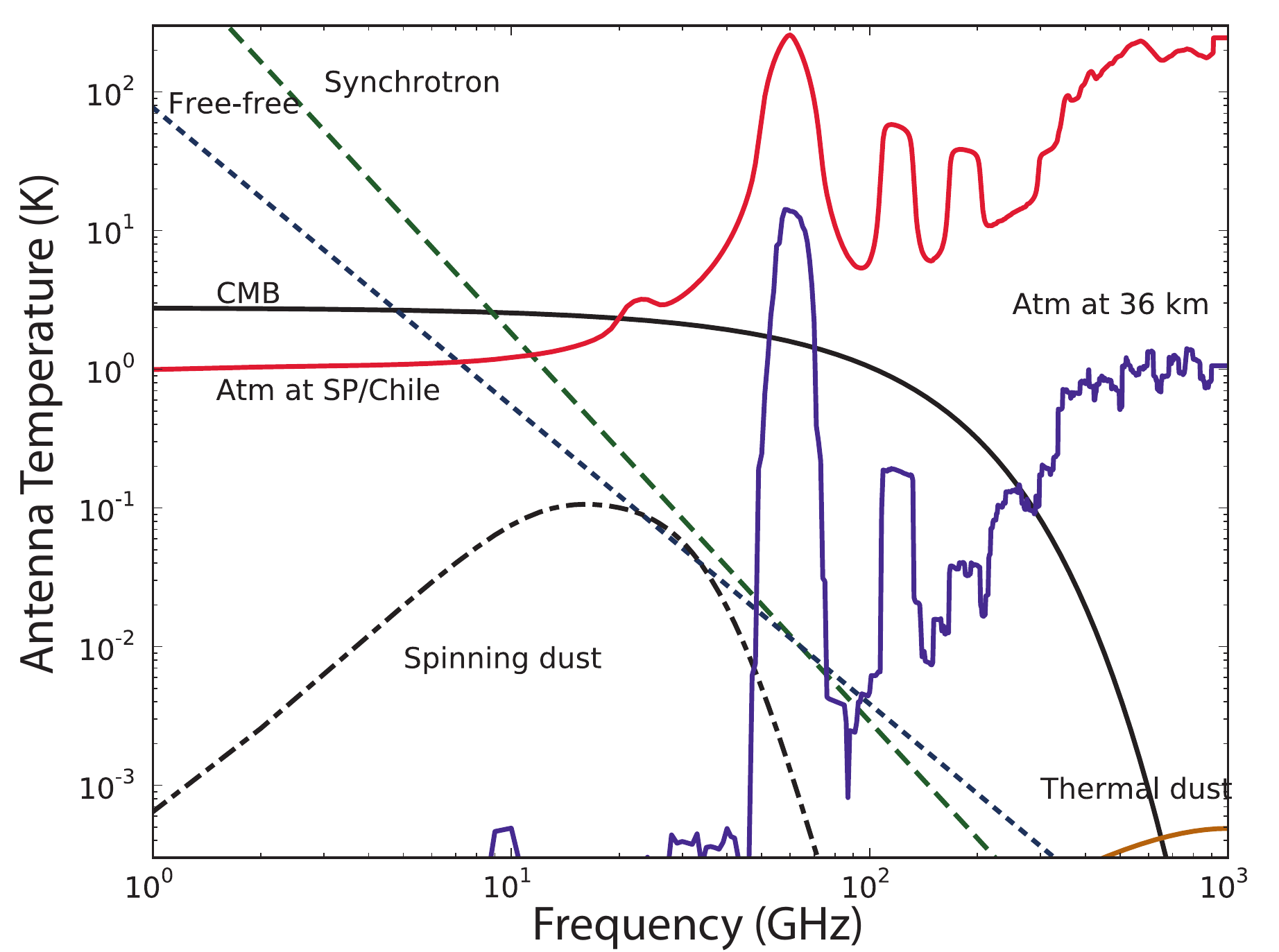}
\vspace{-0.1in}
\caption{\small  Sources of sky emission between 1 to 1000 GHz for a region of sky near a galactic latitude of roughly $20^\circ$.
The flat part of the CMB spectrum (solid black), roughly below 30~GHz, is called the Rayleigh-Jeans portion.
A Rayleigh-Jeans source with frequency independent emissivity would be indicated by a horizontal line on this plot. 
The synchrotron emission (long dash, green) is from cosmic rays orbiting in galactic magnetic fields and is polarized. 
Free free emission (short dash, blue) is due to
``breaking radiation" from galactic electrons and is
not polarized.  The amplitude of the spinning dust (dash dot, black), is not well known. 
This particular spinning model comes from \citet{Ali-Haimoud2009}. 
The standard spinning dust emission is not appreciably polarized. Emission from dust grains (brown, solid), which is more 
intense than the CMB above $\sim$700~GHz, is partially polarized.
The atmospheric models are based on the ATM code (\cite{pardo/etal:2001}), they use the US standard atmosphere,
and are for a zenith angle of $45^\circ$. The Atacama/South Pole spectrum (solid, red) is based on a precipitable water
vapor of 0.5~mm. The difference between the two sites is inconsequential for this plot. The atmospheric spectra have been averaged
over a 20\% bandwidth. The pair of lines at 60 and 120 GHz are the oxygen doublet and are also prominent at balloon altitudes
(solid, purple). The lines at 19 and 180 GHz are vibrational water lines. The finer scale features are from ozone.
\label{fig:spectrum_tant} }
\vspace{-0.3in}
\end{center}
\end{figure}

Figure~\ref{fig:spectrum_tant} shows the antenna temperature of the sky from 1 to 1000 GHz for a region at a galactic latitude of
roughly 20 degrees.
Ignoring emission from the atmosphere, synchrotron emission dominates celestial emission at the low frequency end and
dust emission dominates at high frequencies. These galactic emission components may be different by an order of magnitude
depending on galactic longitude. The CMB radiation dominates emission between about 20 and 500~GHz.
The experimental challenge is, however, to measure spatial fluctuations in the CMB
at parts in $10^{6}$ or $10^{7}$ of the level, a couple of orders of magnitude
below the bottom of the plot. The polarization signals are lower than the temperature anisotropy by a factor of ten and they too
beckon to be measured to percent-level precision.
The instrumental passbands, typically 20-30\%, are chosen to avoid atmospheric emission lines or to help identify
and subtract the foreground emission.

The basic picture in Figure~\ref{fig:spectrum_tant} has remained the same for over thirty years \citep{weiss:1980} though over the past
decade there has been increasing evidence for a new component of celestial emission in the 30 GHz region
(e.g., \citet{kogut/etal:1996, deOliveira1997, leitch/etal:1997}). This new component is spatially correlated with dust emission and
has been identified with emission by tiny grains of dust that are spun up to GHz rotation rates, hence
it has been dubbed ``spinning dust."  A variety of mechanisms have been proposed for spinning up
the grains~\citep{draine/lazarian:1998b,Draine1999}.  Still, though, it is not clear that the source is predominantly spinning dust.
Understanding this emission source is an active area of investigation.

\subsection{Instrument Resolution}
\label{sec:angularresolution}

The resolution of a CMB telescope is easiest to think about in the time reversed sense. We imagine that a detector element emits radiation.
The optical elements in the receiver direct that beam  to the sky or onto the primary reflector. The size of the beam at the primary optic
determines the resolution of the instrument. Such a primary optic can be a feed horn that launches a beam to the sky, a lens,  or a 
primary reflector.
The connection between the spatial size of the beam at the primary optic and the resolution can be understood through the
Fraunhofer's diffraction relation (e.g., \citet{born/wolf,hecht:OPTICS}),
\begin{equation}
\psi(\theta) \propto  \int_{apt} \psi_a(r) e^{kr\sin{\theta}\cos{\phi}} rdrd\phi,
\end{equation}
where $\psi_a(r)$ is the scalar electric field (e.g., one component of the electric field) in the aperture or on the primary optic, $\psi(\theta)$ is the angular distribution of the scalar
electric field in the far field ($d>>2D^2/\lambda$) and $k=2\pi/\lambda$. The integral is over the primary reflector or, more generally, the aperture.
For simplicity we have taken the case of  cylindrically symmetric illumination with coordinates $r$ and $\phi$ for a circular aperture of
diameter $D$, although a generalization is straightforward. The normalized beam profile is then given by $B(\theta)=|\psi(\theta)|^2/|\psi(0)|^2$.
That is, if the telescope scanned over a point source very far away, the output of the detector as measured in power would have this profile
as a function of scan angle $\theta$. Equation 1 gives an excellent and sometimes sufficient
estimate of the far field beam profile.

To be more specific, let us assume that the aperture distribution has a Gaussian profile so that the integrals are simple.
That is, $\psi_a(r)=\psi_0e^{-r^2/2\sigma_r^2}$. We also assume that $\psi_a$ is negligible at $r \geq D/2$ so that
we can let the limit of integration go to infinity.
This is the ``large edge taper'' limit.  In reality, no aperture distribution can be Gaussian and some are quite far from it.
The integral evaluates to $\psi_0\sigma_r^2e^{-\sigma_r^2k^2\sin^2(\theta )/2}$.
For small angles, $\theta\sim\sin(\theta)$, and we find from the above that $B(\theta)=e^{-\theta^2/2\sigma_B^2}$ where
$\sigma_B=\lambda/\sqrt{8}\pi\sigma_r$. In angular dimensions, the beam profile is most often characterized by a full width at half maximum,
or twice the angle at which $B(\theta)= 1/2$. We denote this as $\theta_{1/2}$ and find
$\theta_{1/2}=\sqrt{8\ln(2)}\sigma_B=\sqrt{\ln (2)}\lambda/\sigma_r\pi$. We see the familiar relation that the beam width is proportional to the
wavelength and inversely proportional to the size of the illumination pattern on the primary reflector with a pre-factor that depends on the geometry.
For this far-field Gaussian profile the beam solid angle is $\Omega_B=\int B d\Omega = 2\pi\sigma_B^2$.

\begin{figure}[htb]
\begin{center}
\includegraphics[width=5.0in]{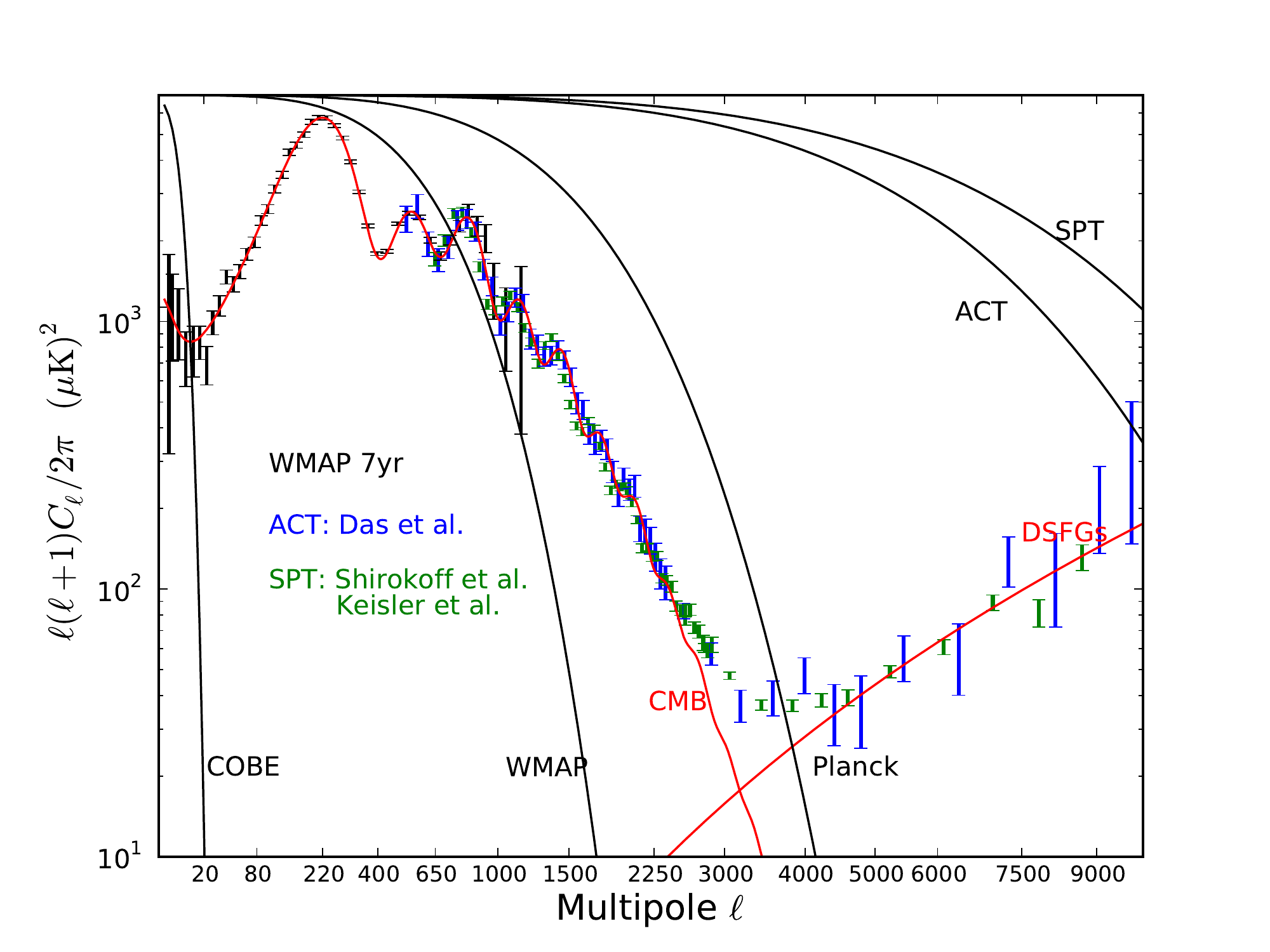}
\caption{\small Current best published measurements of the CMB temperature power spectrum (data points, \cite{komatsu/etal:2011,shirokoff/etal:2011,das/etal:2011, keisler/etal:2011}) and a $\Lambda$CDM cosmological model
(solid red, up to $\ell=3000$). The model power spectrum for $\ell>3000$ is due Poisson noise from
confusion-limited dusty
star forming galaxies (DSFGs) at 150~GHz.
The x-axis is scaled as $\ell^{0.45}$ to emphasize the middle part of the anisotropy spectrum. Gaussian approximations to the
window functions are shown for COBE ($7^{\circ}$), WMAP ($12^\prime$), Planck ($5^\prime$),
ACT ($1.4^\prime$, \cite{swetz/etal:2011}),
and SPT ($1.1^\prime$, \cite{schaffer/etal:2011}). The large size of the WMAP error bars near $\ell=2$ and 1000
are due to ``cosmic variance" and finite beam resolution, respectively.
\label{fig:pspec}
}
\end{center}
\end{figure}

The natural ``observable" for anisotropy measurements is the angular power spectrum for the following reason.
When the distribution of the amplitudes of the fluctuations
is Gaussian, as it apparently is for the primary CMB, {\it all} information about the sky is contained in the power spectrum.  If there are
correlations in the signal,  for example if the cooler areas had a larger spatial extent than the warmer areas or discrete sources of emission were
clustered together, then higher-order statistics would be needed to fully describe the sky.  Even in this case, the power spectrum is the best
first-look analytic tool for assessing the sky. Searches for ``non-Gaussianity" are an active area of research. While there are many possible sources of non-Gaussianity,
the primary CMB anisotropy appears to be Gaussian to the limits of current measurements (e.g., \citet{komatsu/etal:2011}).
A snapshot of the latest measurements of the power spectrum is shown in Figure~\ref{fig:pspec}.

The instrument resolution as expressed in the power spectrum is obtained from the Legendre transform of $B^2(\theta)$.
To appreciate this, we take a step back and describe the connection between the observable, that is the angular power spectrum,
and the antenna pattern of the instrument.
Because the CMB covers the full sky, it is most usefully expressed as an expansion in spherical harmonics. The monopole term ($\ell=0$)
has been determined by COBE/FIRAS to be $T_{CMB}=2.725\pm0.001$~K (\citet{fixsen/mather:2002}, plotted in
Figure~\ref{fig:spectrum_tant}). The dipole term ($\ell=1$) is dominated by the peculiar velocity of the solar system with respect to the cosmic reference frame.
As we are primarily concerned with cosmological fluctuations, we omit these terms from the expansion and we write the fluctuations as
\begin{equation}
\delta T(\theta,\phi)=\sum_{\ell\geq 2, -\ell\leq m \leq\ell} a_{\ell m}Y_l^m(\theta,\phi).
\label{eq:dt}
\end{equation}
To the limits of measurement, the CMB fluctuations appear to be statistically isotropic (e.g., \cite{basek/etal:2006}): they
are the same in all directions and thus have no preferred $m$ dependence. The overall variance of the CMB fluctuations is then given by
\begin{equation}
<\delta T^2(\theta,\phi)>=  \sum_{\ell\geq 2}\frac{2\ell+1}{4\pi} <|a_{\ell m}|^2>
= \sum_{\ell\geq 2}  \frac{2\ell+1}{2\ell(\ell+1)} \frac{\ell(\ell+1)}{2\pi}C_\ell \equiv
\sum_{\ell\geq 2}  \frac{2\ell+1}{2\ell(\ell+1)} \cal{B}_\ell,
\label{eq:var}
\end{equation}
where the factor of $2\ell+1$ comes from the sum of the $m$ values, all of which have the same variance, and the $4\pi$
comes from averaging over the full sky. Generally $C_\ell$ is called the power spectrum, but in cosmology the term is just as frequently used for $\cal{B}_\ell$.
These quantities are the primary point of contact between theory and measurements.
Cosmological models provide predictions for $C_\ell$; experiments measure temperatures on a patch of the sky and provide an
estimate of $C_\ell$.
The quantity  most often plotted is $\cal{B}_\ell$
\footnote{The factor of $\ell(\ell+1)/2\pi$ \citep{bond/efstathiou:1984}, as opposed to the possibly more natural $\ell(2\ell+1)/4\pi$ \citep{peebles:1994},
is derived from the observation that the cold dark mater model, without a cosmological constant, approaches $\ell(\ell+1)$ at small $\ell$ for a scalar
spectral index of unity. Needless to say, the model that gave rise to the now-standard convention does not describe Nature. Another choice would
be $(\ell+1/2)^2$ because the wavevector $k\rightarrow\ell+1/2$ at high $\ell$. There is not a widely agreed upon letter for the plotted power spectrum.
We use $\cal{B}$ for both ``bandpower" and
J. R. $\cal{B}$ond who devised the convention.  The term bandpower refers to averaging the $\cal{B}_\ell$ over a band in $\ell$.}.
It is the fluctuation power per logarithmic interval in $\ell$. The x-axis of the power spectrum is the spherical harmonic index $\ell$.
As a rough approximation, $\ell\approx 180/\theta$ with $\theta$ in degrees.

The process of measuring the CMB with a beam of finite size acts as a convolution of the intrinsic signal (eq.~\ref{eq:dt})
with the beam function, $B(\theta)$. The finite resolution averages over some of the smaller angular scale fluctuations
and thereby reduces the variance given in eq.~\ref{eq:var}. By Parseval's theorem, a convolution in one space corresponds
to  a multiplication in the Fourier transform space. In our case, because we are working on a sphere with symmetric beams,
Legendre transforms, as opposed to Fourier transforms, are applicable. We may think of the square of the Legendre
transform of $B(\theta)$, $B_\ell^2$, as filtering the power spectrum. (There is one power of $B$ associated with one temperature map.)
The transform of $B(\theta)$ is given by
\begin{equation}
B_\ell = 2\pi \int B(\theta) P_\ell(\cos\theta)d\cos(\theta) = B_0e^{\ell(\ell+1/2)/2\sigma_B^2 },
\label{eq:bl}
\end{equation}
where $P_\ell$ is a Legendre polynomial and $B_0$ is a normalization constant.

A Gaussian random field is fully described by the two-point correlation function, $C(\theta )$ (the Legendre transform of the power spectrum),  which gives the
average variance of two pixels separated by an angle. The variance given in eq.~\ref{eq:var}
is the angular correlation function evaluated for zero angular separation between pixels.
The general relation, including the effects of measuring with a beam of finite resolution,  is
\begin{equation}
C_{meas}(\theta) = <\delta T_{meas}(\theta_1,\phi_1)\delta T_{meas}(\theta_2,\phi_2) > = \sum_{\ell\geq 2} \frac{2\ell+1}{4\pi}C_\ell P_\ell(\theta) W_\ell .
\label{eq:ctheta}
\end{equation}
Here $W_\ell$ is the ``window function.''  In this expression, the
angle $\theta$ goes between
directions ``1'' and ``2''.~\footnote{We follow common notation but note that in eq.~\ref{eq:dt}, $\theta$ is a coordinate on the sky;
in eq.~\ref{eq:bl}, $\theta$ is the angular measure of the beam profile with a $\theta=0$ corresponding to the beam peak;
and in eq.~\ref{eq:ctheta}, $\theta$ is the angular separation between two pixels on the sky. }
 The window function encodes the effects of the finite resolution. For a symmetric beam,
$W_\ell=B_\ell^2$.
Figure~\ref{fig:pspec} shows approximations to the  window functions, assuming Gaussian shaped beams, for COBE, ($\theta_{1/2}=7^\circ$), WMAP($\theta_{1/2}=12^\prime$), Planck($\theta_{1/2}=5^\prime$), ACT($\theta_{1/2}=1.4^\prime$), and SPT($\theta_{1/2}=1.1^\prime$). One immediately sees the relation between the resolution and how well one can determine the power spectrum. For example, COBE, which we discuss in more detail below, was limited to large angular scales (low $\ell$) because of its relatively low angular resolution.

\subsection{Throughput and Modes}
\label{sec:throughput}

One of the key characteristics  for any optical system is the ``throughput" or \'etendue  (or ``A-Omega") of the system.  It is a measure of the total
amount of radiation that an optical system handles. Using Liouville's theorem,
which roughly states that the volume of phase space is conserved for a freely evolving system, one can show that $A\Omega$
is conserved for photons as long as there is no loss in the system. This means that at each plane of the system the integral of
the product of areal ($dA$) and angular distribution ($d\Omega$) of
the radiation is constant. For example, let's assume that the effective angular distribution of a beam launched from a primary optic of radius
$r$ is a top hat in angular extent with an apex angle defined as $\theta_{1/2} = 2 \theta_0$, analogous to the $\theta_{1/2}$ definition for Gaussian distributions above. Then one obtains a throughput of
\begin{equation}
\label{eqn:aomegatop}
A \Omega = 2 \pi^{2}r^{2} (1 - \cos \theta_0) \simeq \frac{\pi^{2}D^{2}\theta_{1/2}^{2}}{16},
\end{equation}
where the approximation holds for $\sin\theta_0 \simeq \theta_0$.
If the angular distribution is a Gaussian with width $\sigma_{B}$ ($\theta_{1/2} = \sqrt{8\ln 2} \sigma_{B}$) then
\begin{equation}
\label{eqn:aomegagauss}
A \Omega \simeq 2\pi^{2} r^{2} \sigma_B^2 \simeq \frac{\pi^{2}D^{2}\theta_{1/2}^{2} }{16 \ln 2} \approx \frac{\pi^{2}D^{2}\theta_{1/2}^{2} }{11}.
\end{equation}
Here we also assume that $\sigma_{B}$ is small, so that the integration over the angular pattern gives appreciable contributions only for $\sin \sigma_{B} \simeq \sigma_{B}$.
If the primary reflector has an effective diameter of 1~m and the beam has $\theta_{1/2} =  0.15^\circ$ then the throughput is 0.04~cm$^{2}$sr (assuming
Equation~\ref{eqn:aomegatop}).
Let's say this radiation is focused down to a feed with an effective collecting area of 1~cm$^2$. Conservation of throughput implies that now $\theta_{1/2}=13^\circ $.
In other words, as you squeeze down the area the radiation has to go through by focussing or concentrating, the solid angle increases.
While $A\Omega$ is conserved for any lossless optical system, it has a specific value for a system that supports only
a ``single mode" of propagating radiation:
\begin{equation}
\label{eqn:lambda2}
A\Omega=\lambda^2,
\end{equation}
where $\lambda$ is the wavelength of the radiation and $A$ is the effective area of the aperture.
We discuss modes below. This relation, which may be derived from the Fraunhofer integral
and conservation of energy (as in \citet{born/wolf},  Section 8.3.3.),
is a generalization of familiar results from diffraction theory. For example, Airy's famous expression that the angular diameter of the spot size
from a uniformly illuminated aperture is $2.44\lambda/D$, where $D$ is the aperture diameter, is equivalent to
Equation~\ref{eqn:lambda2}\footnote{The Airy beam profile is given by
$B_n(\theta )= [2J_1(x)/x]^2$ where $x=\pi D\sin (\theta )/\lambda$ and $J_1$ is a Bessel function. The value of
$1.22\lambda/D$ is the angular separation between the maximum and the first null. For small angles,
$\theta_{1/2}=1.03\lambda/D$. The total solid angle is $2\pi \int B_n(\theta )\sin(\theta )d\theta$.
To make the integral simple and avoid considering the difference between projecting onto a plane versus a sphere,
we consider the limit of small $\theta$. Then, $\Omega= 8\lambda^2/\pi D^2\int_0^\infty[J_1(\pi Dx/\lambda)]^2x^{-1}dx=\lambda^2/A$. }.
The relations above give a handy conversion between the system's effective aperture, the angular extent of the beam
and the frequency of interest for single mode optical systems.
Combining Equations~\ref{eqn:aomegagauss} and~\ref{eqn:lambda2} we obtain
\begin{equation}
\label{eqn:beamsize}
\theta_{1/2, rad} = 1.06 \lambda/D,
\end{equation}
where $D$ is the effective illumination on the primary reflector.

In the above treatment we brought in the concept of a single propagating mode of radiation. A mode is a particular spatial pattern of the
electromagnetic field. 
Radiation propagation in a rectangular waveguide of height $a/2$ and width $a$ gives a familiar example.
For frequencies less than a cutoff, $\nu_c<c/2a$, no electromagnetic radiation can propagate down a  waveguide of length longer than a few $\lambda$. For  $c/2a<\nu_c<\sqrt{5/4}c/a$ only the TE$_{10}$ mode of radiation propagates;  at frequencies just above 
just above $c/a$ the TE$_{10}$ and TE$_{01}$ can propagate. Above $\sqrt{5/4}c/a$ the
TE$_{10}$, TE$_{11}$, and TM$_{11}$ modes  are free to propagate. With  the geometry of a  cylindrical waveguides of diameter $d$ the lowest frequency mode is
the TE$_{11}$ (which supports two polarizations) with a cut-off frequency $\nu_c = c/1.7d$, and the next modes are TM$_{01}$ and TE$_{21}$ which turn on at frequencies that are 1.31 and 1.66 higher, respectively, than the lowest.

Experimentally, the selection for operating in a single mode is typically achieved by having a waveguide somewhere along the
light path, typically at the entrance to the
detecting element\footnote{In a close packed array this may be approximated having the pixel size smaller than $\lambda$. Such a spatial mode would support two polarizations.}. The waveguide is
essentially a high-pass filter, selecting the lowest frequency that can pass through the system. An additional
low-pass filter then rejects frequencies at which the second and higher modes are propagating. Experimenters have been using single modes
because these systems have particularly well behaved and calculable beam patterns.
If a second mode were added, say by operating at a higher frequency so that both the TE$_{11}$ and
TM$_{01}$ propagated (for a cylindrical waveguide), one would receive more signal, an advantage, but the beam pattern of
the combination of modes would be different, likely more complex
compared to the single mode illumination, and there would likely be increased spill over the edge of the primary.

Consider a radio receiver that observes a diffuse Planckian source of temperature $T$ through a telescope.
The surface brightness is given by
\begin{equation}
S_\nu(T) = \frac{2h\nu^3}{c^2(e^{h\nu/kT}-1)}\rightarrow \frac{2\nu^2}{c^2} kT,
\end{equation}
where $h$ is Planck's constant, $k$ is Boltzmann's constant, and $S_\nu$ is measured in ${\rm W/m^2srHz}$. The expression on
the right is the surface brightness in the Rayleigh-Jeans limit. The power that
makes it through to the detector is given by
\begin{equation}
P=\frac{1}{2}\int_{\Omega}\int_{\nu} \epsilon (\nu , \theta , \phi) A_e(\nu)S_\nu(\theta ,\phi )B(\nu ,\theta ,\phi ) d\Omega d\nu,
\end{equation}
where the factor of 1/2 comes from coupling to a single polarization, $A_e$ is the effective area and
$\epsilon$ is the transmission efficiency of the instrument. For clarity of discussion we will henceforth assume that
the transmission efficiency is unity. If $S_\nu$ is uniform across the sky, we are in the Rayleigh-Jeans limit ($h\nu<<kT$),
and $A_e$ and $B_n$ are relatively independent of frequency over  a small bandwidth (commonly achieved), then
\begin{equation}
P=\frac{1}{2}\int_{\nu} A_e(\nu) 2\frac{\nu^2}{c^2} k T  \int_{\Omega}B_n(\nu, \theta, \phi ) d\Omega d\nu=
kT\int_{\nu}  \frac{A_e(\nu) \Omega }{\lambda^2}  d\nu  = kT\int_{\nu} d\nu = kT\Delta\nu.
\end{equation}
Thus, each mode of radiation delivers $kT\Delta\nu$ of power to the detector. If there is a second mode in the system
that is supported in this bandwidth then it also contributes
$kT\Delta\nu$ of power.  It is possible, even likely, that different modes are supported over different but overlapping bandwidths.

Increasing the amount of celestial power on one's detector is an advantage when trying to detect a faint signal like the CMB.
The trade off is between control of the optical properties of the system and collecting power onto the detector. Note that using a
larger telescope does not increase the detected
power if one detects only a single mode. A larger telescope merely increases the resolution.
In a bolometric system, one can to a certain extent control the number of
modes that land on the detector. For example, one can place the absorbing area at the base of a ``light collector" or Winston cone
\citep{welford/winston:1978}. An approximation to the number of
modes in the system is then found by beam mapping to determine $\Omega_B$, measuring the band pass to find the
average wavelength $\lambda_a$, and dividing
by the collecting area of the input optics. This gives the number of modes as $\alpha_m=A\Omega/\lambda_a^2$. This is only an
approximation because it
assumes knowledge of the aperture distribution (for the collecting area) and that all modes couple to the
detector with the same efficiency. We use $\alpha$ because
often this quantity is not an integer. Although formally modes come in integer sets, not all modes couple equally to the detector
output. In the early days of CMB bolometry muti-moded systems were often used.
As detectors became more sensitive, the field moved toward single-moded bolometric systems as pioneered in the
White Dish experiment \citep{tucker/etal:1993}.
This led to more precise knowledge of the beams.
To a good approximation, the current generation of bolometric
CMB instruments all operate single-moded  (with the first mode of propagation).
However, there are modern examples of multi-moded systems though they are not used for the primary CMB bands. They include
the 345~GHz band on Boomerang \citep{jones:2005} and Planck's 545 and 857 GHz bands
\citep{ade/etal:2010, maffei/etal:2010} where there are just a few modes. In these cases, the coupling of the radiation in the bolometer's
integrating cavity is in practice not possible to compute accurately.
Interest in multi-moded systems has returned with at least one satellite proposal for an instrument called PIXIE for measuring the CMB polarization in a
massively over-moded system \citep{kogut/etal:2011}. The PIXIE concept is based on the observation
that the signal improves as the number of modes, $n_m$,  but the noise degrades only as $\sqrt{n_m}$ in the photon limited
noise regime (see below). Thus, S/N improves as $\sqrt{n_m}$.

\subsection{Noise}
\label{sec:noise}

\begin{figure}[tbh]
\begin{center}
\includegraphics[width=4.0in]{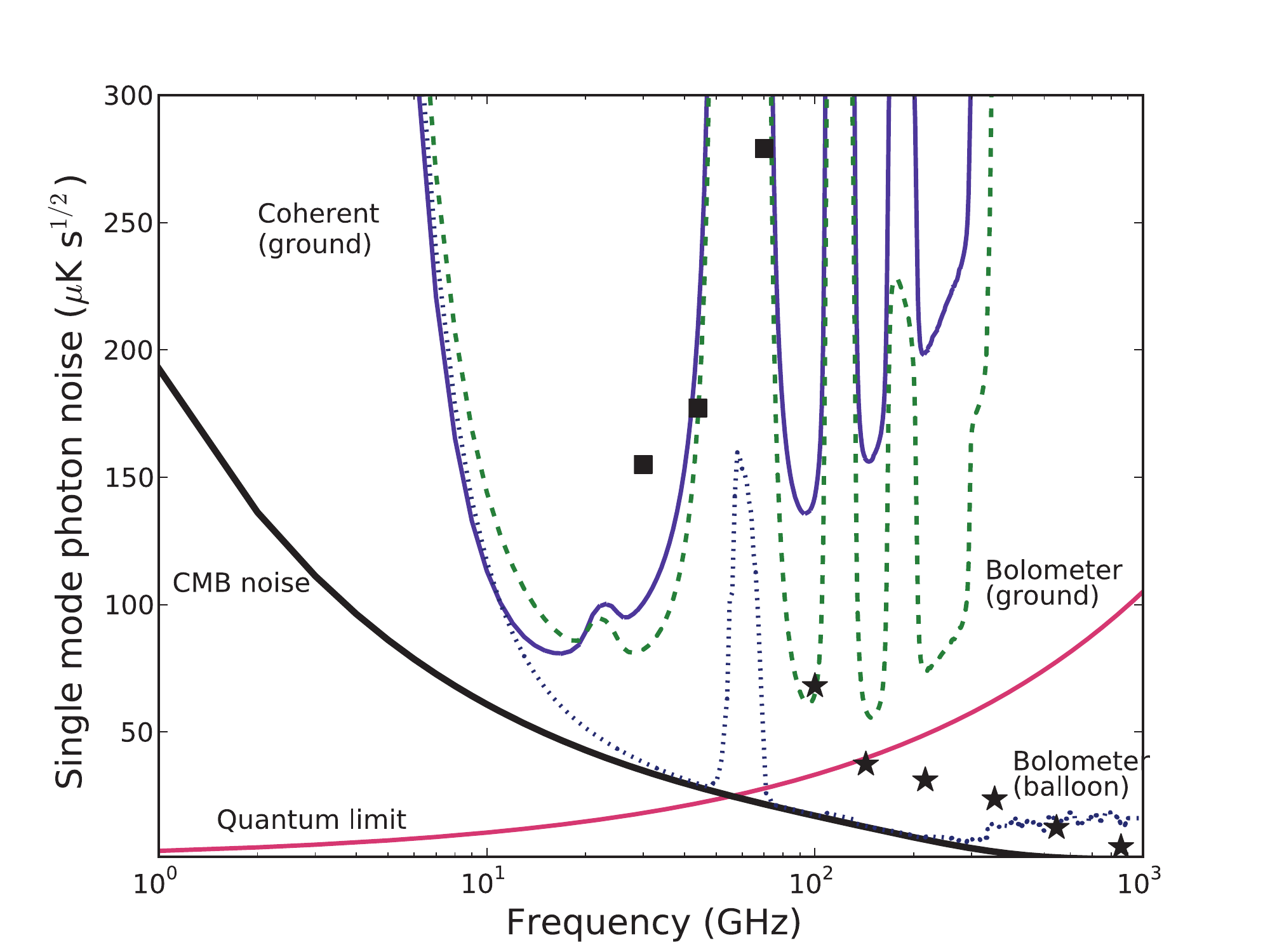}
\caption{\small Photon noise from 1 to 1000 GHz for a single mode of radiation for a 20\% bandwidth in frequency.
The CMB noise (solid, thick black) is for a region of sky without any other foreground emission. It sets a fundamental limit over most  of this frequency range (the far infrared background, not shown, sets the limit near 1 THz). Noise from atmosphere (Chile or South Pole, zenith
angle of 45$^\circ$) for bolometers
(dash, green) is lower than for coherent receivers (solid, purple) except below $\sim$20~GHz. Bolometers on balloon (dot, blue) are 
limited by CMB noise between 80 and 200~GHz. The atmospheric noise shown is due
to thermal emission and does not include 
contributions from turbulence, changes in column density,  or water vapor, which can increase the noise many-fold. Also 
shown are reported coherent receiver (square) and bolometer (star) noise for the Planck satellite, both adjusted 
as discussed in the text and both for total intensity.  The Planck bolometers are close to the fundamental noise limit.
\label{fig:noise}
}
\end{center}
\end{figure}

The choice of an optical system and its location is intimately connected with the desired noise performance.
There are a number of contributing noise sources that depend on the type of detector, how it is biased, and on its environment
(see e.g., \cite{mather:1982,pospieszalski:1992}). For this review we concern ourselves primarily with the photon noise from the sky because it sets the ultimate detection limit.
We first consider bolometric or ``direct'' detectors which detect the total power and destroy all phase information
in the incident field.
Equation \ref{eq:photonnoise} gives the photon noise power on the detectors per mode (e.g., \citet{zmuidzinas:2003}) as:
\begin{equation}
N^2(\nu)\tau = \frac{\Delta\nu}{ \eta(\nu) (k\Delta\nu )^2}  (h\nu )^2n(\nu )[1+\eta(\nu )n(\nu )],
\label{eq:photonnoise}
\end{equation}
where $\tau$ is the integration time, $\Delta\nu$ is the bandwidth,
$n(\nu )$ is the occupation number (power in a mode divided by $h\nu$), and $\eta(\nu )$ is the quantum efficiency which we take to be unity. We have approximated integrals by multiplying by a bandwidth of $\Delta\nu$. Because each mode of radiation delivers a power of $kT\Delta\nu$ one may convert from the ${\rm Ws}^{1/2}$ to ${\rm Ks}^{1/2}$ by dividing by $k\Delta\nu$. The left-hand term in the expression is the Poisson term and the right-hand term accounts for the correlations between the arrival times of the photons. When there are multiple modes in the system, one cannot simply assume that the above holds for each mode. One must take into account 
the correlations between the photon noise in each mode (\citet{lamarre:1986, richards:1994, zmuidzinas:2003}).

For coherent detector systems, one first amplifies the incident electric field while retaining phase information. After multiple stages of amplification, mixing, etc. one at last records the power in the signal.  Because the amplitude and phase are measured simultaneously, and these quantities do not commute, quantum mechanics sets a fundamental noise limit of
$N\sqrt{\tau}=h\nu/k\sqrt{\Delta\nu}$. In practice, the best systems achieve three times the quantum limit over a limited bandwidth and in ideal conditions. A good estimate of the  noise limit is:
\begin{equation}
N(\nu)\sqrt{\tau} = \frac{3(h\nu/k) + T_{sky}}{\sqrt{\Delta\nu}},
\end{equation}
where $T_{sky}$ is the antenna temperature of the incident radiation.

In Figure~\ref{fig:noise} we show the noise limit for a single-moded detector with 20\%  bandwidth at a high altitude ground-based site (e.g., the South Pole or the Atacama Desert) and at a typical balloon altitude of 36~km.   We also show the noise level for the current generation of detectors on the Planck satellite. The quoted sensitivities for the bolometric detectors \citep{planck-HFI-inst:2011} (two polarizations combined and all noise terms in the high frequency limit) are adjusted up to
a 20\% reference bandwidth. The quoted sensitivities for the coherent detectors \citep{planck-LFI-inst:2011} (all noise terms in the high frequency limit) have been adjusted down to a 20\% reference bandwidth.

Advances in bolometric detectors  have reached the point where the intrinsic noise is near the noise limit set by 
the photon noise. Thus, to improve sensitivity,  one wins more quickly by adding detectors as 
opposed to improving the detector noise. This is one of the motivations behind large arrays of detectors, and their 
associated large fields of view.  In both cases one can also win by increasing the number of modes.

\subsection{Polarization Terminology}
\label{sec:polarizationterminology}

There is a well developed terminology for describing polarization.  Imagine a telescope beam that points at a
single position on the sky and feeds a detector that can measure the amplitude and phase of  a partially coherent electric field.
Because the electric field is a vector in a plane, it can be completely specified by measuring its horizontal, $E_x(t)$,  and vertical, $E_y(t)$,
components at each instant. To measure the intensity of the field, one averages the detector outputs over time.
The polarization properties of the field, assuming they are relatively constant, are completely specified with the coherency matrix:
\begin{equation}
\twobytwo{< E_xE_x^*>}
{< E_xE_y^*>}{< E_yE_x^*>}
{< E_yE_y^*>}  \propto
\frac{1}{2} \twobytwo{I}{0}{0}{I} +
\frac{1}{2} \twobytwo{Q}{U}{U}{-Q} +
\frac{i}{2} \twobytwo{0}{V}{-V}{0}
\label{eq:stokes}
\end{equation}
where the ``*'' denotes a complex conjugate and the average is taken over time. The coherency matrix can also be represented by means
of Stokes Parameters $I,\, Q,\, U$ and $V$, as shown in the right hand side of Equation~\ref{eq:stokes}. The polarization has the symmetries
of a spin-two field.
The proportional sign indicates that the Stokes parameters, which represent intensities,  are reported in Kelvins.
The total intensity in the radiation is the trace of the matrix. Stokes $Q$ is the intensity of the horizontal polarization minus the 
vertical. Stokes $U$
is the in-phase correlation between the two components of the field minus the $180^\circ$ out-of-phase correlation. If the incident radiation
was pure Stokes $Q$, and one rotated the field by $45^\circ$ in the x-y plane, then the output would be pure $U$.
Stokes $V$ measures circular polarization. However, the CMB is expected to be only linearly polarized. (See \cite{zal+seljak:1997} and
\cite{kamionkowski/etal:1997} for a discussion and formalism.) Although
deviations from this prediction are of great interest, they are beyond the scope of this article.

As the beam is scanned across the sky, one makes maps of $I$, $Q$, and $U$. Of course, the values of $Q$ and $U$ depend on
the specification of a coordinate system. However, the $Q$ and $U$ maps may be transformed into
`E modes and B modes'. The advantage of these modes is that they are independent of the coordinate system and, for the CMB, are
directly related to different physical processes in the early universe \citep{kamionkowski/etal:1997,zal+seljak:1997} .
The E-modes correspond to a spin-two field with no curl and originate primarily from density perturbations in the early
Universe. This E-mode signal has been detected by a number of
instruments. The B-modes correspond to a spin-two field with no divergence and can originate from tensor-type
physical processes such as gravity waves that are predicted to have been generated by
an inflationary epoch as close at $10^{-35}$ seconds after the big bang.  To date the B-mode signal has not yet 
been discovered.  If the B-modes are of sufficient amplitude to be detected,
their impact on cosmology and physics would be enormous. Not only would the discovery significantly limit the number of models that
could describe the early universe, but it would mark the first observational evidence of gravity operating on a quantum scale.

At large angular scales, $\ell \ltsim 100$, B-modes may result from inflationary gravitational waves,
and from galactic foreground emission. At higher $\ell$ multipoles the primary contribution to the B-mode spectrum is
from E-modes being gravitationally lensed so that they produce a B-mode component. 
The level of primordial (or inflationary) B-modes, is quantified in terms of a parameter $r$,
the ratio of the variance of density perturbations to tensor perturbations. Predictions for $r$ vary over
many orders of magnitude. Currently observations give $r<0.21$ (95\%)~\citep{keisler/etal:2011}, a limit coming from {\it temperature} anisotropy
and other cosmological probes (rather than polarization)~\footnote{The inflation-generated gravity waves also contribute
to the temperature anisotropy and thus can be constrained by such measurements.}.
When translated into temperature units
in Figure~\ref{fig:pspec}, this  becomes a faint ${\cal B} \ltsim 150 \times 10^{-9}$~K for $\ell \sim 90$.  The experimental
challenge is to make accurate and precise polarization measurements at the level of few tens of nano-K.

\section{Ground- and Balloon-Based systems}
\label{sec:groundballoon}

In this Section we guide the reader through the set of considerations facing a designer of a CMB telescope. Once a
resolution is chosen, one must consider whether to use a reflective or refractive system, a combination,
or perhaps only a feed horn. The information we provide is informed by the history of the field. We focus on a number of core design
elements, some of which have found use in multiple
CMB experiments. Perhaps the largest difference in telescope design between CMB and other applications is that in CMB work the edge tapers on ambient temperature optics are kept low. We discuss the advantages and disadvantages of working on a balloon-borne platform. After the turn of the millennium
emphasis in the field turned toward large-throughput systems and polarization-sensitive experiments; both topics are discussed toward
the end of the Section.

\subsection{General Considerations}

Following from the sky toward the detector, CMB optical hardware typically includes some or all of the
following elements: reflectors, lenses, band defining filters, amplifiers, detectors, 
and feed horns or antennas. The order of some of these elements may vary somewhat.  Intermediate
elements, such as a vacuum window are generally not considered part of the optical system although they do have to be
considered as an optical element during the design of the system.

As discussed in Section~\ref{sec:introduction},
the throughput is an overall measure of the amount of light that can be collected by the optical system. However, for a
given aperture size, and therefore for telescopes with the same resolution, the performance of the optical system is more usefully
measured in terms of the `diffraction limited field of view' (DLFOV),
which is that portion of the focal surface across which the optical performance is diffraction limited.
Two related measures are commonly used to determine whether the optical performance is diffraction limited: the Strehl ratio and
the {\it rms} wavefront error.
A system that provides a Strehl ratio larger than 0.8 ({\it rms} wavefront error that is less
than $\lambda/14$) at a particular
field point is generally considered diffraction limited at that field point. The minimum DLFOV is the area on the focal plane with a Strehl ratio
larger than 0.8.  However, some optical systems are optimized for higher Sterol ratios and therefore define the DLFOV as a smaller area enclosing a higher value.

The large majority of balloon and ground based CMB telescopes until the early 2000s illuminated the sky with either
feed horns or employed a combination of feed horns and reflective optical systems.
At the end of this article we present a table of instruments that have published measurements of
the primary CMB polarization or anisotropy. In the
simplest feed-horn-only system, the far-field beam shape, its angular size, and the band-width are determined by the shape of
the feed-horn and the wave-guide components attached to it. The theory and design of feed-horns of various types is quite mature,
is discussed in a number of publications, and is subject of ongoing research~\citep{clarricoats/olver:1984,olver/etal:1994,balanis}.
We can use an approximate empirical relation for corrugated feeds, $\theta_{1/2,deg} = 90\lambda/D$, to show that
practical considerations limit optical systems that use only feeds to have coarse resolution.
A horn with a reasonably large effective aperture diameter of $\sim$15~cm produces a beam size of  $\sim6^\circ$ and $\sim1^\circ$
at 30 and 150 GHz, respectively. More complete studies show that it is difficult to produce a beam
with $\theta_{1/2}<2.5^\circ$ at 150~GHz with a corrugated structure of reasonable volume and tolerances \citep{lin:2009}.

\begin{figure}[htb]
\begin{center}
\includegraphics[width=0.45\linewidth]{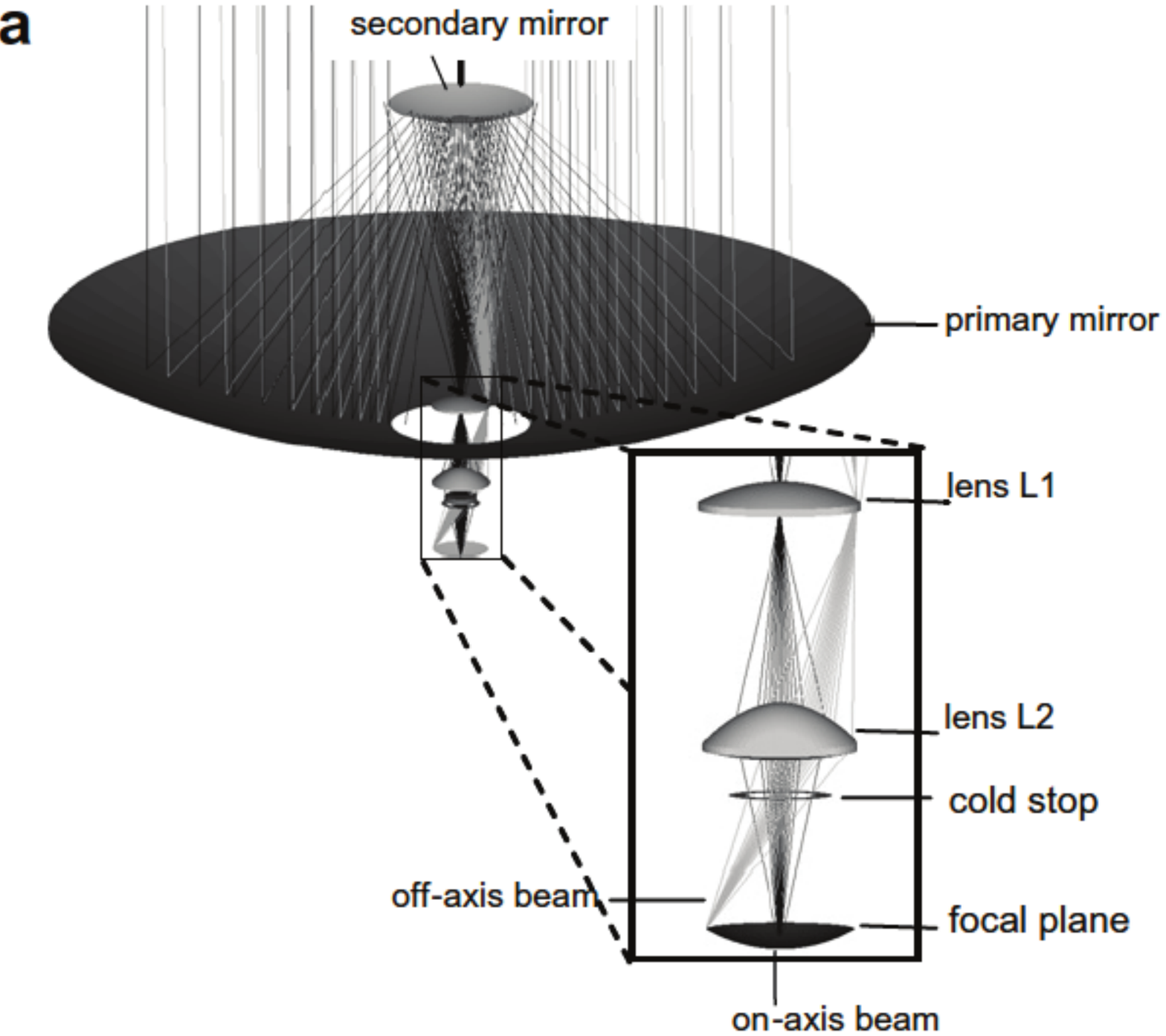}
\includegraphics[width=0.45\linewidth]{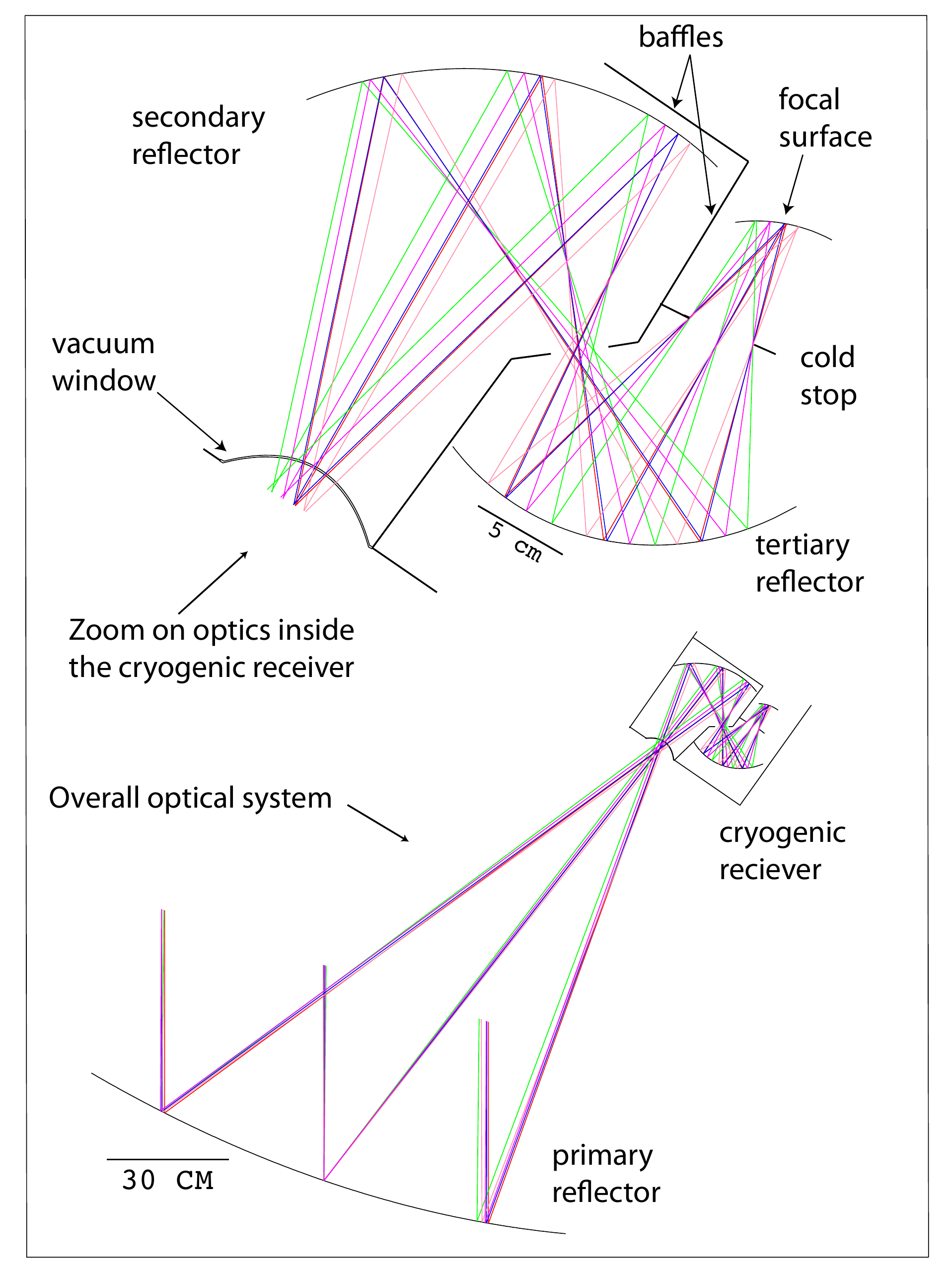}
\caption{\small Examples of centered (left) and decentered (right) CMB optical systems. The centered, two-reflector Cassegrain system
is that of the ground-based QUAD instrument (Figure from \citet{osullivan/etal:2008}).
The primary reflector diameter was 2.6~m. The optical elements after the secondary, shown in the inset are all
maintained at a temperature of 4~K.  The decentered, 3-reflector Gregorian system is that of the MAXIMA \citep{rabii/etal:2006}
balloon-borne experiment.
The primary is an off-axis section of a parabola with a diameter of 1.3~m. Two additional reflectors were maintained inside
the cryogenic receiver (only part of which is shown) at a temperature of 4~K; see zoom. \label{fig:centered} }
\end{center}
\end{figure}

There are several design choices for reflector-based systems. In a `centered' (or sometimes `on-axis') optical system the
reflectors are made from central portions of the typically conic sections of revolution that describe the surface shape. As a consequence,
such systems naturally have central obscurations. Examples include the White Dish telescope~\citep{tucker/etal:1993},
and the QUAD optical system~\citep{hinderks/etal:2009}, shown in  Fig.~\ref{fig:centered}, which consisted of a 1.2~m,
and 2.6~m parabolic primaries, respectively, and hyperbolic secondary in a Cassegrain configuration (QUAD also used lenses internal to the
cryostat as part of the optical train).

In a decentered (sometimes also called `off-axis', or `offset') optical system off-axis portions of the conic sections are used, the reflectors
are not centered on each other's
axis of symmetry, and there is no self obscuration by the reflectors. An example of a decentered optical system from the MAXIMA 
experiment~\citep{hanany/etal:2000} is shown in Fig.~\ref{fig:centered}.
For fixed entrance aperture diameter, centered systems are more compact compared to de-centered systems. However, they have
lower aperture efficiency and are more prone to scattering of radiation to side-lobes
caused by either diffraction from the edges of the central
obscuration or by beam-intercepting supports of the secondary reflector(s).
Decentered systems that have an intermediate focal point are typically easier to baffle compared
to centered systems, and are thus less prone to stray radiation. Most CMB telescopes to date use decentered systems;
see Table ~\ref{tab:experiments}.

Practical considerations limit the use of a refractive-only system. One is the weight of the lens. The density of ultra-high molecular
weight polyethylene, a popular material for millimeter wave lenses because of its low absorption, is 0.95 g/cm$^{3}$, making a 1~meter
diameter lens weigh around 100~kg.
In contrast, the 1.3 meter diameter MAXIMA reflector weighed 11~kg, and the 1.4 meter diameter
WMAP primary weighed only 5~kg. The Archeops primary, which was 1.8 m $\times$ 1.5 m, and was made of 6061 aluminum (rather
than specialty materials, as the former two examples), weighed less than 50~kg.
Another advantage of reflectors is their achromaticity.
Many CMB instruments operate with multiple frequencies simultaneously, which facilitates the discrimination of foregrounds sources from the
CMB signal. Eliminating the $\sim 4-30$\% reflection per refracting surface, for polyethylene and silicon, respectively,
requires anti-reflection coatings that operate
over a correspondingly broad range of frequencies, assuming that the different frequencies share the same light train.
In contrast, metal-based reflectors can have nearly unity reflectance between few MHz and several THz.
To date,  there have been less than a
handful of CMB refractive-only systems. They used polyethylene
lenses of $\sim$30~cm diameter
and hence operated at degree-scale resolution, and with one or at most two frequencies per optical train.
An advantage of a refractive-only system is that it typically provides a large DLFOV with no obscurations and in a
relatively compact package.

A common theme for all CMB telescopes is controlling and understanding the sidelobes.  Invariably
antenna patterns have non-vanishing response beyond the design radius of the optical elements. This ``spill-over" is quantified
by the edge-taper, which is the level of illumination at the edge of the optic relative to the center of the beam, typically quoted in dB.
The best practice is to ensure that such spill-over radiation finds itself absorbed on a cold surface with stable temperature. How cold and how stable?
Sufficiently cold such that the total power coupled through the spill-over is small compared to the total load on the
detectors, and stable compared to the time scales of the largest sky scans.  Through scattering, some of the spill-over radiation can
find its way to the sky, potentially coupling astronomical sources that are away from the main beam, or to the ground
causing spurious signals as the telescope scans the sky. Three techniques have
been used most widely to control sidelobes, sometimes in combination with each other: (i) Use of specifically-shaped feed horns to launch the
beams from the detector element into the rest of the optics (or to the sky).
As mentioned earlier, the theory of the beam patterns produced by feed horns
is quite developed and measurements are in good agreement with predictions.  Thus one can design the
antenna pattern to have only low levels of spill-over. (ii) Use of an aperture stop to control and define the illumination on the primary reflector.
A millimeter-wave black and often cold aperture stop is placed at a location along the optical path that has an image of the primary. Adjusting the
diameter of the stop effectively controls the illumination and edge taper on the primary optic. In addition, radiation on
the wings of the beam, at levels below the edge taper on the stop, is intercepted by cold surfaces.
(iii) Use of shields and baffles to absorb and redirect spill-over radiation.

\subsection{Balloon- vs. Ground-based Systems}
\label{sec:balloonground}

High instantaneous sensitivity is the primary driver to mount a CMB experiment on a balloon-borne platform.
This improved sensitivity over a ground-based system is a consequence of two distinct elements: lower atmospheric emission and higher
atmospheric stability. There is significantly less atmospheric power loading at balloon altitudes
compared to ground, leading to lower photon noise; see Figures~\ref{fig:spectrum_tant} and \ref{fig:noise}.
At 150~GHz the atmospheric loading is about 100 times smaller at balloon altitudes than on the ground.
This implies that if balloon-based systems control detector noise and emission from the
telescope, their fundamental noise limit could be the CMB itself.  As a generic example, assume a telescope
similar to MAXIMA's~\citep{rabii/etal:2006}
with one ambient and two aluminum reflectors cooled to 4~K.
For simplicity, let's assume a flat passband between 120 and 180 GHz. Since the instrument
operates in the single mode limit at an effective wavelength of $\lambda = 0.2$~cm,
the throughput per feed horn is 0.04~cm$^{2}$sr (equation~\ref{eqn:lambda2}). Thus the expected power from the CMB incident on the telescope is 0.9~pWatt~\footnote{1 pWatt = $10^{-12}$ Watt.}.  In comparison,
the power emitted by a 250~K ambient (at float) temperature reflector, is 0.6~pWatt.
This value assumes the 150 GHz bulk emissivity
of aluminum of 0.13\%. The {\it two} cold reflectors contribute together a negligible 0.01~pWatt.

This example illustrate several issues. Emission from even a single ambient temperature reflector on a balloon platform
is not negligible compared to the power from the CMB. Emission from a telescope with two ambient temperature reflectors,
as in for example EBEX~\citep{reichborn/etal:2010},
would give rise to loading higher than the CMB. These calculations assume the lowest aluminum
emissivity,  that is the bulk emissivity.  If the actual surface has an effective emissivity of 0.5\%, because of, for example,
a layer of contamination, the power produced by a single warm reflector is 2.1~pWatt, which is more than twice the power from the CMB.

An effective way to mitigate emission is to maintain optical elements at low temperatures. This
technique is particularly useful for optical systems with small apertures. It has been used with
the ground-based BICEP~\citep{aikin/etal:2010} experiment
and the balloon-borne FIRS~\citep{page/etal:1994} and Arcade~\citep{singal/etal:2011} instruments.
Other experiments  had one or two ambient temperature reflectors, but maintained subsequent optical components at cryogenic temperatures,
see for example Figure~\ref{fig:centered}.

In addition to higher photon loading, {\it variations} in atmospheric emission on the ground, essentially due to emission
from transiting clouds of water vapor, are a source of increased noise at low temporal frequencies.
The $1/f$ knee of this extra noise, that is, the low-frequency point at which the
higher frequency white noise level doubles,
varies with observing site on Earth and with specific atmospheric conditions.
The combination of low frequency noise and higher loading makes CMB observations
above $\sim$250~GHz difficult from anywhere on Earth.
Observations {\it at large angular scales} are
a challenge at almost any frequency because of the spatial structure of the atmosphere\footnote{Spatial
turbulence in the atmosphere is parametrized in terms of a Kolmogorov
spectrum~\citep{tatarskii:1961,church:1995,lay/halverson:2000}
that depends on the spatial wave number $q$ as either
$q^{-11/3}$ or $q^{-8/3}$ depending on whether the turbulent layer is three- or two-dimensional. Thus at large angular scale,
low $q$, atmospheric fluctuations can be quite large.}.
The only ground based experiment to report anisotropy measurements at angular scales larger
than $\sim8^\circ$ ($\ell \ltsim 20$) is Tenerife~\citep{davies/etal:1987,watson/etal:1992}, which operated at 10 and 15~GHz.
As shown in Figure~\ref{fig:spectrum_tant}, atmospheric emission
drops at lower microwave frequencies in part making the Tenerife measurement possible. The significantly lower atmospheric emission
on a balloon-borne platform and in particular the absence of water clouds, essentially eliminates atmospheric emission
as a significant source of low frequency noise.

The situation is different for measurements of the polarization of the CMB.
One may think of a polarization measurement of Stokes $Q$ or $U$ as the difference in intensities of two polarization
states\footnote{For bolometric systems on can simply imagine the difference of two intensity measurements.
For coherent or interferometric systems, the square law detector outputs the product of two differently polarized electric fields.}.
If the two measurements are done simultaneously, or quickly relative
to the time scale of atmospheric turbulence, the measurement is immune to fluctuations in atmospheric emission {\it if} the atmosphere
is not polarized.
Zeeman-splitting of oxygen in Earth's magnetic field~\citep{weiss:1980} polarizes atmosphere emission
near the strong oxygen lines at 60 and 118~GHz. However it has
been shown that the atmospheric linear polarization is negligible compared to the levels expected for
either CMB E or B-modes~\citep{keating/etal:1998}.
Circular polarization near the lines is significantly higher than linear~\citep{hanany/rosenkranz:2003}, and conversion of
atmospheric circular to linear polarization in the instrument is a source of concern.
Calculations of the effect of atmosphere polarization as a function of {\it spatial scale}
depend on knowledge of Earth's magnetic field over the corresponding spatial scales.
Available information~\citep{igrf2010} suggests that spatial variations
in the field are limited to large angular scales, roughly above $10^\circ$. Thus, experiments probing polarization anisotropy at smaller angular
scales should not be affected by atmospheric polarization. For this reason, there is a relative abundance of experiments probing polarization from
ground-based instruments at frequencies below 250~GHz. Despite these comments, we note that large angular scale polarization
($\ell \ltsim 30$, $\theta \gtsim 6^\circ$) has so far been measured only by the WMAP satellite.

While the higher instantaneous sensitivity on a balloon-borne platform is appealing, it comes with particular
optical-system and total integration-time trade-offs.
To date the largest CMB reflector mounted on a balloon-platform was the 2.2~m primary of the BEAST
experiment~\citep{meinhold/etal:2005}. It was made with carbon-fiber technology and weighed 8~kg. A similar reflector made
using aluminum would likely weigh
in excess of 50~kg.  Using significantly larger telescopes is challenging and thus
resolution has a practical upper limit for a balloon-borne platform. Although BEAST operated at 40~GHz with
$\theta_{1/2} = 23^\prime$, one can imagine using a similar telescope to achieve
$\theta_{1/2} = 6^\prime$ at 150~GHz\footnote{For example, the BLAST balloon payload~\citep{pascale/etal:2008}, which had frequency bands
between  600 and 1200~GHz, had a centered Cassegrain system with a 2~meter aperture primary providing a resolution of $30^{\prime\prime}$
at the highest frequency.}. To take full advantage of the balloon-platform, though, it would be
advantageous to cool the large reflector
to well below ambient temperature to minimize radiation loading of the detectors. 
So far, this has not been achieved.
Finally, the duration of a balloon flight is $\sim$1 day (for launches in north America) to $\sim$20 days (for two circumnavigations in
a long duration flight in Antarctica)~\footnote{Currently, record duration for a science payload in an Antarctic flight is close to 42 days~\citep{seo/etal:2008}.
NASA is developing capabilities for flights of 100 days.}, and a balloon experiment can typically be launched once every 1--2 years. In
contrast, ground-based observatories have the potential for significantly longer continuous integration times.

\subsection{Arrays of Detectors and Increases in DLFOV}
\label{sec:arrays}

Improvements in detector sensitivity throughout the 1990s placed new demands on optical systems. The sensitivity
of individual detectors approached the photon noise limit (see Figure~\ref{fig:noise}),
earlier for ground based experiments for which the atmosphere is a strong source of emission, and then even for balloon-borne payloads.
Improved experiment sensitivity could only be realized
by using arrays of detectors in the focal plane, and thus by increasing the DLFOV.
Small-sized focal plane arrays were used in the late 1990's
by MAXIMA (16 elements), BOOMERANG (16 elements, \cite{crill/etal:2003}), QMAP (6 elements, \cite{deoliveiracosta/etal:1998}) and
Toco (8 elements, \cite{miller/etal:1999}).
The detector elements were bolometers that used neutron transmutation doped germanium (MAXIMA, BOOMERANG),
high electron mobility transistors, or HEMTs (QMAP, Toco), and superconductor-insulator-superconductor, or SIS, mixers (Toco).
The push for large focal plane
arrays accelerated in the early 2000, after the launch of WMAP, and when Planck was slated for launch later in the decade.
The main focus in CMB studies evolved toward measurements of the faint polarization of the CMB and of small scale anisotropy,
and it was broadly recognized that significantly higher mapping speeds can only be achieved by implementing large arrays of detectors.

To achieve higher throughput, CMB telescope designers turned
to on-axis refractive systems and to new reflective optical systems that included additional corrections of aberrations.
In addition, modern detector arrays with hundreds to thousands of elements are primarily based on superconducting technologies (such as transition-edge sensor (TES)
bolometers, e.g., \citet{irwin/hilton:2005}), and are thus fabricated on flat silicon wafers.
This requires optical systems to have flat focal planes and, depending on the coupling to the detectors, can require
that the focal planes are image-space telecentric,
namely that the focal surface is perpendicular to all incident chief rays.
The use of large focal plane arrays also prompted designers to be more conservative in
their definition of DLFOV and strive for Strehl ratio values higher than the accepted minimum of 0.8.
Higher Strehl ratios are generically associated with smaller beam asymmetries, which
simplifies data analysis by reducing variations in the window functions across the focal plane array. What Strehl
value should one strive for?  We are not familiar with a quantitative study of the effects of minimum Strehl ratio on specific
beam asymmetry nor its effects on the systematic error budget for any experiment. Designers of large arrays
have generally attempted to provide for Strehl ratios larger than 0.9.

The advent of large focal plane arrays has stimulated research and innovation into the specific coupling of the electromagnetic
radiation into bolometric TES arrays. A variety of coupling approaches have been developed, including: feedhorn, contacting or immersion
lens,
phased-antenna, and filled focal plane arrays (e.g.,~\cite{padin/etal:2008,obrient/etal:2008,kuo/etal:2008,niemack/etal:2008,yoon/etal:2009}).
The optical coupling of the detector array can have a substantial impact on the total system throughput and can drive critical optical design decisions.
Here we provide a brief overview of the tradeoffs between filled and feedhorn-coupled focal planes following the
analysis of~\cite{griffin/bock/gear:2002}. Section~\ref{sec:largeground} provides an example comparison between
the optical coupling techniques used in the ACT and SPT focal planes.

For an instrument with a fixed DLFOV that is fully populated with detectors, a filled focal plane array of bare
detectors can provide $\sim3$ times faster mapping speed of an extended source (like the CMB) and
with $0.5 F \lambda$ spacing\footnote{Note that parameterizing the detector spacing or the feedhorn aperture in units of the focal
ratio, $F$, times the wavelength, $\lambda$, provides sufficient information to approximate the aperture efficiency, spillover efficiency,
and other relevant optical quantities for estimating the mapping speed.} $\sim3.5$ times faster mapping speed of point sources
than a feedhorn coupled array with $2 F \lambda$ spacing. For an instrument that is limited by readout technologies to a fixed
number of detectors, feedhorn coupled arrays can provide the fastest mapping speed by increasing the FOV until the feedhorns
are spaced by $\sim 2 F \lambda$. In practice, large detector arrays to date have generally operated between these extremes
with feedhorn coupled arrays spaced between $1$ to $2 F \lambda$ and filled arrays spaced
between $0.5$ to $1 F \lambda$.

The differences in instrument design requirements for filled focal plane arrays compared to feedhorn (or other beam forming element)
arrays are substantial. The Gaussian illumination of feedhorns provides well understood beam properties (and potentially
single-moded coupling), strong rejection of stray light from sources outside the main beam, and the horn itself provides a Faraday
enclosure around the detector. In contrast, filled detector arrays are exposed, can couple to radiation from approximately $\pi$~steradian,
and therefore require a cryogenic stop cooled to $\sim1$~K to prevent blackbody radiation within the cryostat from dominating the detector noise.
Individual detectors in a filled focal plane are smaller and have lower optical loading than feedhorn coupled detectors, making it more
difficult to achieve photon limited noise performance and requiring readout of more detectors. The great potential advantage of filled arrays
is of course the opportunity to maximize the throughput and mapping speed of detector arrays that fill the available field of view.

Another approach to maximizing the use of the DLFOV is to increase the optical bandwidth of the optics and have multiple detectors
operating in different frequency bands within each focal plane element.  This can be achieved by use of dichroic beam splitters
in the optical path that illuminate independent detector arrays, or by optical coupling techniques being developed to make
``multi-chroic'' detectors by separating the frequency bands using superconducting filters integrated into the detector array
(e.g.,~\cite{obrient/etal:2008,schlaerth/etal:2010,mcmahon/etal:2012}). Advantages of multi-chroic detector arrays compared to
dichroic beam splitters include: more compact optics designs, fewer optical elements, a smaller detector array footprint, and the ability to fill the entire DLFOV in high optical throughput systems.

\subsection{Refractor-Based Optical Systems}
\label{sec:refractor}

\begin{figure}[htb]
\begin{center}
\includegraphics[width=4.5in, angle=270]{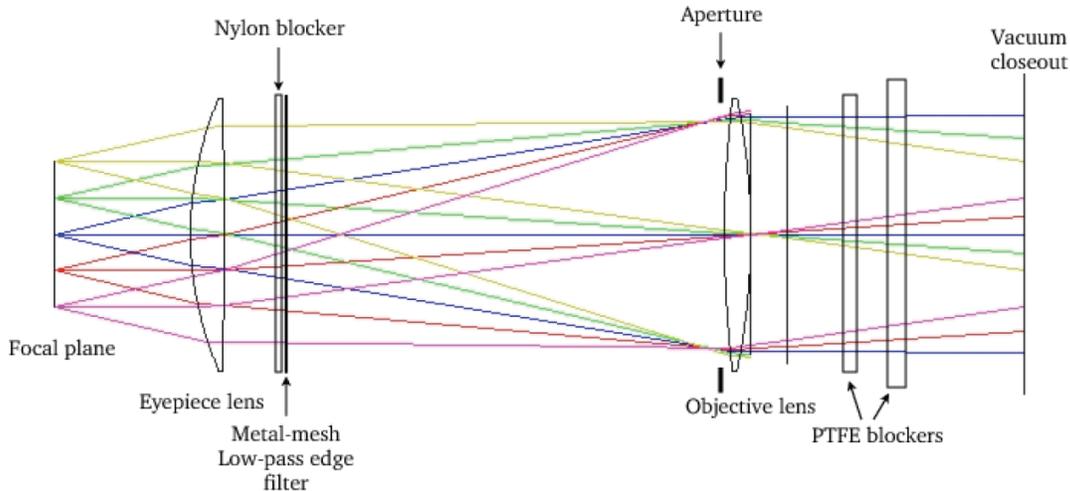}
\caption{\small The all-refractive optical system of BICEP (Figure is from \citet{aikin/etal:2010}). The vacuum window is on the right. The
two lenses and other filters are all maintained at a temperature of 4~K. \label{fig:bicepoptics} }
\end{center}
\end{figure}

In refractive
systems, the absence of a central obscuration, combined with the on-axis nature of the optics provides large easy gains in DLFOV.
For example, a NASA study for a CMB polarization space-based mission~\citep{bock/epic:2008} presented an optical design that
had an aperture of 30~cm, total throughput of 40~cm$^{2}$sr, Strehl ratio larger than 0.99 over a FOV of $15.3^\circ$ in diameter
and resolution of $0.57^\circ$ at 135~GHz. The optical performance was achieved with two polyethylene lenses that were pure conic sections.
To our knowledge, BICEP was the first CMB experiment to {\it implement} a refractive-only system; the focal
plane was indeed telecentric, as shown in Figure~\ref{fig:bicepoptics}. The throughput was 34~cm$^{2}$sr~\footnote{Throughput calculations
assume Equation~\ref{eqn:aomegatop} taking the entrance aperture diameter and the total FOV available by the optical
system of the experiment. A significantly smaller throughput
value can be obtained by taking the throughput per detector element and multiplying by the number of detectors implemented. We opt for
the first version because our primary interest in this chapter is in the overall optical design independent of the choice
of detector spacing on the focal plane.} and it had resolution of $0.6^\circ$ at 150~GHz~\citep{yoon/etal:2006}.
The optical design approach of the BICEP system is also used in the ground-based Keck-Array~\citep{sheehy/etal:2010}
and the balloon-borne SPIDER~\citep{filippini/etal:2010}
instruments. They use 5 and 6 independent receivers, respectively, each of which is a fully refractive optical system similar to BICEP's,
to increase the total throughput of the entire system.

To achieve a combination of high throughput  {\it and} high resolution other experiments have
implemented a combination of reflective fore-optics (e.g., primary and secondary reflectors) and refractive back-optics. The
roles of the lenses is to further correct aberrations and to produce telecentricity. Cases in point are EBEX,
Polarbear~\citep{arnold/etal:2010}, ACT, and SPT.

\subsection{Reflector-Based Optical Systems}
\label{sec:reflector}

The large majority of CMB telescopes to date are based on decentered reflecting
systems~\footnote{\citet{korsch:1991} and~\citet{love:1978} give thorough reviews of reflecting optical systems.}.
Most designers have found that systems with up to three powered reflectors\footnote{The term ``powered" reflectors refers to reflectors
with focusing properties, rather than flat reflectors used to only fold the path of the beam.} are sufficient to provide the DLFOV necessary for
their experiments. Minimizing the number of reflectors makes the system more compact, simpler to assemble, and easier to analyze in terms
of mis-alignment errors.

With essentially any ray tracing program it is fairly straight forward to show that the DLFOV produced by a single reflector -
typically, but not always an off-axis section of a parabola - is rather limited. The primary limiting aberration is
astigmatism \citep{chang/prata:2005}. Nevertheless a combination of feed horn and a single reflector
was used with the Saskatoon experiment and later for QMAP and TOCO, which used the same
telescope configuration and managed to pack 6 and 8 element array of detectors in the focal plane, respectively.
With two reflector systems, the classical configurations are either a Cassegrain (parabolic
primary and hyperbolic secondary) or a Gregorian (parabolic primary and elliptical secondary). We don't know of strong preferences between
the two systems from an image quality point of view. However the Gregorian system offers advantages in that the intermediate focus point
between the primary and secondary reflectors is a convenient location for baffling (see for example the decentered system in
Figure~\ref{fig:centered}), and that for a fixed aperture size the Gregorian has been
shown to be more compact~\citep{brown/prata:1994}. Likely for these reasons most CMB experiments that use reflectors opt for the
Greogrian configuration.

The primary source of aberrations in both Gregorian and Cassegrain systems are astigmatism and coma. Therefore a good
design starting point are those designs that provide some cancelation of either of these aberrations, at least at the center
of the FOV. In an aplanatic Gregorian, making the primary slightly
elliptical and the secondary a more eccentric ellipsoid than in the classical Gregorian cancels coma.
The Planck optical system is based on an aplanatic Gregorian design.

Dragone has described designs that cancel astigmatism (D1), and both astigmatism and coma (D2) at the center of the FOV of either
Gregorian or Cassegrain systems~\citep{dragone:1982,dragone:1983}. In both cases Dragone uses the concept of a single equivalent
paraboloid that has the same antenna pattern properties (at the center of the FOV) as the two reflector system. He finds modifications
to this equivalent system that cancel aberrations and then translates these modifications back to the two reflector system. In a
Gregorian D1 system the modification is conceptually simple, it is a relative tilt between the
axes of symmetry of the secondary ellipsoid and the primary parabola.\footnote{\citet{graham:1973} first suggested introducing such a
tilt in a decentered Cassegrain telescope to eliminate the cross polarization introduced by the asymmetrical configuration of the two
reflectors. Subsequently \citet{mizuguchi/yokoi:1974} and~\citet{mizuguchi/yokoi:1975}, and later
others (e.g.~\citet{mizuguchi/ak/yokoi:1976,mizuguchi/ak/yok:1978,dragone:1978}), made the idea
more quantitative, expanded it to a Gregorian system, and to a system with more than 2 reflectors. (Some papers are published with Mizugutch in place of Mizuguchi.)
The primary motivation in these studies remained the elimination of cross-polarization
at the center of the FOV. Thus a decentered Cassegrain or Gregorian optical system with a tilt between the axes of
symmetry of the primary and secondary is sometimes referred to as a ``Mizuguchi-Dragone telescope." In a series of publications
in the early 1980's Dragone analyzes {\it aberrations} in decenetered reflecting system. It so happens that
the tilt that cancels cross-polarization also cancels astigmatism. \label{footnote:mizuguchi} }
In D2 Dragone derives additional corrections to the shape
of the reflectors such as to cancel coma. Variants of D1 have become a popular starting point
for several CMB designers. \citet{hanany/marrone:2002} compared the performance of several optical designs including the
classical Gregorian, the aplanatic
Gregorian, the D1 and the D2 and showed that the D2 provides the largest DLFOV.

To our knowledge, ACME was the first telescope based on a Gregorian Dragone design~\citep{meinhold/etal:1993}.
The reflectors for EBEX and Polarbear are exact D1 designs. The WMAP design started from a D1 design and Planck's design is
inspired by D1 in that it is an aplanatic but with the addition of the D1 tilt between the symmetry axes of the primary
and secondary reflectors.

A third reflector, or equivalently refracting elements, are typically added to the base two-reflector system to achieve additional
requirements of the optical system, such as to produce a distinct aperture stop, to further cancel aberrations and thus increase the DLFOV,
or to make the focal surface telecentric. In MAXIMA, a cold aperture stop is placed at an image of the primary that is formed by the
tertiary reflector; see Fig.~\ref{fig:centered}. In BOOMERANG the tertiary reflector {\it is} the aperture stop.
In ACT, EBEX, and Polarbear, lenses that reimage the primary
form the cold stop. In all cases the cold stop is used to control the illumination on the primary reflector.

A useful property of the D1 and D2 designs is that they cancel cross-polarization at the center of the field of view.
This property has been pointed out by earlier authors (see footnote~\ref{footnote:mizuguchi}).
Minimum cross polarization is useful for telecommunication systems that can double
the bandwidth by using two distinct polarizations with a single antenna. It is also useful for polarization
sensitive CMB experiments, but this usefulness is limited as will be discussed in the next Section.

\begin{figure}[htb]
\begin{center}
\includegraphics[width=5.0in]{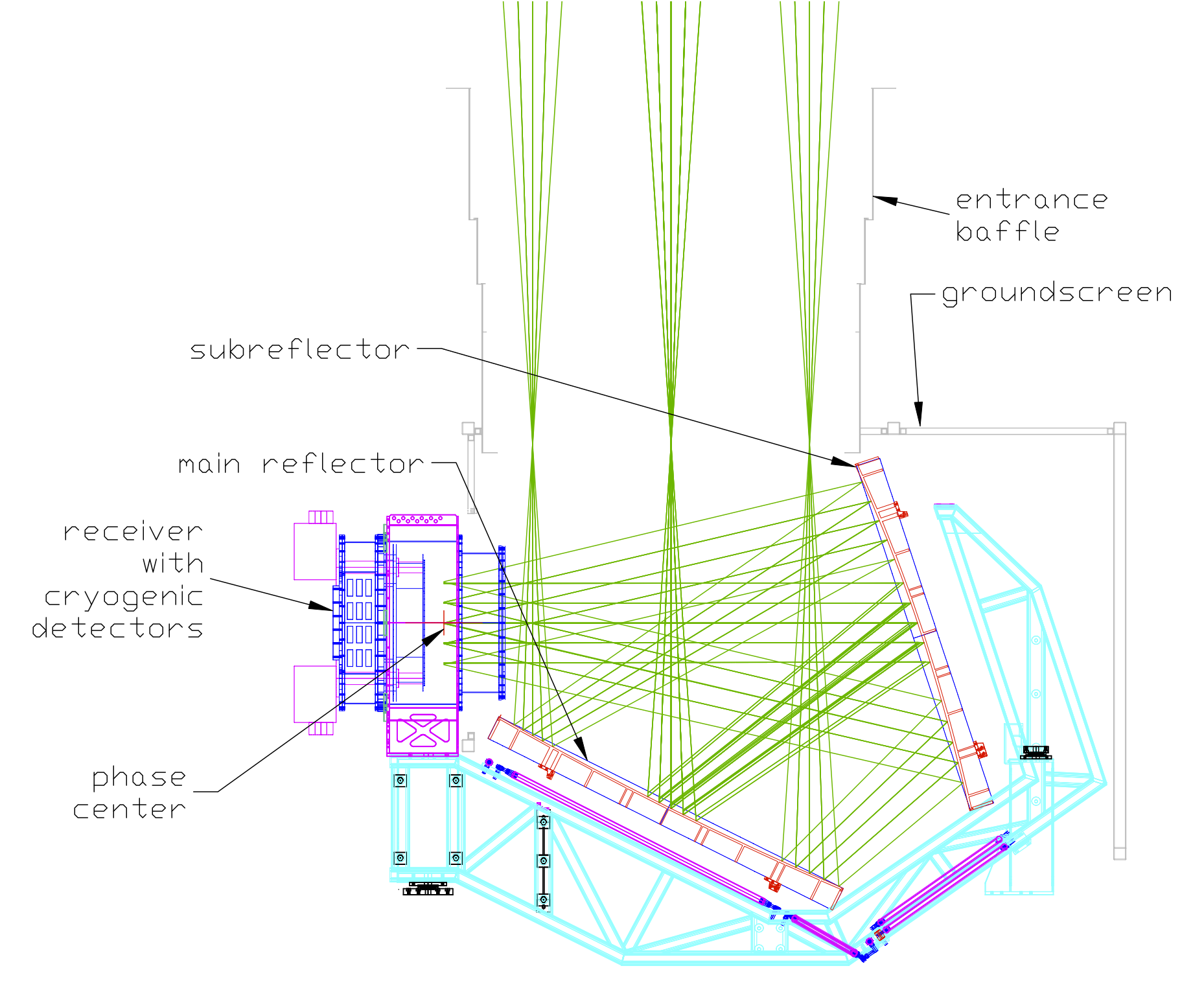}
\caption{\small The crossed-Dragone telescope of the QUIET experiment. The primary reflector is 1.4~m in diameter; in the ray limit it is the
limiting aperture in the system. The system was designed
using physical optics by propagating beams from the focal plane horns to the sky. An apparent entrance aperture is
formed  sky side of the primary, even though there is no physical aperture stop there.
The entrance baffle intercepts side-lobes
that are inherent to the cross-Dragone design, as described in the text.   \label{fig:quiet} }
\end{center}
\end{figure}

An innovative two reflector system that provides a particularly large FOV is described by \citet{dragone:1983b}.
This system, which has become known as the `Crossed Dragone'  configuration~\footnote{This configuration has
also been used with compact antenna test ranges since the 1980s~\citep{olver:1991}.},
began to be adopted by some CMB experimenters in the mid-2000. However,
the first use of the general configuration for the CMB group was by the IAB
experiment \citep{piccirillo:1991,piccirillo/calisse:1993}.
The crossed-Dragone system has a parabolic primary and hyperbolic secondary and is thus essentially a Cassegrain system. It cancels
astigmatism, coma, and spherical aberrations at the center of the field (the spherical aberrations are produced because of the
correction for coma). An exact implementation of the design is shown in Figure~\ref{fig:quiet}. 
Variants of the crossed Dragone system and detailed design procedures are described by~\cite{chang/prata:1999}.
In addition to the large DLFOV provided by this system, the focal plane is nearly telecentric and the cross-polarization
is small. \citet{tran/etal:2008} compared the performance of the Gregorian Dragone (of the D1 variety) and crossed Dragone systems. They
found that the DLFOV provided by the crossed Dragone is about twice that provided by the Gregorian Dragone.

A crossed-Dragone system was developed for CLOVER~\citep{piccirillo/etal:2008}, a UK-based CMB experiment that has since
been canceled. It was also proposed as an alternative optical system in an initial concept study, called EPIC, for a future CMB
polarization satellite~\citep{bock/epic:2008}. In a subsequent round the team had baselined the crossed-Dragone as its flagship
design~\citep{bock/epicim:2009,tran/etal:2010}.   A very large throughput of 908~cm$^{2}$sr was obtained at 100~GHz with an aperture diameter
of 1.4~m. The ambitious design accommodated a total of 11,000 detectors in an elliptical focal plane with long and short
axes of 1.5, and 1~m, respectively, and Strehl ratios larger than 0.8 over this entire, nearly flat focal surface.
The first CMB experiment that has made measurements with a crossed
Dragone system was QUIET~\citep{imbriale/etal:2011}; see Figure~\ref{fig:quiet}.
The telescope throughput was $\sim$530~cm$^{2}$sr with a 1.4~m primary reflector, and 0.22~deg resolution at 90 GHz.
The ground-based ABS~\citep{essinger-hileman:2011}, which operates at $\sim$145~GHz, uses smaller, 60~cm reflector primary;
both primary and secondary are maintained at LHe temperature.
In the ABS design the entrance aperture is somewhat sky-side of the primary reflector, rather than on
the reflector itself. This modification, which was later
adopted by the EPIC team, is used to better define the illumination on the primary reflector and to control side-lobes, with minimal
penalty in optical performance.

Despite's its appealing features, the crossed Dragone system has several challenges which are described in \citet{tran/page:2009}.
One challenge is that the system's aperture stop is at the very front effectively limiting
the resolution for systems in which this stop needs to be cold. Another challenge is the compactness of the system.
Both the secondary reflector and focal surface are very close to the incoming bundle of rays causing diffraction side lobes on
the edges of the reflectors and leading
to physical packing difficulties when implementing a cryogenic focal plane. \citet{tran/etal:2010} analyze the issue
of sidelobes in more detail in the context of the EPIC Mission Concept report.

\subsection{Polarization Properties}
\label{sec:PolarizationProperties}

Interest in measurements of the polarization of the CMB has motivated interest in the polarization properties of telescopes and other
optical components. This is a relatively new area of research, for which only limited experience is available.

The susceptibility of an instrument to polarimetric systematic errors depends on the
experimental approach to the polarization measurements, and a detailed discussion is beyond the scope of this chapter. However,
considering the optical system alone one can generalize as follows: systematic errors are minimized if the antenna pattern
is completely independent of the polarization state being probed, if such an antenna pattern is independent of field position in
the focal plane, and if the incident polarization orientation is not altered by the optical system.
These desirables are never fully satisfied and a body of literature has evolved around quantifying the systematic errors
induced by non-idealities (e.g., \citet{hhz:2003,odea/etal:2007,shimon/etal:2008,meng/etal:2011}).

As a starting point, one models the
antenna patterns in each of the two polarization states $i=a, b$ as an elliptical Gaussian~\citep{hhz:2003}
\begin{equation}
B_{i}(A_{i}, \hat{n}, \vec{b_{i}}, e_{i}) =  A_{i} \, \exp \left[  - \frac{ 1 }{ 2 \sigma_{i}^{2} } \left( \frac{ (n_{1} - b_{i1})^{2} }{ (1+e_{i})^{2} }  + \frac{ (n_{2} - b_{i2})^{2} }{ (1-e_{i})^{2} } \right) \right],
\end{equation}
where $\vec{b}$ is an offset between the beam center and a nominal direction $\hat{n}$ on the sky, $\sigma$ is the
mean beam width, and $e$ is the ellipticity.
It has become common to quantify beam-induced errors in terms of differential gain $g$
\begin{equation}
g \propto \frac{ A_{a} - A_{b} }{ (A_{a} + A_{b})/2 }  ,
\label{eqn:relgain}
\end{equation}
differential pointing $\vec{d}$
\begin{equation}
\vec{d} \propto \frac{ \vec{b_{a}} - \vec{b_{b}} }{2} ,
\label{eqn:relpoint}
\end{equation}
differential ellipticity
\begin{equation}
q \propto \frac{ e_{a} - e_{b} }{2} ,
\label{eqn:relellip}
\end{equation}
and cross polarization.
We use the `proportional to' notation above because authors differ on the normalization of the different quantities. The physical interpretation
of these errors is straight forward. Each of the $Q$ and $U$ Stokes parameters that quantifies the polarization content of incident radiation
is formed by differences of intensities between two orthogonal polarization states. A `differential gain' systematic error arises when the overall
beam size, or the antenna gain is different between any two orthogonal states.\footnote{Differential gain also arises when other factors in the system
affect gain between two polarization states, for example when two independent detectors that are sensitive to the two polarization states have different
responsivities.}  `Differential pointing' and `ellipticity' arise when there is a difference in the centroid of the beams,
and their ellipticities, respectively. It is
not difficult to see that respectively these effects couple the temperature anisotropy, its gradient and its second derivative into the
polarization measurement (see also~\citet{yadav/su/zal:2010}). These errors are of the general category of `instrumental polarization',
in which unpolarized radiation becomes
partially polarized by the instrument. It is a different category from cross-polarization, also called `polarization rotation', which acts only on polarized
light, and its physical effect as far as the CMB is concerned is to mix E and B-modes (see Section~\ref{sec:polarizationterminology}).
All of the errors should be minimized through the design of
the optics, or calibrated with an accuracy commensurate with the goal of the experiment.

Both instrumental polarization and polarization rotation occur because reflection and absorption, and hence transmission, of light depend on
the polarization state of the incident radiation, on the angle of incidence, and, where relevant, on the materials making
lenses or other optical elements (such as a half-wave plate, vacuum window, or reflectors). As an example, consider the
effects of a standard $\sim$0.05~mm thick vacuum window made of $\sim$9~cm diameter polypropylene,
which has an index of refraction close to 1.5. This will have high mm-wave transmission. Such a window, which is similar
to the one used on the MAXIPOL polarimeter, is naturally bowed by few cm into the cryostat
because of differential pressure.  MAXIPOL had an array of 16 photometers with an intermediate focus near this window and the angle of
incidence of rays for an edge photometer spanned 20 to 55 degrees.
Differential reflection between the two polarization states is about 1\%, which generates
a 1~$\mu$K spurious, scan-synchronous polarized signal from the $\sim$100~$\mu$K {\it rms} CMB anisotropy. This spurious signal was of no
consequence for the analysis of the MAXIPOL data, but is one example of an instrumental polarization
that may be a contaminant for future, higher sensitivity experiments.
A much larger signal of $\sim$12~mK is generated by the 0.5\% emissivity
of the 250~K primary reflector~\footnote{A 1.2 K nearly unpolarized signal is generated by 
the 0.5\% emissivity. This signal is differentially polarized at the 1\% level.}. 
However, if that signal is stable it can removed in analysis as an overall offset.
Of course, reflection from the asymmetric primary reflector, which is
an off-axis section of a parabola, also polarizes unpolarized light, but this effect is very small and, again, if the temperature is relatively constant this
overall polarized offset, or even if it is slowly drifting, can be removed in analysis.

As mentioned above, the systematic errors associated with the optics are either negligible compared to statistical errors in the measurement, or they
need to be calibrated with an accuracy that makes them negligible. The faintness of the B-mode signal makes it susceptible
to the various systematic errors and this is our focus in the following discussion. In terms of absolute magnitude,
the most challenging effects are differential gain and polarization rotation.
Differential gain is challenging because it couples the temperature
anisotropy directly into the polarization measurement\footnote{The 2.7~K CMB monopole can also lead to
a polarized signal through instrumental polarization.
However the magnitude of this signal is a constant across the observation and since essentially all CMB polarimeters are differential, they are not
sensitive to this overall offset.}. A residual differential gain of less than 0.1\% is necessary if the leakage from temperature anisotropy is to be
less than 10\% of the B-mode predicted power spectrum with $r=0.1$~\citep{zaldarriaga:private}.
Because polarization rotation can mix E and B-modes, an uncorrected rotation of $0.3^\circ$
of the incident polarization by the instrument gives rise to a spurious B-mode that is a factor of 10 below the B-mode signal from lensing. This is the
level that should reasonably be targeted by an instrument that intends to {\it detect} the B-mode from lensing and not
be limited by this systematic effect.
A rotation of $1.3^\circ$ ($0.4^\circ$) gives rise to a spurious B-mode that is a factor of 10 below the cosmological Inflationary
signal with $r=0.1 \, (0.01)$~\citep{zaldarriaga:private}.

The cross Dragone and the on-axis refractive-only systems appear to provide the lowest levels of the systematic errors discussed above.
The EPIC team has quantified the expected level of instrumental polarization in their proposed crossed Dragon
telescope~\citep{tran/etal:2010} and showed that
the telescope nearly satisfies mission goals.  Exceptions included differential gain, which is apparently caused by
differential attenuation due to the finite
conductivity of the aluminum surface, and differential rotation, which was up to $0.6^\circ$  at the edge of the focal plane~\citep{johnson:private}.
Therefore both would need to be calibrated and corrected if the mission is to achieve its goals of setting a limit on $r=0.01$.
\citet{imbriale/etal:2011} show measurements from QUIET's crossed Dragone that demonstrate the absence
of cross polarization at the center of the FOV.

A thorough analysis of beam-induced systematics for a refractive system was presented in conjunction with the
release of data from BICEP~\citep{takahashi/etal:2010}. All measured systematic effects were smaller than benchmarks
calculated for $r=0.1$. Many measured
effects were also acceptable for $r=0.01$ with the two exceptions: `relative gain' and differential
pointing\footnote{Although \citet{takahashi/etal:2010}
define relative gain slightly differently than in Equation~\ref{eqn:relgain} the underlying physics is the same.}. The origin of
the relative gain mismatch is likely the detection system and not BICEP's centered refractive optics.
The observed differential pointing is an intriguing effect also observed by the QUAD
experiment~\citep{hinderks/etal:2009}. The centroids
of beams corresponding to two polarization sensitive bolometers, which {\it do} share the same optical path, were offset relative to each other.
The teams speculated birefringence in the high density polyethylene lenses, among other causes~\citep{pryke:private}.
The effect was not of any consequence for the
analysis of the experiments' data, and the BICEP team reports sufficient sensitivity to remove the effect if it had been necessary.
If it persists, then future refractive telescopes searching for $r$ values smaller than
about 0.1 will also need to characterize the effect and subtract it during the analysis of the data.

\citet{tran/etal:2008} compare the polarization performance of the cross- and Gregorian systems. They show that in every respect the
cross-Dragone has superior polarization properties compared to the Gregorian system. For example, at a field point $3^\circ$  away from
the center of the focal plane of the Gregorian-Dragone system the cross-polar response is -25~dB
below the co-polar response, whereas for a similar aperture cross-Dragone the response is -50~dB. We note, however, that once
the need arises to calibrate the effects induced by the optical system, the difference in performance between the Gregorian and crossed
systems may become inconsequential.

\section{Large ground based telescopes, ACT and SPT}
\label{sec:largeground}

\begin{figure}[htb]
\begin{center}
\includegraphics[width=5.0in]{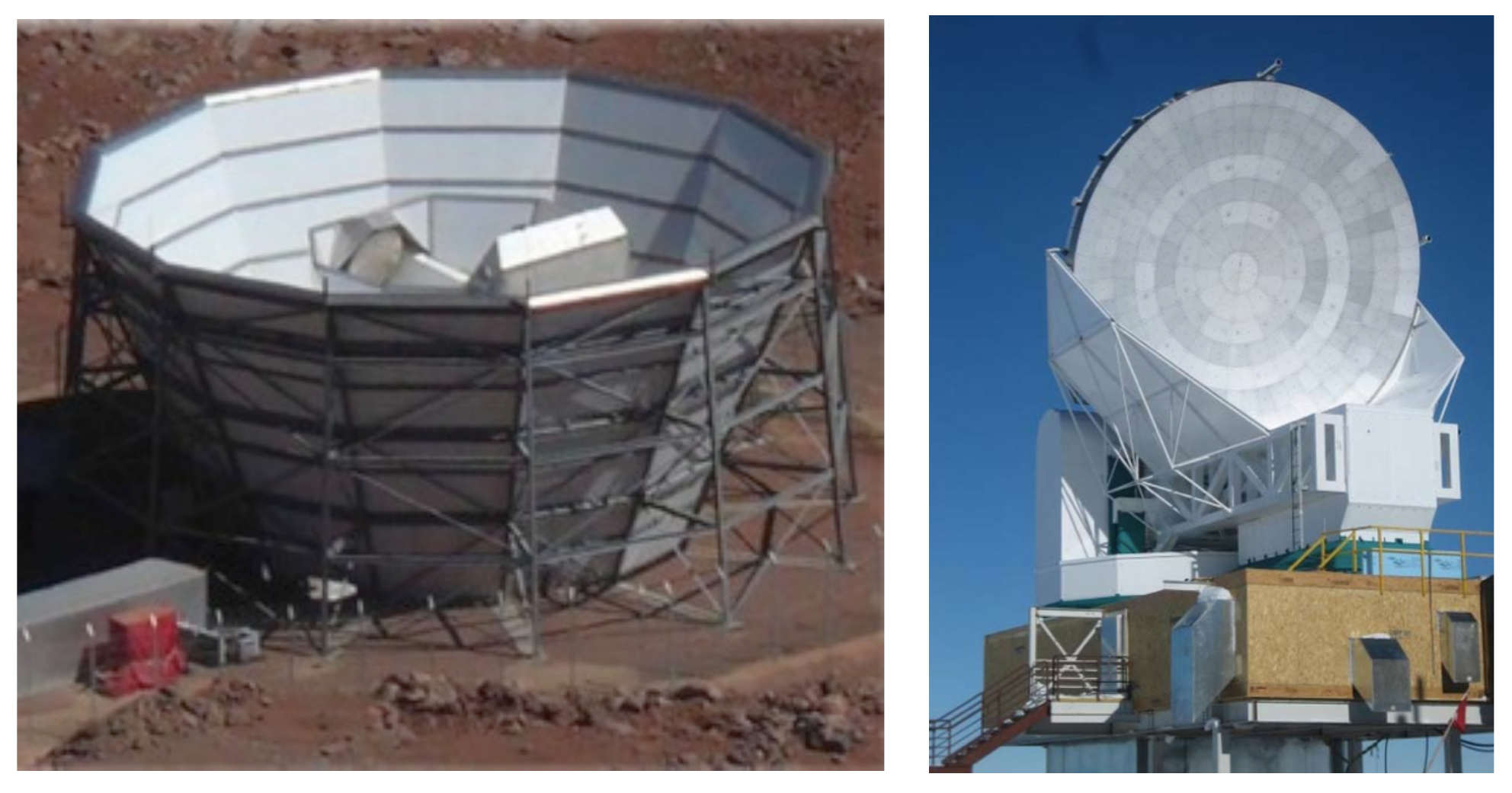}
\caption{\small {\it Left}: The ACT 6 m telescope in northern Chile. The telescope is inside the 13 m tall ground screen. The secondary is just visible near the center of the ground screen.  {\it Right}: The SPT 10 m telescope at the South Pole. Most of the circular primary reflector is visible, while the cryogenic secondary reflector and receiver are housed inside the white structure below and to the right of the primary. 
{\it \small (Photos courtesy of ACT and SPT Collaborations.)} \label{fig:ACT_SPT_photo}}
\end{center}
\end{figure}

With six-meter and ten-meter primary reflectors, the ACT \citep{fowler/etal:2007,swetz/etal:2011} and SPT \citep{carlstrom/etal:2011}
(Figure~\ref{fig:ACT_SPT_photo}) are currently the largest CMB survey telescopes providing the highest resolution. Reviewing them
in more detail is instructive because the designs incorporate the state of optics knowledge for ground based instruments as of mid-2000.

There are a number of reasons to probe the CMB with high resolution. As can be seen in Figure~\ref{fig:pspec},  arcminute resolution, and hence
a window function that extends to $\ell>5000$,  is required to determine the contribution of point
sources and other secondary sources of anisotropy (not shown)
to the damping tail of the primary CMB anisotropy.  It is also advantageous to have minimal change in the beam over
$\ell<3000$, the region where the primary anisotropy dominates.  Even with its six-meter primary,  the ACT beam suppresses the
fluctuations by a factor of three in power by
$\ell=6000$. Lastly, one can discover galaxy clusters via the thermal Sunyaev-Zel'dovich (SZ) effect \citep{sunyaev/zeldovich:1980} -- the
inverse Compton scattering of CMB photons with hot electrons in ionized gas -- which requires $\sim$1~arcminute resolution in bands near 150 GHz.

The design of the ACT \& SPT optical systems are shown in Figures~\ref{fig:ACT_SPT_opt} and~\ref{fig:ACT_SPT_rec}. They are both
based on off-axis Gregorian telescopes. The SPT secondary mirror is maintained at cryogenic temperatures and an additional lens focuses
the light onto the telecentric focal plane. In ACT the secondary mirror is maintained at ambient temperature. At the entrance to the
cryogenic receiver the focal surface provided by the
the Gregorian telescope is split into three distinct light trains, each dedicated to a single band of electro-magnetic frequencies.
An off-axis Gregorian telescope was chosen for both systems because  of the guiding principles listed in Table~\ref{tab:ACT_SPT_guiding},
which are also discussed in Section~\ref{sec:reflector}.  In addition,  it is easier to implement
co-moving ground shields and to minimize spillover to the surroundings on the sides of an off-axis Gregorian.

\begin{table}[htdp]
\begin{center}
\begin{tabular}{l}
\hline
$\bullet$\ Clear aperture (off-axis optics) to minimize scattering and blockage.\\
$\bullet$\ Fast primary focus to keep the telescope compact and enable fast scanning.\\
$\bullet$\ Large (FOV $\sim 1^\circ$) and fast ($F\sim 1$) diffraction-limited focal plane.\\
$\bullet$\  Space for structure and a cryogenic receiver near Gregorian focus.\\
\hline
\end{tabular}
\caption{\small \label{tab:ACT_SPT_guiding}
Guiding principles of the ACT and SPT designs.}
\end{center}
\end{table}

\subsection{Detailed Optical System Comparison}

\begin{figure}[htb]
\begin{center}
\includegraphics[width=5.0in]{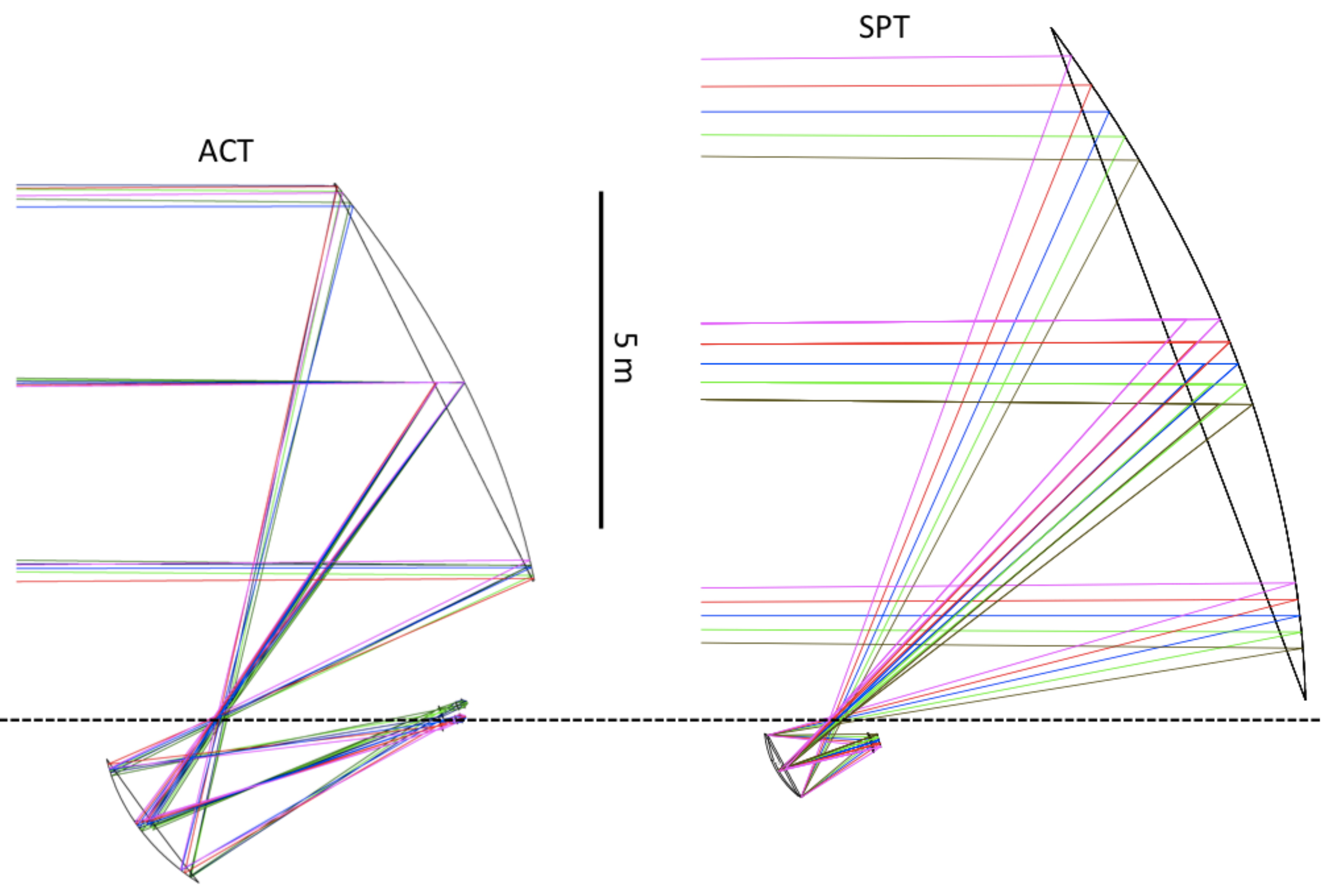}
\caption{\small Ray trace diagram of the ACT and SPT optical systems.  Figure~\ref{fig:ACT_SPT_rec} shows an enlarged view of the parts
of the systems that are maintained at cryogenic temperatures. The distribution of rays across the primary reflectors is
indicative of the proximity of the cold stop to an image of the primary reflector. Having a stop at an image of the primary in ACT enables
more uniform illumination of the primary (Table~\ref{tab:ACT_SPT_compare}). The dashed line shows the conic axis of the primary reflector
of both telescopes, which is also the conic axis of the secondary reflector for ACT. The SPT secondary reflector conic axis is tilted by $25.3^{\circ}$
to minimize aberrations and cross-polarization and maximize the FOV, which also moves the Gregorian focus below the
primary axis, providing space for the receiver.  The lower edge of the ACT primary is further off-axis than SPT, which provides space for
the receiver without tilting the secondary reflector.  (SPT ray trace courtesy of N. Halverson.)
\label{fig:ACT_SPT_opt}}
\end{center}
\end{figure}

The SPT is a standard Gregorian-Dragone design (D1, in the language of Section~\ref{sec:reflector}) \citep{padin/etal:2008}. The SPT Gregorian
configuration and the primary diameter, parabolic shape, and surface accuracy (20 $\mu$m rms) were designed to accomplish a wide
range of millimeter and sub-millimeter science goals.  The initial ACT design
started from a standard Gregorian-Dragone (D1), however, the conic constants of both ACT reflectors were then numerically 
optimized to maximize the DLFOV
area by minimizing the transverse ray aberration across the focal plane~\citep{fowler/etal:2007}\footnote{We note that this optimization approach 
in which the conic constants of both reflectors are simultaneously optimized across a flat focal plane is different from the optimization discussed
in \citet{hanany/marrone:2002}, where the focal plane radius of curvature, defocus and tilt were optimized to obtain the largest DLFOV.}.
This leads to an aplanatic-like solution in which the primary reflector becomes elliptical instead of parabolic.
The ACT reflector optimization converged to a 150 GHz DLFOV at the Gregorian focus of $\sim370$ cm$^2$~sr with
near zero tilt of the secondary reflector axis, which led to the decision to align the
ACT primary and secondary reflector axes to simplify the system alignment (Figure~\ref{fig:ACT_SPT_opt}).

\begin{table}[htdp]
\begin{center}
\begin{tabular}{|l|c|c|}
\hline
  & ACT & SPT\\
\hline
$D_{p}$ (m) & 6 & 10\\
$D_{i}$ (m) & 5.6 & 7.5\\
$D_{s}$ (m) & 2 & 1\\
$F_{G}$ & 2.5 & 1.3 \\
\hline
Temperature Receivers & & \\
$F_{temp}$ & 0.9 & 1.3 \\
$F \lambda_{150}$ & 0.5 & 1.7\\
$A \Omega_{R_{temp}}$ (cm$^2$ sr) & 40 & 105 \\
$A \Omega_{D_{eff}}$ (relative) & $\sim$2.5 & 1 \\
Min. Strehl$_{150}$ & 0.97 & 0.89 \\
FWHM$_{150}$ (arcmin) & 1.37 & 1.15\\
\hline
Polarization Receivers & & \\
$F_{pol}$ & 1.4 & 1.3 \\
$F \lambda_{150}$ & 1.4 & 1.6\\
$A \Omega_{R_{pol}}$ (cm$^2$ sr) & 180 & 140 \\
\hline
\end{tabular}
\end{center}
\caption{\small \label{tab:ACT_SPT_compare}Comparison of some ACT and SPT optical parameters.
The top section gives the main telescope parameters; the middle section gives the properties of the receivers deployed prior to 2012;
and the bottom section is for the polarization-sensitive receivers. $D_p$, $D_i$, and $D_s$ are the diameters of the primary reflector,
the illumination of the primary, and the secondary reflector, respectively.
$F_G$, $F_{temp}$, and $F_{pol}$ refer to the approximate focal ratio at the Gregorian, temperature receiver, and polarization receiver focii, respectively.
$F \lambda_{150}$ refers to the approximate 150 GHz detector spacing or feedhorn aperture for each receiver.
$A \Omega_{R_{temp}}$ and $A \Omega_{R_{pol}}$ refer to the approximate throughput of the ``Temperature Receivers'' (for all three arrays)
and ``Polarization Receivers'' for each telescope at the detector focus, and $A \Omega_{D_{eff}}$ is an estimate for the relative effective
throughput of the different detector array technologies following the prescription of \cite{griffin/bock/gear:2002}.  For the temperature receivers,
the  minimum 150 GHz Strehl ratios and the measured 150 GHz beam full-width-half-maximum, FWHM$_{150}$,
are also provided~\citep{swetz/etal:2011,schaffer/etal:2011}.}
\end{table}

The secondary reflectors of ACT and SPT meet substantially different design requirements.  The SPT secondary was designed 
to meet the science goals of the first generation camera, and has a short focal ratio ($F_G\approx1.3$) to couple to a 
close-packed feedhorn array at the Gregorian focus. The entire secondary is cooled to $\sim10$~K inside a vacuum vessel 
and is used as the system stop. Because this stop is not quite at an image of the primary different detectors in the focal 
plane illuminate different sections of the primary mirror. This is one reason that  only 
$\sim 7.5$~m of the $10$~m SPT primary reflector is illuminated by each feedhorn (Table~\ref{tab:ACT_SPT_compare}, 
Figure~\ref{fig:ACT_SPT_opt}). To ensure that the secondary reflector could be aligned accurately inside the cryogenic receiver, 
it was machined as a single piece of metal, which limited the size to $\sim 1$~m diameter - the largest diameter 
that could easily be machined as a single element.

The ACT secondary reflector focal ratio ($F_G\approx2.5$) was selected as a balance between smaller re-imaging optics and minimizing beam expansion because of the need to closely pack neighboring stacks of optical elements at multiple cryogenic stages (including a vacuum window, filters at 300~K, 40~K, and 4~K, and a lens at 4~K) near the Gregorian focus. The secondary diameter was chosen to be $\sim2$~m as a balance between taking advantage of the increase in diffraction limited throughput that a larger secondary provides and minimizing the mass far from the telescope center-of-mass (to enable fast scanning) and cost. The combination of the aplanatic design, a larger secondary, and a higher $F$ result in the ACT having substantially greater diffraction-limited throughput at the Gregorian focus, than the SPT; however, the ACT design requires an additional set 
of reimaging optics because the secondary reflector cannot easily be incorporated into a cryogenic receiver (like the SPT secondary).  This results in increased thermal emission from the ACT secondary, and constrains use of the ACT Gregorian DLFOV because of the difficulty of building a compact, low-loss vacuum window as large as the DLFOV ($\sim0.7$~m diameter).

One characteristic of optimized aplanatic designs, like ACT, is that the Gregorian focal plane becomes more perpendicular to the conic axis of the reflectors, which is an advantage for on-axis systems, but results in a less image-space telecentric focal plane for off-axis systems. For example, in the ACT design the angle-of-incidence of the chief ray at the center of the Gregorian focal plane is $18.7^{\circ}$. A telecentric focal plane was not a requirement for the ACT receiver, and at higher $F$ numbers this ``focal plane tilt'' is reduced, which led to an acceptable level of residual 5-$8^{\circ}$ tilts at the filled-focal-plane arrays used in the ACT receiver.  However, coupling to a tilted focal plane does require custom re-imaging optics, which generally prevents installing receivers from other telescopes without re-configuring the receiver optics to match the ACT focus.  In addition, when considering designs for future instruments that may require a different $F$, it is straightforward to analytically calculate the new secondary parameters to match a parabolic primary, like SPT. It has previously been stated that in an aplanatic Gregorian design, the range of secondary focal lengths is limited~\citep{padin/etal:2008}, however, the ACT design demonstrated that numerical optimization techniques can be used to adjust the focal length of the secondary with the design of an ellipsoidal primary.  As the secondary $F$ is increased (decreased), the DLFOV throughput increases (decreases), which is the same behavior as a classical Gregorian with a parabolic primary.  The limits of changing $F$ in an optimized aplanatic Gregorian design (ACT) relative to a classical Gregorian (SPT) are not clear and warrant further study.  Because of the focal plane tilt and numerical optimization requirements of changing the secondary $F$, in Table~\ref{tab:ACT_SPT_compare_2} we characterize the SPT design as ``easily reconfigurable'' but not the ACT design.

\begin{figure}[htb]
\begin{center}
\includegraphics[width=6.0in]{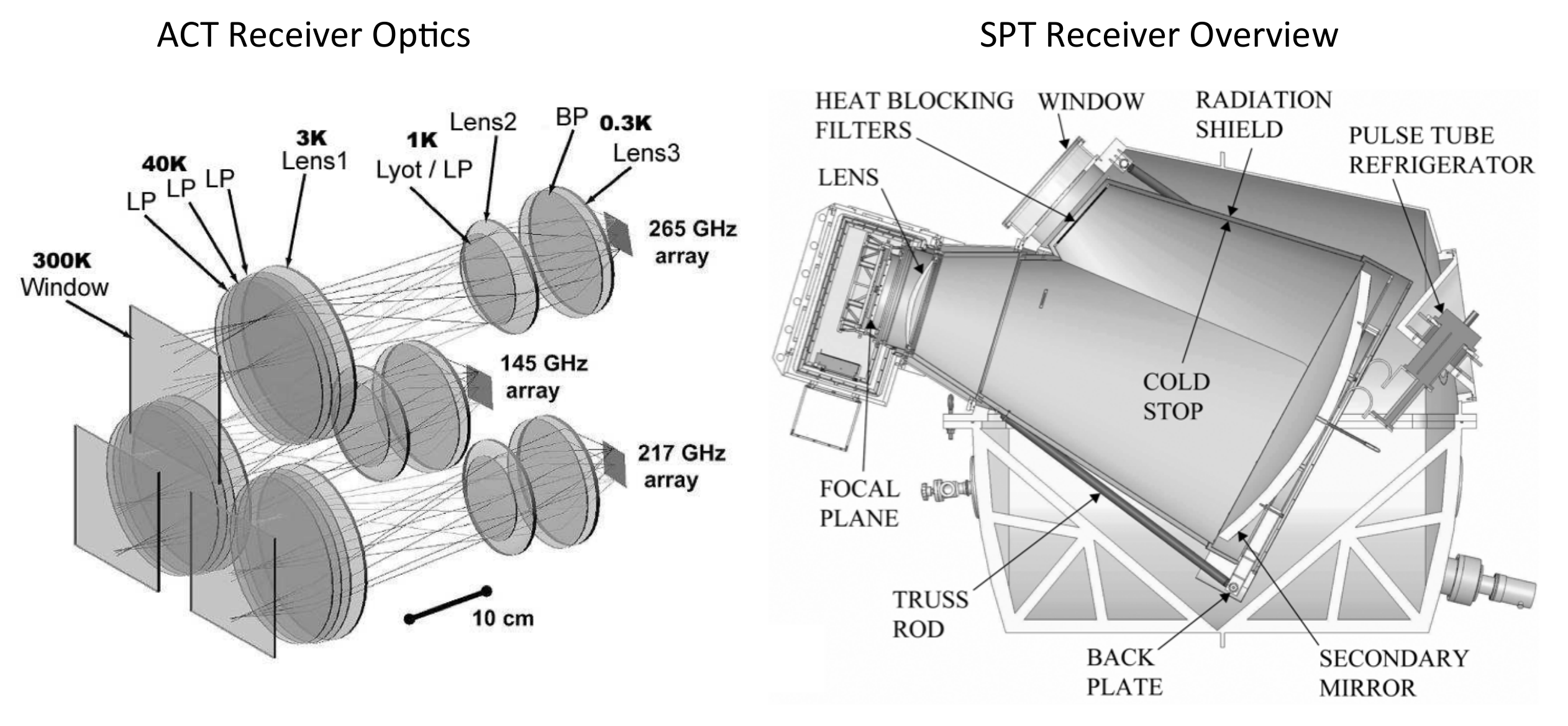}
\caption{\small {\it Left}: The ACT receiver optics include three independent optical paths, each for a different frequency band with its own 
vacuum window,  filters, and set of three silicon lenses \citep{fowler/etal:2007}.  {\it Right}: An overview of the SPT receiver, which includes the 
cryogenic secondary reflector and a single HDPE lens \citep{padin/etal:2008}.  The secondary mirror is the aperture stop of the system;
spill-over past the secondary is intercepted by cold surfaces. 
\label{fig:ACT_SPT_rec}}
\end{center}
\end{figure}

Both the ACT and SPT teams began observing with multi-frequency receivers to measure the CMB and to search for galaxy clusters via the SZ effect in 2008, and are planning to deploy new polarization-sensitive receivers in 2012-13. Here we briefly compare the first-generation receiver optics (Figure~\ref{fig:ACT_SPT_rec}) and detector coupling, then discuss the planned upgrades.  The SPT receiver has two sections that can be operated independently, which is beneficial for testing both systems and for upgrades that will use the same secondary reflector. The first section includes the vacuum window, thermal blocking filters, cryogenic secondary reflector, and the majority of the cold stop.  The second section contains a lens, band-defining filters, and the detector array, which includes detectors at $\sim100$, 150, and 220 GHz. Some advantages of this design include: reduced emission from the secondary reflector, a small-aperture vacuum window at the $F\approx1$ primary focus, and a large stop surface, which minimizes diffraction at the stop. The SPT detectors are coupled to the optics via a flat array of conical feedhorns, so a single high-density polyethylene ($n \approx 1.5$) lens is used to slightly speed up the focus and to improve the coupling to the feedhorns by making the focal plane more telecentric.  The feedhorn coupled array includes 966 feeds with 4.5~mm apertures, resulting in $\sim 1.7 F \lambda$ apertures at 150 GHz.

The ACT temperature receiver includes three independent optical paths that each operate at a different frequency band: 148, 218, 277 GHz.  Each optical path has an independent vacuum window, filters, three silicon ($n\approx3.4$) lenses, and detector array, which makes defining the bandwidth and anti-reflection coating the optical elements relatively easy compared to receivers that use common optical elements for multiple frequency bands.  The first lens in each optical path creates an image of the primary reflector, which is used as the system stop and allows illumination of $>90$\% of the primary reflector (Table~\ref{tab:ACT_SPT_compare}). The stop surface is cooled to 1~K, which is required to minimize the background optical loading on the filled detector arrays from spillover onto the stop. The following pairs of lenses create a fast focus ($F\approx 0.9$) onto the three filled detector arrays, each comprised of $1024$ square bolometric detectors with $1.1$~mm pitch~\citep{niemack/etal:2008}, or roughly $0.5 F \lambda$ at 150 GHz.

Section~\ref{sec:arrays} provides an overview of the tradeoffs between filled focal plane arrays (used in ACT) and feedhorn coupled arrays (used in SPT).
The close packing of the detectors in ACT led to having $\sim3$ times more detectors, despite the ACT receiver having less than half the optical
throughput of the SPT receiver (Table~\ref{tab:ACT_SPT_compare}).
We estimate that the ACT 150 GHz filled array could have $\sim2.5$ higher mapping speed than a feedhorn array similar to the SPT design
filling the same FOV (listed as $A \Omega_{D_{eff}}$ in Table~\ref{tab:ACT_SPT_compare}).
Scaling the mapping speed ratio by the total instrument throughput, $A \Omega_{R_{temp}}$, provides an estimate of the relative mapping speeds
of the two instruments of $MS_{SPT}/MS_{ACT} \approx 1.05$. This estimate includes assumptions about the instrument optical loading conditions,
does not include detector noise, and assumes that 150 GHz detectors fill the same fraction of $A \Omega_{R_{temp}}$ of both instruments.
Based on the mapping speed estimate, the increased throughput of the filled detector arrays on ACT largely compensates for the
smaller $A \Omega_{R_{temp}}$.  This suggests that filled detector arrays hold promise for maximizing the mapping speed of future instruments;
however, significant development is needed to scale the readout and fabrication of filled arrays for systems with larger $A \Omega$.
In addition to simplifying instrument requirements as discussed above, feedhorns (and other beam forming detector coupling techniques)
have advantages in terms of minimizing systematic effects in polarization measurements.

\begin{table}[htdp]
\begin{center}
\begin{tabular}{|l|c|c|}
\hline
  & ACT & SPT\\
\hline
Primary shape & Ellipsoid & Paraboloid \\
Easily reconfigurable & No & Yes \\
Stop type & Primary image & Secondary reflector \\
Cold stop Temperature & 1 Kelvin & 10 Kelvin\\
Refractive optics & 3 Silicon lenses per array & HDPE lens\\
Temp. array coupling & Filled focal plane & Conical feedhorns\\
Pol. array coupling & Corrugated feedhorns & Corrugated \& profiled feedhorns\\
\hline
\end{tabular}
\end{center}
\caption{\small \label{tab:ACT_SPT_compare_2} Comparison of some ACT and SPT design features.}
\end{table}%

\subsection{ACTPol and SPTPol}

The ACT and SPT polarization-sensitive receivers, henceforth ACTPol \citep{niemack/etal:2010} and SPTPol \citep{mcmahon/etal:2009}, both use corrugated feedhorns (as well as 90 GHz profiled feedhorns in the case of SPTPol) to couple to the detector arrays. This is done because of the excellent polarization properties of corrugated/profiled feedhorns.  SPTPol uses a nearly identical optics layout to the SPT layout, but the used FOV area has been increased by $\sim$33\%  by surrounding the central 150 GHz detectors with individual 90 GHz feedhorns.  ACTPol uses a similar optics concept to ACT (3 independent optics paths with 3 silicon lenses each), however, the diameter of most of the optical components is roughly two times larger, and the detector arrays are circular instead of square, which leads to a factor of 4.5 increase in throughput at the final focus (Table~\ref{tab:ACT_SPT_compare}).  Unlike the original ACT lenses, the orientation of the ACTPol lenses is optimized to create image-space telecentric focal planes to couple to the flat feedhorn arrays.

The ACT design approach included maximizing the throughput of the DLFOV at the Gregorian focus, but the DLFOV does not necessarily limit the usable FOV. For example, the edges of the ACTPol lenses that are furthest from the boresight are placed outside the Gregorian DLFOV, however, the lenses improve the image quality in this region to be strongly diffraction-limited (Strehl ratios $> 0.9$). Adding a tertiary reflector to a design like SPT is another approach for increasing the DLFOV. Thus far, the throughput of the telescope receivers has generally been designed to match the requirements of the superconducting detector arrays that were feasible to manufacture and read out at that time.  As superconducting detector array technologies continue to increase in size and decrease in cost, higher throughput optical systems will be required to illuminate them.  These examples suggest that modified secondary, tertiary or re-imaging optics may allow substantial increases in the useable FOV for these telescopes in the future.

\section{Interferometers}
\label{sec:interferometers}

Interferometers are a natural choice for measuring the anisotropy in
the CMB.  The correlation of the outputs from two antennas, called a visibility, is
just a Fourier component of the product of the sky and the
response of a single antenna. The single antenna response is called the
primary beam, and sets the interferometer's field of view.  It should
not be confused with the response of the pair of telescopes
(or more generally, of an array of telescopes) to a point source,
called the synthesized beam.  Thus the visibility is directly
relatable to the CMB power spectrum. In the following we imagine that
the region of sky we cover is small enough that we may consider it
flat so that the expansion of the temperature field in $Y_{\ell m}$ spherical harmonics
can be replaced by Fourier modes.

\begin{figure}[htb]
\begin{center}
\includegraphics[width=3.15in]{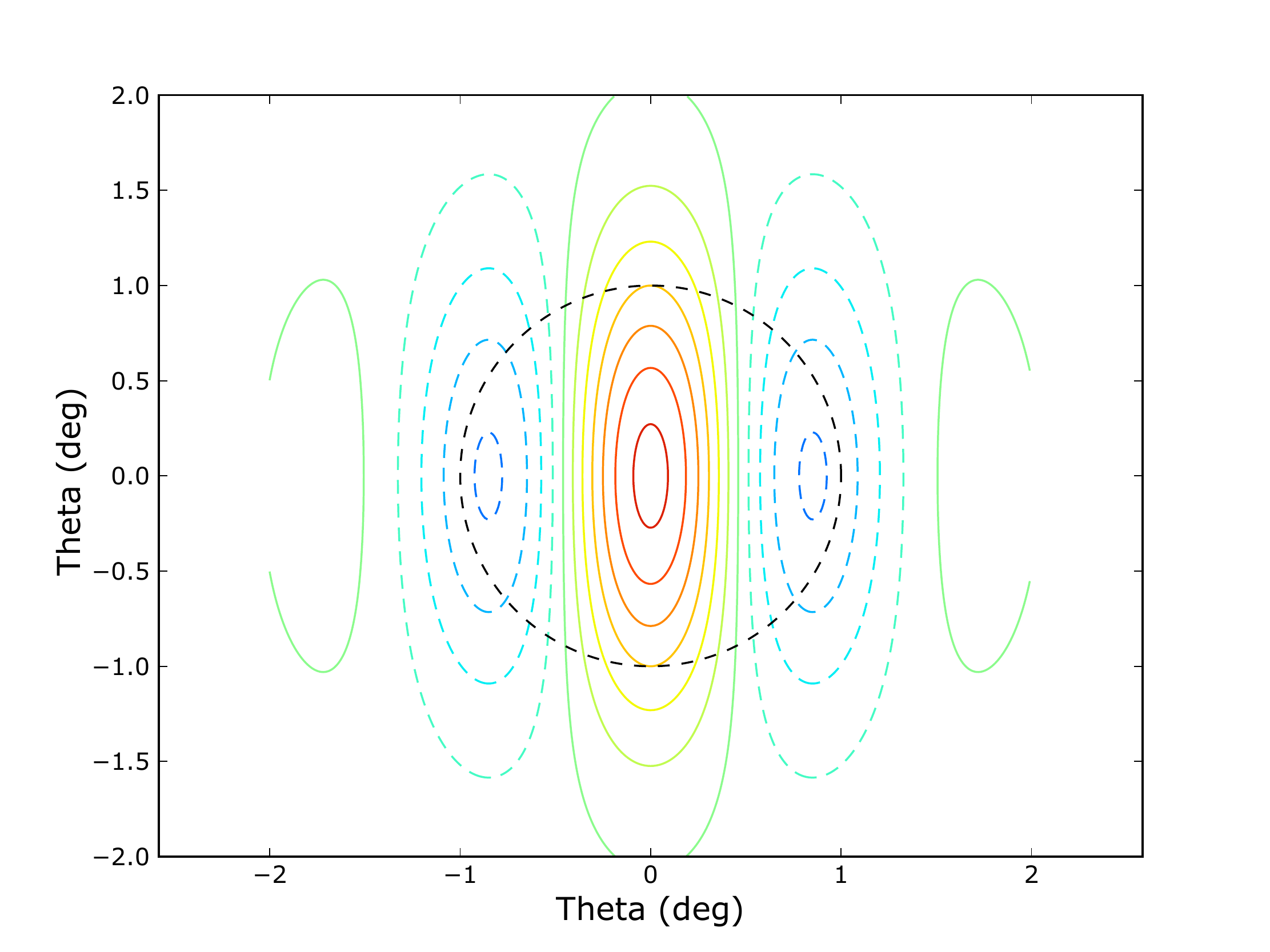}
\includegraphics[width=3.15in]{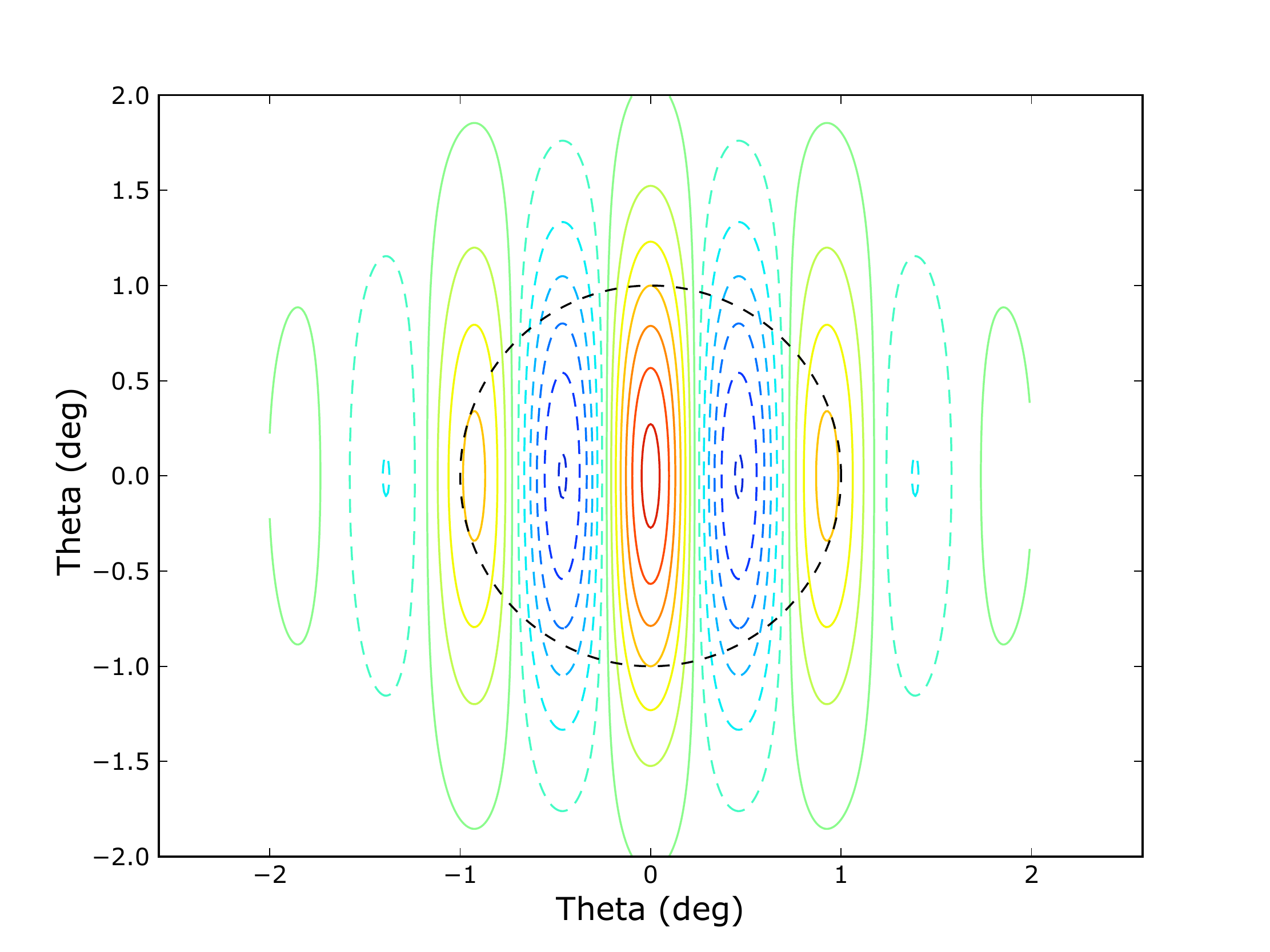}
\caption{\small Instantaneous interferometer beams.
Consider a 30 cm diameter aperture with a Gaussian profile with $\sigma_r=6.3$~cm (Section~\ref{sec:angularresolution}) and
$\lambda=1~$cm. The resulting beam has $\theta_{1/2} = 2^\circ$. This is indicated by the dashed circular lines in the figures. If two such apertures are placed so that the center-to-center separation
is 30 cm (30$\lambda$) then one obtains the instantaneous beam pattern shown on the left. The dashed lines indicate negative lobes
and the solid lines positive lobes. The output of the interferometer is then the integral
of the product of this beam pattern and the sky. One can see that only spatial fluctuations that resemble this pattern will give a non-zero output.
If the center-to-center separation is increased to 60$\lambda$, then one obtains the instantaneous pattern on the right. Note that the
extent of the pattern is still determined by $\theta_{1/2}$ from one antenna.
 \label{fig:intbeams} }
\end{center}
\end{figure}

Consider the case of two identical antennas (or telescopes) of diameter D that are 
separated by distance L. A particular configuration of the antennas, set by the spatial 
separation L is called a `baseline'. When L=2D the antennas are in a `compact
configuration'. The instantaneous beam pattern on the sky would
resemble that of a double slit pattern for two wide slits. A representation is shown in Figure~\ref{fig:intbeams}. A visibility is the
instrument response of the sky times this instantaneous beam.  As the
telescopes are moved apart, the envelope of the pattern remains that of
the beam pattern of a single telescope although the number of fringes
inside the envelope increases. With multiple measurements with baselines
of different lengths and orientations, one may fill out the ``U-V"
(or loosely Fourier) plane with visibilities.  To compute the
power spectrum, one  averages the variance over annuli in the U-V plane.
To make a map of the sky, one then transforms the visibility map to real
space. There is no reason to constrain oneself
to the envelope of the beam, many images like the one in Figure~\ref{fig:intbeams} can be
mosaicked together to probe spatial wavelengths longer than that of the
beam size and to increase the resolution of the visibility spacing.

In some sense interferometers are the opposite of more conventional
mapping schemes. With real-space maps, the size of the beam sets the
resolution and the scan size of the instrument sets the size of the
field one observes. With interferometers, the separation of the
telescopes determines the resolution and the transform of the antenna
illumination pattern sets the field size (before any mosaicking).
With real-space methods, a map is made and the power spectrum is
obtained from its Fourier transform. With interferometers, the power
spectrum is fairly directly measured from the visibilities.  The
real-space maps can be obtained from the transform of the
visibilities, though they are rarely used for CMB science.

Interferometers have been used to measure the CMB anisotropy since the mid-1980s
(e.g., \citet{knoke/etal:1984} and reviewed in~\citet{partridge:1995}), although most of the early efforts were aimed at arc-minute angular
scales or smaller because they used the VLA. The first interferometric observation of the primary CMB was made
with a dedicated two-element correlation receiver~\citep{timbie/wilkinson:1988}.
Results from CAT, when combined with results at lower $\ell$, gave
evidence for the existence of an acoustic peak at $\ell \simeq 200$~\citep{osullivan/etal:1995,scott/etal:1996,baker/etal:1999}.
Anisotropy measurements by the 13 antennas of the
Cosmic Background Imager interferometer at 31~GHz~\citep{pearson/etal:2003}, together with results from
ACBAR~\citep{kuo/etal:2004} and WMAP helped break cosmic parameter degeneracies.
The DASI interferometer made the first measurements of the polarization
of the CMB~\citep{kovac/etal:2002}.

Interferometers have a number of advantages over real-space methods.
The spatial filtering of a visibility makes it insensitive to scales
much larger (or smaller) than the fringe spacing, so interferometers
filter out almost all atmospheric fluctuations during the correlation.
They can be set up to measure fine angular resolution easily, simply
by increasing the baseline length.  As one adds elements, the number
of baselines grows as the square of the number of antennas.  The
relative response to different spatial wavelengths (analogous to the
beam of a single-dish telescope) is set by the easily-measured
separation of antennas.  Because of the intrinsic atmospheric
filtering, they do not have to scan rapidly.  The primary disadvantage
is that the cost of correlation in a classical interferometer grows as
the number of dishes squared (though see, for example, the Fast Fourier
Transform Telescope \citep{tegmark/zaldarriaga:2009,tegmark/zaldarriaga:2010}),
for sparsely sampled arrays interferometers are slower at mapping the sky, a
separate cryostat is required for each receiver, the components are
expensive, and that at 150~GHz coherent receivers are not yet a
``commodity" and the mechanical tolerances are tight.  With the advent
of arrays of thousands of bolometers and of order 100-element
polarization-sensitive coherent receivers \citep{gaier/etal:2003},
{\it classically-configured} interferometers have lost much of their appeal
for measuring the CMB.

However, the quest for primordial gravitational waves and the advantages of interferometers have
driven the invention of  new designs that go well beyond the classic configuration.  There is now an international effort
called QUBIC \citep{battistelli/etal:2011, charlassier/etal:2008, timbie/etal:2006} to use arrays of bolometers in a novel optical configuration
to make an interferometer capable of detecting the polarization B-modes.  The instrument observes the sky through an array
of feeds whose signals are then interferometrically combined on two arrays of $\sim$1000
bolometers, with one array per polarization.  As opposed to multiplying the signals from a pair of antennas, in QUBIC
the interfering electric fields are summed and squared by the bolometers. It is a modern version of the Fizeau-style adding configuration.

\section{The CMB Satellites}
\label{sec:satellites}

There have been four satellites with instruments dedicated to measuring the CMB anisotropy:  Relikt \citep{strukov/skulachev:1984}, the COsmic Background Explorer (COBE) \citep{boggess/etal:1992}, the Wilkinson Microwave Anisotropy Probe (WMAP) \citep{bennett/etal:2003d}, and Planck \citep{tauber/etal:2010b}.The frequency coverage, sensitivity, and resolution are given in Table~\ref{tab:satspecs}.  There is a marked improvement over time.

\begin{table}[htdp]
\begin{center}
\footnotesize
\begin{tabular}{|p{0.35cm}p{0.2cm}p{0.9cm}p{0.8cm}|p{0.35cm}p{0.2cm}p{0.9cm}p{0.7cm}|p{0.35cm}p{0.4cm}p{0.9cm}p{0.8cm}|p{0.35cm}p{0.7cm}p{0.9cm}p{0.8cm}|}
\hline
\multicolumn {4}{|c|}{Relikt (1983)} &      \multicolumn {4}{|c|}{COBE (1989)}& \multicolumn {4}{|c|}{WMAP (2001)}&
\multicolumn {4}{|c|}{Planck (2009)}\\
Freq &~N&~~S& ~$\theta_{1/2}$  & Freq   &~N&~~S& ~$\theta_{1/2}$  & Freq   &~~N&~~S& ~$\theta_{1/2}$  &  Freq  &~~N&~~S& $~\theta_{1/2}$ \\
 GHz &&mKs$^{1/2}$  &~deg  & GHz   &&mKs$^{1/2}$&~deg  & GHz   &&mKs$^{1/2}$& ~deg  &  GHz  &&mKs$^{1/2}$& ~deg \\
\hline
&&&&&&&& 22.7 &~~2&0.49  & 0.82$^\circ$ &&&& \\
&&&&&&&&&&&& 28.5  &~~4&0.15& 0.54$^\circ$ \\
&&&& 31.5 &~~2&30 &~ 7$^\circ$ &&&&&&&&\\
&&&&&&&& 33.0 &~~2&0.51  & 0.62$^\circ$ &&&&\\
37.5  &~~1& 25& ~5.8$^\circ$ &&&&&&&&&&&&\\
&&&&&&&& 40.7  &~~4&0.47 & 0.49$^\circ$ &&&& \\
&&&&&&&&&&&& 44.1  &~~6& 0.16& 0.47$^\circ$ \\
&&&& 53  &~~2&11 & ~7$^\circ$ &&&&&&&& \\
&&&&&&&& 60.6  &~~4&0.54 & 0.33$^\circ$ &&&&\\
&&&&&&&&&&&& 70.3 &~12& 0.13& 0.22$^\circ$ \\
&&&& 90 &~~2&16  & ~7$^\circ$ &&&&&&&&\\
&&&&&&&& 93.4  &~~8& 0.58 & 0.21$^\circ$ &&&&\\
&&&&&&&&&&&& 100 &~~8/4P& 0.0173   & 0.16$^\circ$ \\
&&&&&&&&&&&& 143  &~12/4P& 0.0084  & 0.12$^\circ$ \\
&&&&&&&&&&&& 217  &~12/4P& 0.0068  & 0.078$^\circ$ \\
&&&&&&&&&&&& 353  &~12/4P& 0.0055  & 0.074$^\circ$ \\
&&&&&&&&&&&& 545  &~~4& 0.0045 & 0.063$^\circ$ \\
&&&&&&&&&&&& 857  &~~4& 0.0019  & 0.061$^\circ$ \\
\hline
\end{tabular}
\caption{\small \label{tab:satspecs} Characteristics of the four CMB anisotropy satellites giving for each frequency band (Freq) the number 
of science detectors (N)  the sensitivity (S) for the white noise limit and relative to a Rayleigh-Jeans emitter and the angular 
resolution ($\theta_{1/2}$). The noise limits for each detector in a frequency band have been added in quadrature for this estimate. 
To convert to a sensitivity in CMB temperature units, multiply by, for example, 1.03, 1.05, 1.13, 1.3, 1.66, 3.15, 13.8, $\sim170$, 
$\sim 10^{4}$ for the 28.5~GHz to 857~GHz bands on Planck, respectively. The year of the launch is given next to each mission name. 
Relikt lasted 6 months, COBE 4 years, WMAP 9 years, and the Planck HFI 2.5 years. The improvement in sensitivity and the increase 
in complexity over time is evident. For Planck the bands with $\nu \geq 100$~GHz are bolometric. At lower frequencies
they are radiometers with HEMT amplifiers. The notation ``12/4P" means 
that there are a total of 12 detectors of which there are 4 polarization sensitive pairs, each pair sharing the same feed, and a total of 8 feeds. 
For the HEMT bands each detector is a single linearly polarized radiometer; two orthogonally polarized radiometers share a feed. The same 
configuration is use for the 31.5~GHz band on COBE.}
\end{center}
\end{table}

The great benefit of a satellite is the ability to make all-sky maps from a very stable platform.  The stability of space allows one to understand the instrument, especially the noise and systematic effects, in detail. An ideal anisotropy map  would be fully described by simply a temperature and uncertainty for each pixel on the sky with an overall offset removed. In reality, all maps have some degree of correlation that must be accounted for in the most demanding analyses. The source of the correlation could be due to non-ideal aspects of the instrument's noise (e.g., ``1/f" noise), remnants of glitch removal, contamination through the sidelobes, an imperfect differential measurement (for COBE and WMAP), or asymmetric optics.  Multiple techniques have been developed to account for these correlations.

An important factor in making high-fidelity maps is cross-linking. In the limit of a perfectly stable instrument, cross linking is not necessary 
but in reality the gain and offsets of all detectors vary over time. From the point of view of one pixel, a well cross-linked map would have 
scan lines running through in all different directions connecting the pixel in question to those around it.
Ideally the cross-linking takes place on multiple time scales. A set of such measurements for each pixel over the 
full sky provides a strong spatio-temporal filter that allows for the separation of instrumental effects such as varying gains 
and offsets from the true underlying signal.

Cross-linking has an advantage for the optics as well. Because of the premium on size and mass for a satellite, the focal planes 
are packed to the hilt. As a result, the beam profiles are not symmetric let alone Gaussian. 
Perfect cross-linking has the effect of producing an effectively symmetric beam profile (with a coarser resolution that depends on the inherent 
asymmetry) thereby simplifying the analysis. 
As with correlations, the most demanding analyses must take the remaining asymmetries into 
account~\citep{hinshaw/etal:2007,hanson/etal:2010}.

\subsection{Relikt}
\label{sec:relikt}

\begin{figure}[tbh]
\begin{center}
\includegraphics[width=5.5in]{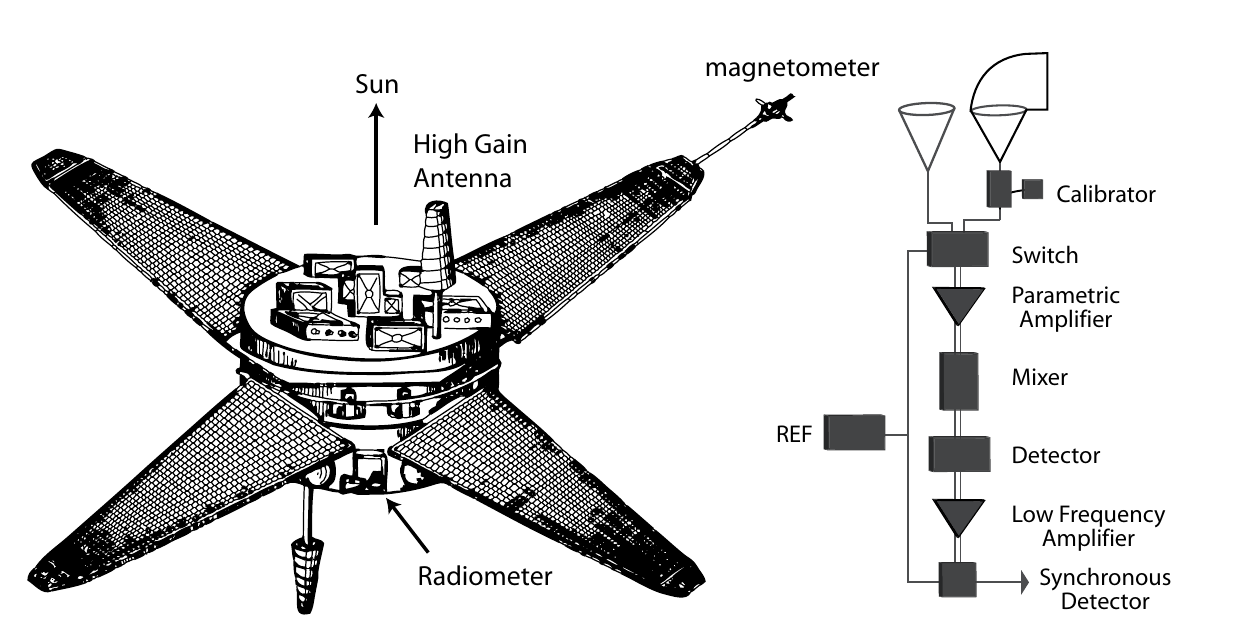}
\caption{\small A line drawing of the Relikt satellite from~\citet{strukov/skulachev:1986}. The Sun is towards the top of the page.
The radiometer is the small box in the center between the bottom two solar panels. The  reference feed points toward the bottom of the
page and the scanning feed points towards the left. The instrument package weighed only 30 kg. 
\label{fig:relikt}
}
\end{center}
\end{figure}

Relikt was the first space-based anisotropy satellite \citep{strukov/skulachev:1984}. It was part of the Soviet space program and was launched on
the Prognoz-9 satellite in 1983, roughly six years before the launch of NASA's COBE.  The microwave radiometer, shown in Figure~\ref{fig:relikt} was
one of a number of instruments on the satellite. Its three objectives were \citep{strukov/skulachev:1986}: `` (1) to determine the angular distribution of
the relic radiation and (in the case of the discovery of anisotropy) to estimate the mean density of matter in the universe; (2) to determine the
distribution of faint extended radio sources on the celestial sphere; and (3) to refine the velocity-vector parameters of the observer's motion with
respect to the reference frame of the relic radiation. '' Outside of detecting the primary anisotropy,  these goals were achieved.

Relikt was a differential instrument although was not symmetric. A reference corrugated feed with $\theta_{1/2}=10^\circ$ 
pointed in the anti-solar direction.  A second scanning feed with
$\theta_{1/2}=5.8^\circ$ was aimed $90^\circ$ to the reference direction and scanned the sky as the satellite rotated.
The satellite was in a highly elongated orbit with a 26.7 day period. The rotation period was 113 seconds.
In certain parts of the orbit the Earth was observed as the beam scanned over it.
After averaging data for a week, the reference beam was stepped by $7^\circ$ in the ecliptic. Thus, all scans overlapped at the ecliptic poles. Nearly the full sky was observed in the six months the satellite was operational \citep{klypin/etal:1992}.

The scanning feed had a corrugated cone feeding an offset parabola to direct the radiation perpendicular to the symmetry axis of the base of 
the cone. Before launch the beam pattern was mapped~\citep{strukov/skulachev:1984} to -80~dB of the peak and a broad sidelobe at 
the -40~dB level for angles $30^\circ$ to $60^\circ$ was measured.  As a result, during the data analysis roughly half the data were cut 
due to possible contamination by emission from Mars and the Moon into the scanning beam sidelobes~\citep{klypin/etal:1992}.  
Nevertheless,  at the time they produced the best measurement of the dipole 
($\ell=1$,~\citet{strukov/etal:1987}), one of their primary goals, and they placed limits on the large angular scale 
anisotropy~\citep{strukov/etal:1988} that were not improved upon until FIRS \citep{meyer/etal:1991} and COBE~\citep{smoot/etal:1991}.

\subsection{COBE}
\label{sec:cobe}

\begin{figure}[tbh]
\begin{center}
\includegraphics[width=3.5in]{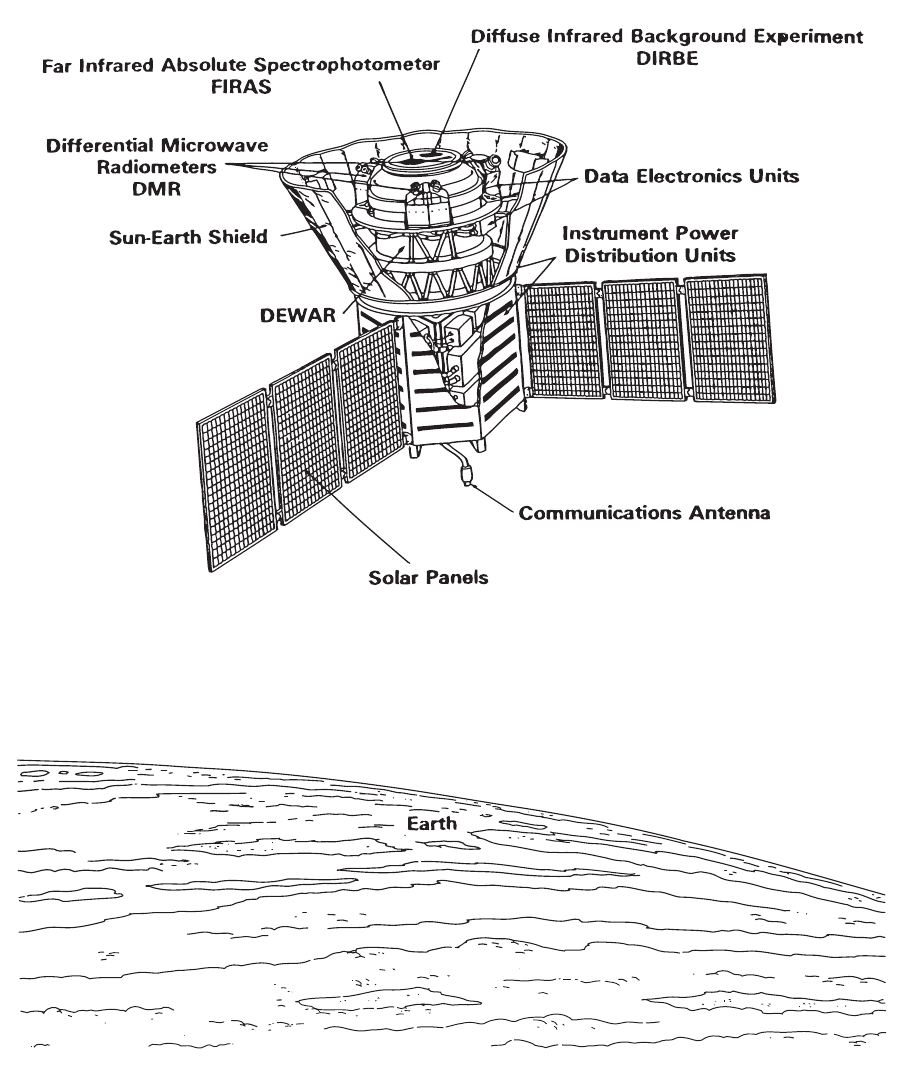}
\caption{\small A line drawing of the COBE satellite~\citep{boggess/etal:1992}.
The three DMR bands are shown. The 31.5 GHz receiver (back right) has a single feed. The two separate radiometers in this band look at the two orthogonal polarizations from one feed. The Sun would be, for example, off to the left in this figure thus illuminating the solar panels. The instruments always look away from and are shielded from the Earth. For scale, the diameter with the deployed solar panels is 8.5m.  The mass of the DMR was 154~kg.
\label{fig:cobe}
}
\end{center}
\end{figure}

The Differential Microwave Radiometer (DMR) instrument \citep{smoot/etal:1990} was one of three aboard the COBE satellite, 
shown in Figure~\ref{fig:cobe}.
The other two were the Far Infrared Absolute Spectrophotometer (FIRAS), which measured the absolute temperature of the CMB
$T_{CMB} = 2.725\pm 0.001$~K~\citep{fixsen/mather:2002}, and the Diffuse Infrared Background Experiment (DIRBE), 
which mapped the IR sky and detected the cosmic infrared background.

\begin{figure}[tbh]
\begin{center}
\includegraphics[width=6.in]{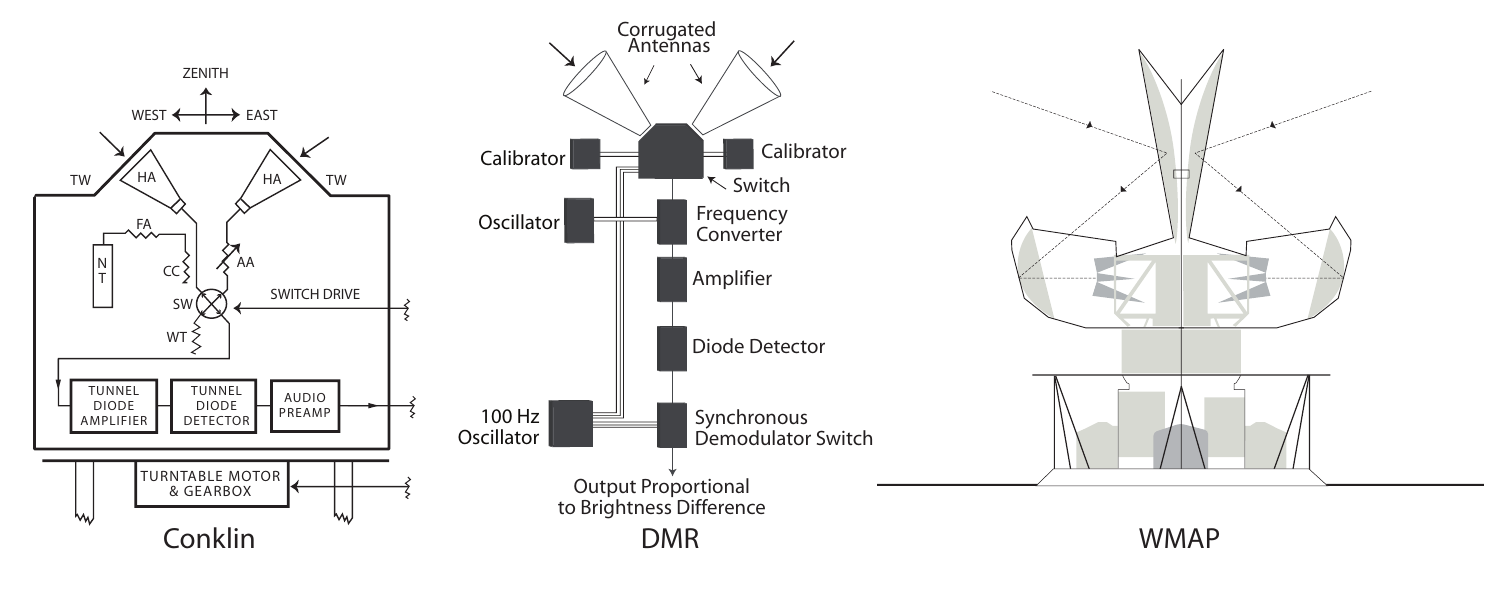}
\caption{\small The evolution of the manifestly differential CMB radiometer over thirty years. The left shows Conklin's radiometer for 
measuring the CMB dipole. The central picture shows the DMR aboard COBE. The angle between the feeds is 60$^\circ$. 
WMAP is on the right.  Figure adapted from~\citet{ftbb:2009}.
\label{fig:cmb_diff_inst}
}
\end{center}
\end{figure}

The COBE satellite was  launched in 1989 into a high inclination polar orbit. The spin axis of the spacecraft always pointed away from the Earth and at
roughly $90^\circ$ from the Sun. At each of the three frequencies given in
Table~\ref{tab:satspecs} there were two receivers like the one shown in Figure~\ref{fig:cmb_diff_inst}.
The radiometers were situated in the satellite so that the feeds observed $+/- 30^\circ$ from the spin axis.
COBE's orbit and scan pattern were a marked improvement over that of Relikt's. Detailed attention was paid to possible contamination by
emission from the Sun, Earth and Moon, and large cuts of the data were not
required \citep{kogut/etal:1992}. The most notable systematic error was due to the affect of the Earth's magnetic field on the Dicke switches.

The DMR optics are especially simple. In each of the six receivers, the sky is viewed through two corrugated feeds with $\theta_{1/2}=7^\circ$ and separated by $60^\circ$. There are, though, only five pairs of feeds. At 31.5 GHz, the two output polarizations of feed pair are sent to two receivers. At the other frequencies, a single polarization from each feed pair is sent to a receiver. The beam patterns were measured before flight to roughly -90~dB from the peak \citep{toral/etal:1989}. The input to the receiver chain is Dicke-switched between the feed outputs at
100 Hz. The 31.5 GHz receiver operated at 300~K; the other two bands were at 140~K. The combined noise level for the three bands was  30~mK-sec$^{1/2}$,  11~mK-sec$^{1/2}$, and 16~mK-sec$^{1/2}$ respectively. 

The DMR instrument was much different in layout than the one on Relikt. As shown in Figure~\ref{fig:cmb_diff_inst}  it took advantage of symmetry.  The first manifestly symmetric anisotropy instrument was designed by Edward Conklin  (\cite{conklin1969thesis,conklin:1969}) with the goal of measuring the CMB dipole from White Mountain, CA.  While there is no apparent evolutionary connection between Conklin, COBE, and WMAP,  the strong appeal of symmetry for making a differential measurement guided the designs
of all these instruments. As we show below, most modern anisotropy instruments, including the Planck satellite, are not
symmetric. In some cases this is driven by the use of bolometers; in others by the fact that the receivers are dual polarized and thus intrinsically differential.

\subsection{WMAP}
\label{sec:wmap}

\begin{figure}[tbh]
\begin{center}
\includegraphics[width=4.in]{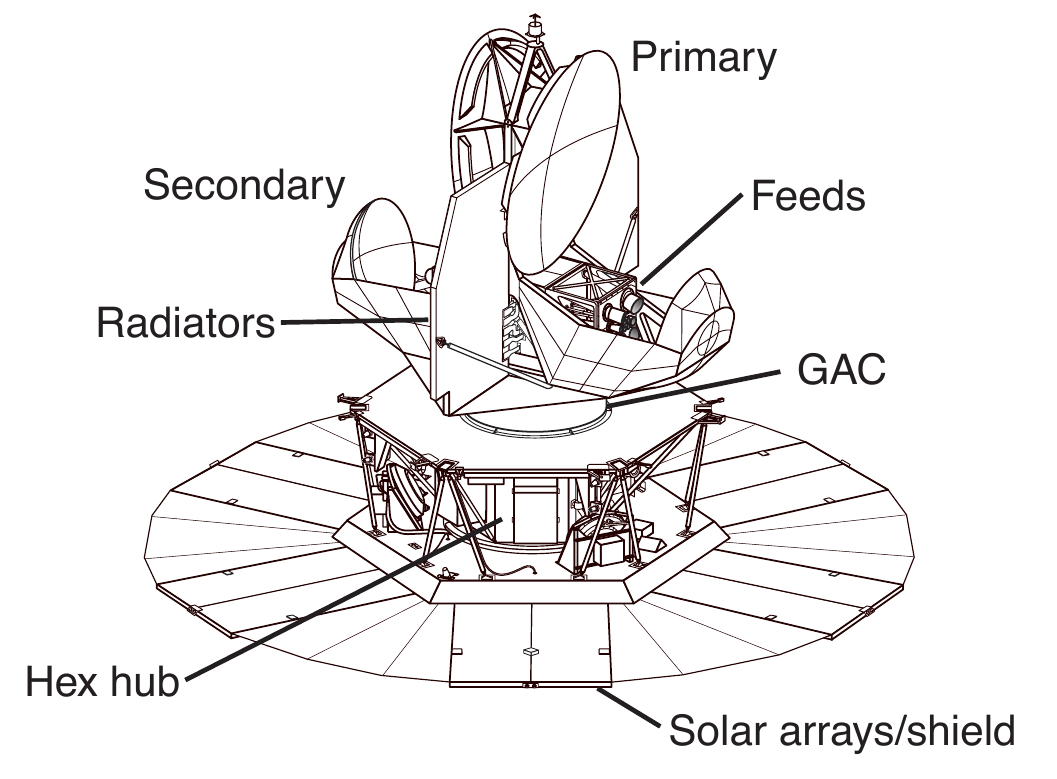}
\caption{\small
Outline of the {\sl WMAP} satellite. The thermal straps at the base of the large radiators provide the
passive cooling for the first stage of the amplifier chain. The overall height is
3.6~m, the mass is 830 kg, and the diameter of the large disk on
the bottom is 5.1 m. Six solar arrays on the bottom of this disk
supply the 400~W to power the spacecraft and instrument.
Thermal blanketing between the hex hub and thermal link provided by the gamma alumina cylinder (GAC), and
between the GAC and radiators, shield the instrument from thermal
radiation from the support electronics and attitude control systems.
\label{fig:wmap}
}
\end{center}
\end{figure}

As shown in Figures  \ref{fig:cmb_diff_inst} and \ref{fig:wmap}, WMAP has similarities to a classic feed plus telescope design. However, unlike the classic system, two telescopes are combined in a back-to-back configuration.   When designing a CMB satellite telescope, there are a number of factors that must be considered:

\begin{enumerate}
\item It is desirable to use an ``offset" design so that the support structure for the secondary does not scatter radiation.
\item In space, one wants to make optimal use of the rocket shroud size. For a given useable focal plane area, the Gregorian
family with its focus between the primary and secondary is especially compact \citep{brown/prata:1994}. For WMAP, multiple designs were considered including the offset Cassegrain, single reflector systems,  and three reflector systems, but the two-mirror Gregorian
was the most optimal.
\item One wants to get as many beams on the sky with as wide a frequency coverage as the technology will permit. In other words
one wants a large DLFOV. To minimize abberations and maintain a DLFOV, the higher frequency feeds are placed
near the center of the focal plane and the lower frequency feeds
on the outside. Asymmetric beam profiles are acceptable as their effects may be incorporated in the data analysis.
For WMAP, the scan strategy \citep{bennett/etal:2003d} has the benefit of symmetrizing the beam profiles.
\item At least two types of modeling code are needed. A fast parametric code is useful for trying many designs. However,
the full response must be computed using physical optics in which the field is solved for at each surface. WMAP used the
Diffraction Analysis of a Dual Reflector Antenna code \citep{YRS:DADRA}  which proved to be sufficiently accurate.
\item The precise reflector shape is chosen to minimize abberations over as large as possible an area.
WMAP started as a classic Gregorian following the Mizuguchi-Dragone condition (see Section~\ref{sec:reflector}
and footnote~\ref{footnote:mizuguchi}). Then the surface was
shaped using proprietary surface shaping software \citep{YRS:DADRA}.
\item The feeds must not be able to view each other or couple to each other. This means that low frequency feeds
are shortened or profiled and that high frequency feeds are lengthened with extra corrugations.
\item One must be able to account for all of the solid angle of the beam in intensity and polarization. The optics are
designed with the full $4\pi$ coverage in mind. Not only are the Earth, Moon, and Sun bright objects in the sidelobes, but
emission from the galaxy must be considered.  For WMAP a
specialized test range was built to make sure that, by measurement,  one could
limit the Sun as a source of spurious signal to $<1~\mu$K level in all
bands. This requires knowing the beam profiles down to
roughly $-45$ dBi (gain above isotropic) or $-105$ dB from the W-band peak. It was found that over much of
the sky, the measured profiles differ from the predictions at the $-50$~dB
level due to scattering off of the feed horns and the structure that
holds them that were not part of the model.  During the early part of the mission,
the Moon was used as a source in the sidelobes to verify in part the ground-based measurements and
models.
\item After launch and before commanding the satellite attitude, the optics have a chance of directly viewing the Sun.
Thus the surfaces must be roughened so that the Sun is not focussed on the feeds. The roughening must not increase the microwave emissivity. In addition, the surface must be emissive at infrared wavelengths so that it can radiatively cool.
This is accomplished by evaporatively coating them with a mixture of silicon monoxide and silicon dioxide.
\item There is a premium on mass and therefore the reflectors are almost always made of light-weight composite material.
\end{enumerate}

As a result of these considerations, the WMAP team settled on a design shown in
Figure~\ref{fig:wmap}~\citep{page/etal:2003,barnes/etal:2002}.
The primary reflectors are 1.4$\times$1.6~m.  The
secondaries are roughly a meter across, though most of the surface
simply acts as a shield to prevent the feeds from directly viewing the
Galaxy. Each telescope focusses radiation onto 10 dual-polarization scalar feeds.
These are shown as triangles in the right panel of Figure~\ref{fig:cmb_diff_inst}.
The primary optical axes of the two telescopes are separated by $141^{\circ}$ to allow
differential measurements over large angles on a fast time scale.
The feed centers occupy a 18~cm$\times$~20~cm region in the focal plane,
corresponding to a $4^{\circ} \times 4.5^{\circ}$ array on
the sky.

At the base of each feed is an orthomode transducer (OMT) that sends the
two polarizations supported by the feed to separate receiver chains.
The microwave plumbing is such that a single receiver chain (half of a ``differencing assembly'',  \citet{jarosik/etal:2003})
differences electric fields with two nearly parallel linear polarization vectors, one from each telescope.

Because of the large focal plane the beams are not symmetric nor
are they Gaussian. In addition, as anticipated,
cool-down distortions of the primary reflectors distort the W-band
and V-band beam shapes.

As noted above, precise knowledge of the beams is essential for accurately computing the
CMB angular spectrum.  For WMAP, one of the most CPU intensive aspects of the data analysis was
modeling the beams.  The goal was to find an antenna pattern that matched the in-flight measurements of Jupiter using
each of the two telescopes, the sidelobe measurements from the Moon, and pre-flight ground-based measurements.
Although intensive beam modeling would have been needed in any case, it was all the more important for WMAP 
because of the cool-down distortions. Due to composite CTE mismatches, the in-flight variations across the center of the primary
were approximately $0.5 - 1$~mm,  as shown in~\citet{hill/etal:2009}. To understand the distortions, we developed a model
in which the surfaces were parametrized by over four hundred Fourier modes~\citep{jarosik/etal:2007,hill/etal:2009}. 
For each set of parameters, the full physical optics solution weighted by the measured passband was computed for all feeds. 
We then compared the physical optics prediction to measurements of Jupiter.
The solution required running for many months using hundred processors on Silicon Graphics Origin 300 machines.
The reduced $\chi^2$ of the fit are typically $<1.1$, suggesting quite a good fit given the overall complexity of 
the system coupled with multiple precise measurements of Jupiter. With the combination of the measurements and the model,
the beam could be characterized at the $-40$ to $-50$ dB level and the beam solid angles was determined to better than
$1$\% \citep{hill/etal:2009}.

\subsection{Planck}
\label{sec:planck}

Planck is significantly more complex and sensitive than WMAP. Thus,  although it was conceived at roughly
the same time, it took longer to build.
Planck combines two different technologies in one focal plane. The low-frequency instrument (LFI) uses coherent 
detectors cooled to 20~K operating between 30 and 70 GHz. The high-frequency instrument uses
bolometric detectors cooled to 0.1~K operating between 90 and 900 GHz. An attractive aspect of the design is that it covers a very 
wide frequency range in one spacecraft.  Much of the telescope optimization that was done for WMAP, and discussed earlier, 
was independently carried out for Planck.  However, because of Planck's wide frequency range and much higher sensitivity  
(Table~\ref{tab:satspecs}), many of the optical system specifications were considerably more demanding than similar ones for WMAP.

Planck is the first CMB satellite to use a single telescope.  This choice was driven by the bolometers in the HFI, the most sensitive detectors aboard Planck. The LFI instrument uses a differential receiver except that one of the inputs is terminated in  a cold load as opposed to the sky as with WMAP. For bolometers,  the analogue to a differential microwave receiver is a Fourier Transform Spectrometer (FTS). Not only would an FTS be more cumbersome and costly to build, but the intrinsically large FTS bandwidth is not amenable to single mode optics or to minimizing the photon background. A single telescope was the natural choice.

The Planck telescope, which is an aplanatic Gregorian~\citep{tauber/etal:2010},
is roughly 20\% larger than WMAP's. Planck's primary is 1.56$\times$1.89~m and the secondary is
1.05$\times$1.10~m. Thus, at a given frequency it has about 20\% more resolution. As with WMAP, the drivers for the design are to get the maximum number of feeds in the focal plane, with the most symmetric beams, in the most compact telescope. Planck had the additional challenge of supporting two different technologies in the focal plane operating at different temperatures.  The in-flight surface accuracy is significantly better than WMAP's.

\begin{figure}[tbh]
\begin{center}
\includegraphics[width=6.in]{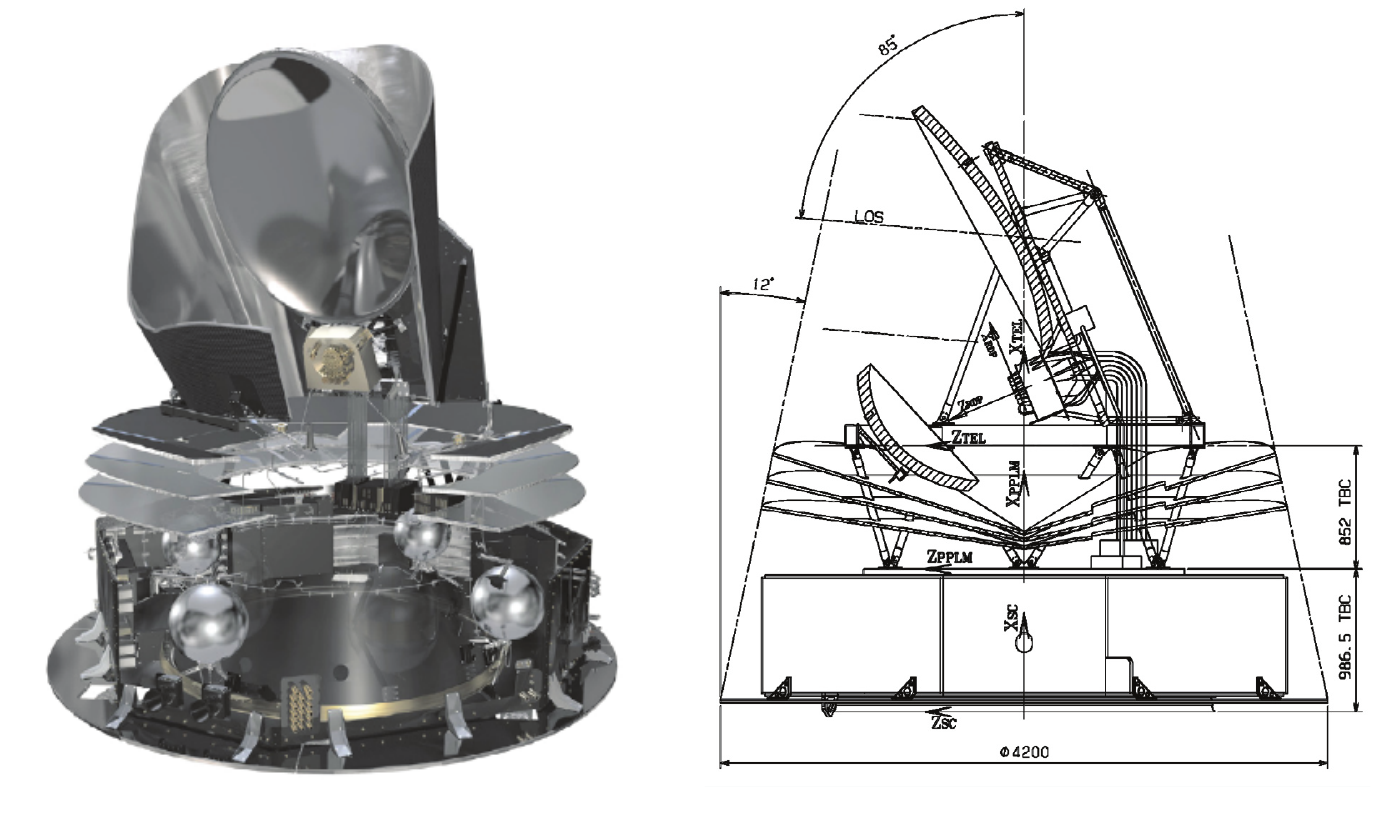}
\caption{\small
A cut-away view and line drawing of Planck. The focal plane is located just below the primary reflector. Planck spins about a
vertical axis. As with WMAP, Planck is located at the second Lagrange point, roughly $1.5\times10^6$~km away from Earth.
In this orientation, the Sun, Earth, and Moon are in the direction of the bottom of the page. Note the open ground shield around the optical system and the relatively large secondary reflector. Figures from \citet{tauber/etal:2010} and \citet{tauber/etal:2010b}.
\label{fig:planck}
}
\end{center}
\end{figure}

Planck HFI observations were completed in early 2012 when the instrument ran out of liquid cryogens as expected.  Early publications from Planck have demonstrated the excellent performance of the instrument, and more results and CMB maps will be released by the Planck team in the coming years.

\subsection{A Future Satellite}

Because a satellite is the ideal platform for measuring the CMB, considerable effort has gone into designing the
next generation instrument. After Planck, there will be little motivation for measuring the primary temperature anisotropy for
$\ell \ltsim 2500$ if the fluctuations turn out to be Gaussian to the limits of Planck's detector noise. However, there is quite a bit more
to be learned from measuring the polarization, and particularly the B-mode, as described
in Section~\ref{sec:polarizationterminology}. At the time of this writing, there are two relatively mature US satellite
concepts: EPIC~\citep{bock/epic:2008,bock/epicim:2009} and PIXIE~\citep{kogut/etal:2011}. These are very different missions.
EPIC is based on the next generation single-moded detectors operating in the limit of CMB photon noise. In the EPIC-IM design about 11,000 transition
edge sensor bolometers are fed by a 4~K cooled crossed-Dragone telescope that provides sensitivity to $\ell \gtsim 1500$.
The detectors are cooled to $\sim$100~mK by means of a continuously
operating adiabatic demagnetization refrigerator.  PIXIE, on the other hand, operates in the limit of many modes that are measured by just four detectors
at the output of a polarizing Fourier Transform Spectrometer (FTS). Its sensitivity to B-mode science is at $\ell \ltsim 200$.
The frequency spectrum of the anisotropy is determined by
scanning the reflectors of the FTS much in the same was as was done by COBE/FIRAS in its absolute measurement on the sky and by BAM \citep{tucker/etal:1997} in its search for the spectrum of the anisotropy.

A European collaboration has proposed the COrE mission concept~\citep{core}. It is based on $\sim$1.5 meter diameter rotating half-wave plate as the
first optical element that feeds a two mirror system. The primary reflector is $\sim$1.8~m diameter. The resolution varies between 
23 and 1.3 arc-minutes for 45 and 795~GHz, respectively.  Detection is based on feed horn coupled superconducting detectors to 
maintain control of systematics and achieve single mode coupling with high sensitivity.

\section{Acknowledgements}
The authors are collaborators on just a subset of the instruments discussed in this article. We have learned about other optical
system through papers and talking with colleagues, though all errors are of course ours. We would especially like to thank
Elia Battistelli,
Cynthia Chiang,
Nils Halverson,
Bill Jones,
Akito Kusaka,
Jeff McMahon,
Lucio Piccirillo, 
Jon Sievers, 
Suzanne Staggs,
Ed Wollack,
and Sasha Zhiboedov
for discussions and suggestions that improved this article. Ed Wollack in particular made numerous helpful comments.  We also thank
Chaoyun Bao,
Angela Glenn,
Michael Milligan,
and
Keith Thompson
for help with the figures.


\begin{table}[htdp]
\begin{center}
\tiny
\begin{tabular}{|l|l|c|c|l|c|c|}
\hline
Experiment & Type$^a$ &  N$_{feeds}$$^{b}$ & N$_{det}$$^{c}$ & Optical Design & Plat.$^{d}$ & Reference \\
\hline
Isotropometer & Dicke-switched & 2 & 1 & Feed & Gnd & \cite{wilkinson/partridge:1967}\\
Stanford &  Dicke-switched & 2 & 2 & Parabola  & Gnd  & \cite{conklin:1967}\\
Crawford Hill &  Maser & 1 & 1& Horn/reflector &Gnd &\cite{wilson/penzias:1967} \\
Aerospace &  Coherent & 2 & 2& 4.6 m telescope &Gnd & \cite{epstein:1967}\\
White Mountain &  Coherent & 2 & 2& Dicke-switch 2 feeds  & Gnd & \cite{conklin:1969}\\
Ratan & Coherent & 1 & 1& Parabolic &Gnd & \cite{pariiskii/pyatunina:1971}\\
KaDip &  Dicke-switched & 2 & 2& Feeds &Gnd & \cite{boughn:1971}\\
XBal & Dicke-switched & 2 & 2& Feeds &Bal & \cite{henry:1971}\\
NRAO-P & Parametric amp & 2 & 2&140ft  Greenbank Centered Cass. &Gnd & \cite{pariiskii:1973}\\
Goldstone &  Maser & 1 & 1& 64m Goldstone Centered Cass. &Gnd & \cite{carpenter:1973}\\
Parkes &  Correlation & 2 & 2& 64m Parkes Centered Cass. &Gnd & \cite{stankevich:1974}\\
U2 &  Dicke-switched & 2 & 2& Two corrugated feeds & Plane & \cite{smoot/etal:1977}\\
Testa-Griga &  Bolometers & 1 & 1& Cass &Gnd & \cite{caderni/etal:1977}\\
Greenbank-R &  Parametric amp & 2 & 2&  Cass &Gnd & \cite{rudnick:1978}\\
MIT &  Bolometers & 2 & 2& Two 0.4m sph refl + flat + lightpipe &Bal & \cite{muehlner:1977}\\
KKaQBal &  Dicke-switched & 2 & 2& Feeds &Bal & \cite{cheng/etal:1979}\\
PolCMB &  Coherent (P)&1 &1 & First pol & Gnd  & \cite{nanos:1979}\\
KPRO &  Coherent &1 &2 & 11m Cas & Gnd  & \cite{partridge:1980}\\
Convair &  Bolometer & 1&  1&  lens/feed on  FTS with chopper & Plane&\cite{fabbri/etal:1980a} \\
DBal &  Bolometer &1 & 1 & Lens/feed & Bal&\cite{fabbri/etal:1980b}  \\
NRAO91 &  Coherent & 2 &2 &     91m NRAO & Gnd & \cite{ledden/etal:1980}\\
OVRO40 &  Coherent   &2 &2 & OVRO 40m &  Gnd & \cite{seielstad/etal:1981}\\
GBank-UW &  Coherent & 1& 1 & 140ft GB & Gnd & \cite{uson/wilkinson:1982}\\
MaserBal &  Maser &2 &1  & Dicke switch  &Bal & \cite{fixsen/etal:1983}\\
WBal &  Mixer & 1& 1 & Dicke-switched chopper &   Bal & \cite{lubin/etal:1983}\\
JodrellBank & Coherent &2 & 2 & Prime focus of 100ft MkII telescope  &Gnd  & \cite{lasenby/davis:1983}\\
Relikt &  Parametric amp &2 & 2 & Two feeds  &Sat & \cite{strukov/skulachev:1984} \\
NCP &  Coherent &2 &2  & Dicke switch w/ feed-fed parabolas  & Gnd& \cite{mandolesi/etal:1986}\\
Tenerife &  Coherent & 2& 2 & Two feeds with chopping plate  &Gnd & \cite{davies/etal:1987}\\
IAB-I &  Bolometer & 1& 1 & 1m parabola   &Ant & \cite{dalloglio/debernardis:1988}\\
MITBal2 &  Bolometer & 2& 4 & Horns and chopper  &Bal & \cite{halpern/etal:1988}\\
OVRO & Maser & 2& 1 &  OVRO 40 m  & Gnd & \cite{readhead/etal:1989}\\
SKInt &  Mixer & 2& 2 &2 feeds  & Int & \cite{timbie/wilkinson:1990}\\
FIRS &  Bolometer& 1& 4& Single cryogenic horn/lens & Bal & \citet{page/etal:1990} \\
ARGO &  Bolometer  & 1& 4 & 1.2 m Centered Casegrain   & Bal& \citet{debernardis/etal:1990}\\
SP/ACME   & SIS Mixer & 1 &  1  &  1m Decentered Gregorian & Ant& \citet{meinhold/lubin:1991}\\
COBE &  Coherent  & 10& 12 & Feeds    & Sat& \citet{smoot/etal:1991} \\
SP/ACME   & HEMT & 1 &  1  &  1m Decentered Gregorian & Ant& \citet{gaier/etal:1992}\\
19GHz &   Maser & 1& 1 &  Feed plus lens  & Bal& \citet{boughn/etal:1992} \\
MAX & Bolometer & 3  & 3 & Same as SP/ACME &   Bal&  \citet{alsop/etal:1992}  \\
IAB-II & Bolometer & 1  & 1 & 0.45 m Decentered Cassegrain &    Ant &  \citet{piccirillo/calisse:1993}\\
White Dish &  Bolometer & 1 & 1 & 1.2 m Centered Cassegrain & Ant & \citet{tucker/etal:1993} \\
SASK & HEMT & 1 & 1   & 1.2 m Off-axis parabola  & Gnd& \citet{wollack/etal:1993} \\
MSAM &  Bolometers  & 1& 4   &  1.4m Decentered Cassegrain & Bal& \citet{cheng/etal:1994} \\
PYTHON &  Bolometers  & 4  & 4 & Off-axis parabola & Ant& \citet{dragovan/etal:1994}\\
CAT &Coherent  & 3 & 6  &   Int & Gnd& \citet{osullivan/etal:1995} \\
BAM & Bolometer & 2 & 2  &  FTS with Off-axis parabola  & Bal& \citet{tucker/etal:1997}\\
SuZIE &  Bolometer  &6 & 6 & CSO   & Gnd & \citet{ganga/etal:1997} \\
IAC-BAR &  Bolometer  & 4& 4 & 0.45m Decentered Para/hyper  & Gnd & \citet{piccirillo/etal:1997} \\
QMAP & HEMT & 3 &  6 & Feed + parabolic reflector &  Bal & \citet{deoliveiracosta/etal:1998} \\
Toco &  HEMT/SIS & 5 &  8 & Feed + parabolic reflector &  Gnd & \citet{miller/etal:1999} \\
JB-IAC & HEMTs & 2 &  2 & Feed plus reflector &  Int & \citet{dicker/etal:1999} \\
HACME &  HEMT & 1 & 1 & Decentered Greg. & Bal & \citet{staren/etal:2000} \\
Viper &  HEMT & 2 & 2 & Decentered Aplanatic Greg. with chopper & Ant & \citet{peterson/etal:2000}\\
RING5M &  HEMT & 2 & 1 & 5.5m/40m OVRO Centered Cass. & Gnd & \citet{leitch/etal:2000} \\
BOOMERANG &  Bolometers & 16 & 16 & Decentered Greg.  with tertiary & Bal & \cite{debernardis/etal:2000} \\
MAXIMA & Bolometers & 16 & 16 & Decentered Greg. with tertiary & Bal & \citet{hanany/etal:2000} \\
PIQUE & HEMT (P)& 1&2 & 1.2m Off-axis parabola & Gnd &  \citet{hedman/etal:2001}  \\
POLAR & HEMT (P)& 1&2 & Cryo feed & Gnd &  \citet{keating/etal:2001}  \\
CBI &  HEMTs & 13 &  13 & Centered Cass. &  Int & \citet{padin/etal:2001} \\
DASI &  HEMTs & 13 &  13 & Feeds  &  Int & \citet{halverson/etal:2002} \\
Archeops &  Bolometers & 21 &  21 & Decentered Greg. &  Bal & \citet{benoit/etal:2003} \\
COMPASS &  HEMT (P) & 1 & 2 & 2.6m Centered Cass. & Gnd & \citet{farese/etal:2003} \\
VSA & HEMTs  & 14 & 14 & Feeds+ Off-axis parabolas & Int  &  \citet{grainge/etal:2003}\\
WMAP & HEMTs  & 20 & 40 & Decentered Greg. & Sat &  \citet{bennett/etal:2003d}\\
Acbar & Bolometers  & 16 & 16 & Decentered Aplan. Greg. w/ chopper  & Ant  &  \citet{kuo/etal:2004}\\
BEAST &  HEMT &  8 &  8 & 2~m Decentered Greg. &  Bal & \citet{meinhold/etal:2005} \\
CAPMAP & HEMT (P)& 16&32 & 7m Decentered Cassegrain & Gnd &  \citet{barkats/etal:2005}  \\
MINT & Mixers  & 4 & 4 & 30 cm Cass. & Int  &  \citet{fowler/etal:2005}\\
MAXIPOL &  Bolometer  (P) & 16   & 16   & Decentered Greg. & Bal & \citet{johnson/etal:2007}  \\
QUAD &  Bolometer (P) & 31  &  62  & Centered Cass. w/lenses & Ant  & \citet{ade/etal:2008} \\
WMPol &  HEMT (P) & 4  &  3  & 2.2~m Decentered Greg. & Ant  & \citet{levy/etal:2008} \\
SPT & TES Bolometer  &  966 & 966 & Decentered Greg. w/lens & Gnd & \citet{staniszewski/etal:2009} \\
BICEP & Bolometer (P) & 49  & 98  & Centered refractive  & Ant  &  \citet{chiang/etal:2010} \\
ACT & TES Bolometer  &  Planar & 3072 & Decentered Greg. w/lenses  & Gnd & \citet{fowler/etal:2010}  \\
QUIET & HEMTs  (P) &  110  &  110  &  Cross-Dragone & Gnd & \citet{bischoff/etal:2011} \\
Planck & HEMT/Bol (P) & 47  & 74  &  1.6m aplanatic Gregorian & Sat & \citet{tauber/etal:2010}\\
\hline
\multicolumn{7}{l}{ $^{a}$ Detector technology. (P) indicated a design specifically for polarization.} \\
\multicolumn{7}{l}{ $^{b}$ Number of feeds} \\
\multicolumn{7}{l}{ $^{c}$ Number of independent detectors} \\
\multicolumn{7}{l}{ $^{d}$ Platform. Gnd = ground; Bal = balloon; Sat = satellite; plane = airplane; Ant = Antarctica; Int = interferometer} \\
\hline
\end{tabular}
\caption{\tiny CMB polarization and anisotropy experiments with comments on their optical and detector configurations. Much of the information about experiments prior to 2000 is adapted from ``Finding the Big Bang" \citep{ftbb:2009}. We include only instruments with astrophysical results as this indicates some level of the maturity of the design. Except for Planck, the citations are
for the first astrophysical result from the instrument.
\label{tab:experiments} }
\end{center}
\end{table}%

\nsectionstar{Cross-References}
Other chapters within PSSS which are related to yours. Using the Table of Contents, you can simply list cross-references by chapter title under this heading.

\bibliographystyle{wmap}
\bibliography{hnp_mar26,apj-jour}

\begin{thebibliography}{219}
\expandafter\ifx\csname natexlab\endcsname\relax\def\natexlab#1{#1}\fi

\bibitem[{{Ade} et~al.(2008)}]{ade/etal:2008}
{Ade}, P., et~al. 2008, \apj, 674, 22

\bibitem[{{Ade} et~al.(2010)}]{ade/etal:2010}
{Ade}, P.~A.~R., et~al. 2010, \aap, 520, A11

\bibitem[{{Aikin} et~al.(2010)}]{aikin/etal:2010}
{Aikin}, R.~W., et~al. 2010, in Society of Photo-Optical Instrumentation
  Engineers (SPIE) Conference Series, Vol. 7741

\bibitem[{{Ali-Ha{\"i}moud} et~al.(2009){Ali-Ha{\"i}moud}, {Hirata}, \&
  {Dickinson}}]{Ali-Haimoud2009}
{Ali-Ha{\"i}moud}, Y., {Hirata}, C.~M., \& {Dickinson}, C. 2009, \mnras, 395,
  1055

\bibitem[{{Alsop} et~al.(1992)}]{alsop/etal:1992}
{Alsop}, D.~C., et~al. 1992, \apj, 395, 317

\bibitem[{{Arnold} et~al.(2010)}]{arnold/etal:2010}
{Arnold}, K., et~al. 2010, in Society of Photo-Optical Instrumentation
  Engineers (SPIE) Conference Series, Vol. 7741

\bibitem[{{Baker} et~al.(1999)}]{baker/etal:1999}
{Baker}, J.~C., et~al. 1999, \mnras, 308, 1173

\bibitem[{{Balanis}(2005)}]{balanis}
{Balanis}, D. 2005, Antenna Theory: Analysis and Design, 3rd edn.
  (Wiley-Interscience)

\bibitem[{{Barkats} et~al.(2005)}]{barkats/etal:2005}
{Barkats}, D., et~al. 2005, \apjl, 619, L127

\bibitem[{{Barnes} et~al.(2002)}]{barnes/etal:2002}
{Barnes}, C., et~al. 2002, \apjs, 143, 567

\bibitem[{{Basak} et~al.(2006){Basak}, {Hajian}, \&
  {Souradeep}}]{basek/etal:2006}
{Basak}, S., {Hajian}, A., \& {Souradeep}, T. 2006, Phys. Rev. D, 74, 021301

\bibitem[{Bennett et~al.(2003)}]{bennett/etal:2003d}
Bennett, C.~L. et~al. 2003, Astrophys. J., 583, 1

\bibitem[{{Beno{\^ i}t} et~al.(2003)}]{benoit/etal:2003}
{Beno{\^ i}t}, A., et~al. 2003, \aap, 399, L19

\bibitem[{{Bock} et~al.(2008)}]{bock/epic:2008}
{Bock}, J., et~al. 2008, ArXiv e-prints

\bibitem[{{Bock} et~al.(2009)}]{bock/epicim:2009}
---. 2009, ArXiv e-prints

\bibitem[{{Boggess} et~al.(1992)}]{boggess/etal:1992}
{Boggess}, N.~W., et~al. 1992, \apj, 397, 420

\bibitem[{{Bond} \& {Efstathiou}(1984)}]{bond/efstathiou:1984}
{Bond}, J.~R. \& {Efstathiou}, G. 1984, \apjl, 285, L45

\bibitem[{Born \& Wolf(1980)}]{born/wolf}
Born, M. \& Wolf, E. 1980, Principles of Optics, sixth edn. (Pergamon Press)

\bibitem[{{Boughn} et~al.(1992){Boughn}, {Cheng}, {Cottingham}, \&
  {Fixsen}}]{boughn/etal:1992}
{Boughn}, S.~P., {Cheng}, E.~S., {Cottingham}, D.~A., \& {Fixsen}, D.~J. 1992,
  \apjl, 391, L49

\bibitem[{{Boughn} et~al.(1971){Boughn}, {Fram}, \& {Patridge}}]{boughn:1971}
{Boughn}, S.~P., {Fram}, D.~M., \& {Patridge}, R.~B. 1971, \apj, 165, 439

\bibitem[{{Brown} \& {Prata}(1994)}]{brown/prata:1994}
{Brown}, K.~W. \& {Prata}, A. 1994, IEEE Transactions on Antennas and
  Propagation, 42, 1145

\bibitem[{{Caderni} et~al.(1977){Caderni}, {Fabbri}, {de Cosmo}, {Melchiorri},
  {Melchiorri}, \& {Natale}}]{caderni/etal:1977}
{Caderni}, N., {Fabbri}, R., {de Cosmo}, V., {Melchiorri}, B., {Melchiorri},
  F., \& {Natale}, V. 1977, \prd, 16, 2424

\bibitem[{{Carlstrom} et~al.(2011)}]{carlstrom/etal:2011}
{Carlstrom}, J.~E., et~al. 2011, \pasp, 123, 568

\bibitem[{{Carpenter} et~al.(1973){Carpenter}, {Gulkis}, \&
  {Sato}}]{carpenter:1973}
{Carpenter}, R.~L., {Gulkis}, S., \& {Sato}, T. 1973, \apjl, 182, L61

\bibitem[{{Chang} \& {Prata}(1999)}]{chang/prata:1999}
{Chang}, S. \& {Prata}, A. 1999, Antennas and Propagation Society International
  Symposium, IEEE, 2, 1140

\bibitem[{{Chang} \& {Prata}(2005)}]{chang/prata:2005}
---. 2005, Journal of the Optical Society of America A, 22, 2454

\bibitem[{{Charlassier}(2008)}]{charlassier/etal:2008}
{Charlassier}, R. 2008, ArXiv e-prints

\bibitem[{{Cheng} et~al.(1979){Cheng}, {Saulson}, {Wilkinson}, \&
  {Corey}}]{cheng/etal:1979}
{Cheng}, E.~S., {Saulson}, P.~R., {Wilkinson}, D.~T., \& {Corey}, B.~E. 1979,
  \apjl, 232, L139

\bibitem[{{Cheng} et~al.(1994)}]{cheng/etal:1994}
{Cheng}, E.~S., et~al. 1994, \apjl, 422, L37

\bibitem[{{Chiang} et~al.(2010)}]{chiang/etal:2010}
{Chiang}, H.~C., et~al. 2010, \apj, 711, 1123

\bibitem[{{Church}(1995)}]{church:1995}
{Church}, S.~E. 1995, \mnras, 272, 551

\bibitem[{Clarricoats \& Olver(1984)}]{clarricoats/olver:1984}
Clarricoats, P. \& Olver, A. 1984, Corrugated horns for microwave antennas, IEE
  electromagnetic waves series (P. Peregrinus on behalf of the Institution of
  Electrical Engineers)

\bibitem[{{Conklin}(1969{\natexlab{a}})}]{conklin1969thesis}
{Conklin}, E.~K. 1969{\natexlab{a}}, Ph.D. thesis, Stanford University

\bibitem[{{Conklin}(1969{\natexlab{b}})}]{conklin:1969}
---. 1969{\natexlab{b}}, \nat, 222, 971

\bibitem[{{Conklin} \& {Bracewell}(1967)}]{conklin:1967}
{Conklin}, E.~K. \& {Bracewell}, R.~N. 1967, \nat, 216, 777

\bibitem[{{Crill} et~al.(2003)}]{crill/etal:2003}
{Crill}, B.~P., et~al. 2003, \apjs, 148, 527

\bibitem[{{dall'Oglio} \& {de Bernardis}(1988)}]{dalloglio/debernardis:1988}
{dall'Oglio}, G. \& {de Bernardis}, P. 1988, \apj, 331, 547

\bibitem[{{Das} et~al.(2011)}]{das/etal:2011}
{Das}, S., et~al. 2011, \apj, 729, 62

\bibitem[{{Davies} et~al.(1987){Davies}, {Lasenby}, {Watson}, {Daintree},
  {Hopkins}, {Beckman}, {Sanchez Almeida}, \& {Rebolo}}]{davies/etal:1987}
{Davies}, R.~D., {Lasenby}, A.~N., {Watson}, R.~A., {Daintree}, E.~J.,
  {Hopkins}, J., {Beckman}, J., {Sanchez Almeida}, J., \& {Rebolo}, R. 1987,
  Nature, 326, 462

\bibitem[{{de Bernardis} et~al.(1990)}]{debernardis/etal:1990}
{de Bernardis}, P., et~al. 1990, \apjl, 360, L31

\bibitem[{{de Bernardis} et~al.(2000)}]{debernardis/etal:2000}
---. 2000, Nature, 404, 955

\bibitem[{{de Oliveira-Costa} et~al.(1998){de Oliveira-Costa}, {Devlin},
  {Herbig}, {Miller}, {Netterfield}, {Page}, \&
  {Tegmark}}]{deoliveiracosta/etal:1998}
{de Oliveira-Costa}, A., {Devlin}, M.~J., {Herbig}, T., {Miller}, A.~D.,
  {Netterfield}, C.~B., {Page}, L.~A., \& {Tegmark}, M. 1998, \apjl, 509, L77

\bibitem[{{de Oliveira-Costa} et~al.(1997){de Oliveira-Costa}, {Kogut},
  {Devlin}, {Netterfield}, {Page}, \& {Wollack}}]{deOliveira1997}
{de Oliveira-Costa}, A., {Kogut}, A., {Devlin}, M.~J., {Netterfield}, C.~B.,
  {Page}, L.~A., \& {Wollack}, E.~J. 1997, \apjl, 482, L17

\bibitem[{{Dicker} et~al.(1999)}]{dicker/etal:1999}
{Dicker}, S.~R., et~al. 1999, \mnras, 309, 750

\bibitem[{{Dragone}(1978)}]{dragone:1978}
{Dragone}, C. 1978, AT T Technical Journal, 57, 2663

\bibitem[{{Dragone}(1982)}]{dragone:1982}
---. 1982, IEEE Transactions on Antennas and Propagation, 30, 331

\bibitem[{{Dragone}(1983{\natexlab{a}})}]{dragone:1983}
---. 1983{\natexlab{a}}, IEEE Transactions on Antennas and Propagation, 31, 764

\bibitem[{{Dragone}(1983{\natexlab{b}})}]{dragone:1983b}
---. 1983{\natexlab{b}}, Electronics Letters, 19, 1061

\bibitem[{{Dragovan} et~al.(1994){Dragovan}, {Ruhl}, {Novak}, {Platt}, {Crone},
  {Pernic}, \& {Peterson}}]{dragovan/etal:1994}
{Dragovan}, M., {Ruhl}, J.~E., {Novak}, G., {Platt}, S.~R., {Crone}, B.,
  {Pernic}, R., \& {Peterson}, J.~B. 1994, \apjl, 427, L67

\bibitem[{{Draine} \& {Lazarian}(1998)}]{draine/lazarian:1998b}
{Draine}, B.~T. \& {Lazarian}, A. 1998, \apj, 508, 157

\bibitem[{{Draine} \& {Lazarian}(1999)}]{Draine1999}
---. 1999, \apj, 512, 740

\bibitem[{{Epstein}(1967)}]{epstein:1967}
{Epstein}, E.~E. 1967, \apjl, 148, L157

\bibitem[{{Essinger-Hileman}(2011)}]{essinger-hileman:2011}
{Essinger-Hileman}, T. 2011, Ph.D. thesis, Princeton University

\bibitem[{{Fabbri} et~al.(1980{\natexlab{a}}){Fabbri}, {Guidi}, {Melchiorri},
  \& {Natale}}]{fabbri/etal:1980b}
{Fabbri}, R., {Guidi}, I., {Melchiorri}, F., \& {Natale}, V.
  1980{\natexlab{a}}, Physical Review Letters, 44, 1563

\bibitem[{{Fabbri} et~al.(1980{\natexlab{b}}){Fabbri}, {Melchiorri},
  {Melchiorri}, {Natale}, {Caderni}, \& {Shivanandan}}]{fabbri/etal:1980a}
{Fabbri}, R., {Melchiorri}, B., {Melchiorri}, F., {Natale}, V., {Caderni}, N.,
  \& {Shivanandan}, K. 1980{\natexlab{b}}, \prd, 21, 2095

\bibitem[{{Farese} et~al.(2003)}]{farese/etal:2003}
{Farese}, P.~C., et~al. 2003, New Astronomy Reviews, 47, 1033

\bibitem[{{Filippini} et~al.(2010)}]{filippini/etal:2010}
{Filippini}, J.~P., et~al. 2010, in Society of Photo-Optical Instrumentation
  Engineers (SPIE) Conference Series, Vol. 7741

\bibitem[{{Finlay} et~al.(2010)}]{igrf2010}
{Finlay}, C.~C., et~al. 2010, Geophysical Journal International, 183, 1216

\bibitem[{{Fixsen} et~al.(1983){Fixsen}, {Cheng}, \&
  {Wilkinson}}]{fixsen/etal:1983}
{Fixsen}, D.~J., {Cheng}, E.~S., \& {Wilkinson}, D.~T. 1983, Physical Review
  Letters, 50, 620

\bibitem[{{Fixsen} \& {Mather}(2002)}]{fixsen/mather:2002}
{Fixsen}, D.~J. \& {Mather}, J.~C. 2002, \apj, 581, 817

\bibitem[{{Fowler} et~al.(2005)}]{fowler/etal:2005}
{Fowler}, J.~W., et~al. 2005, \apjs, 156, 1

\bibitem[{{Fowler} et~al.(2007)}]{fowler/etal:2007}
---. 2007, Applied Optics, 46, 3444

\bibitem[{{Fowler} et~al.(2010)}]{fowler/etal:2010}
---. 2010, \apj, 722, 1148

\bibitem[{{Gaier} et~al.(2003){Gaier}, {Lawrence}, {Seiffert}, {Wells},
  {Kangaslahti}, \& {Dawson}}]{gaier/etal:2003}
{Gaier}, T., {Lawrence}, C.~R., {Seiffert}, M.~D., {Wells}, M.~M.,
  {Kangaslahti}, P., \& {Dawson}, D. 2003, \nar, 47, 1167

\bibitem[{{Gaier} et~al.(1992){Gaier}, {Schuster}, {Gundersen}, {Koch},
  {Seiffert}, {Meinhold}, \& {Lubin}}]{gaier/etal:1992}
{Gaier}, T., {Schuster}, J., {Gundersen}, J., {Koch}, T., {Seiffert}, M.,
  {Meinhold}, P., \& {Lubin}, P. 1992, \apjl, 398, L1

\bibitem[{{Ganga} et~al.(1997){Ganga}, {Ratra}, {Church}, {Sugiyama}, {Ade},
  {Holzapfel}, {Mauskopf}, \& {Lange}}]{ganga/etal:1997}
{Ganga}, K., {Ratra}, B., {Church}, S.~E., {Sugiyama}, N., {Ade}, P.~A.~R.,
  {Holzapfel}, W.~L., {Mauskopf}, P.~D., \& {Lange}, A.~E. 1997, \apj, 484, 517

\bibitem[{{Graham}(1973)}]{graham:1973}
{Graham}, R. 1973, in IEEE International Conference on Radar -- Present and
  Future (IEEE),  134--139

\bibitem[{{Grainge} et~al.(2003)}]{grainge/etal:2003}
{Grainge}, K., et~al. 2003, \mnras, 341, L23

\bibitem[{{Griffin} et~al.(2002){Griffin}, {Bock}, \&
  {Gear}}]{griffin/bock/gear:2002}
{Griffin}, M.~J., {Bock}, J.~J., \& {Gear}, W.~K. 2002, Applied Optics, 41,
  6543

\bibitem[{{Halpern} et~al.(1988){Halpern}, {Benford}, {Meyer}, {Muehlner}, \&
  {Weiss}}]{halpern/etal:1988}
{Halpern}, M., {Benford}, R., {Meyer}, S., {Muehlner}, D., \& {Weiss}, R. 1988,
  \apj, 332, 596

\bibitem[{{Halverson} et~al.(2002)}]{halverson/etal:2002}
{Halverson}, N.~W., et~al. 2002, \apj, 568, 38

\bibitem[{{Hanany} \& {Marrone}(2002)}]{hanany/marrone:2002}
{Hanany}, S. \& {Marrone}, D.~P. 2002, \ao, 41, 4666

\bibitem[{{Hanany} \& {Rosenkranz}(2003)}]{hanany/rosenkranz:2003}
{Hanany}, S. \& {Rosenkranz}, P. 2003, New. Ast. Rev., 47, 1159

\bibitem[{{Hanany} et~al.(2000)}]{hanany/etal:2000}
{Hanany}, S., et~al. 2000, \apjl, 545, L5

\bibitem[{{Hanson} et~al.(2010){Hanson}, {Lewis}, \&
  {Challinor}}]{hanson/etal:2010}
{Hanson}, D., {Lewis}, A., \& {Challinor}, A. 2010, \prd, 81, 103003

\bibitem[{Hecht(1987)}]{hecht:OPTICS}
Hecht, E. 1987, Optics (Reading, MA, USA: Addison-Wesley)

\bibitem[{{Hedman} et~al.(2001){Hedman}, {Barkats}, {Gundersen}, {Staggs}, \&
  {Winstein}}]{hedman/etal:2001}
{Hedman}, M.~M., {Barkats}, D., {Gundersen}, J.~O., {Staggs}, S.~T., \&
  {Winstein}, B. 2001, \apjl, 548, L111

\bibitem[{{Henry}(1971)}]{henry:1971}
{Henry}, P.~S. 1971, \nat, 231, 516

\bibitem[{{Hill} et~al.(2009)}]{hill/etal:2009}
{Hill}, R.~S., et~al. 2009, \apjs, 180, 246

\bibitem[{{Hinderks} et~al.(2009)}]{hinderks/etal:2009}
{Hinderks}, J.~R., et~al. 2009, \apj, 692, 1221

\bibitem[{{Hinshaw} et~al.(2007)}]{hinshaw/etal:2007}
{Hinshaw}, G., et~al. 2007, \apjs, 170, 288

\bibitem[{{Hu} et~al.(2003){Hu}, {Hedman}, \& {Zaldarriaga}}]{hhz:2003}
{Hu}, W., {Hedman}, M.~M., \& {Zaldarriaga}, M. 2003, Phys. Rev. D, 67, 043004

\bibitem[{{Imbriale} et~al.(2011){Imbriale}, {Gundersen}, \&
  {Thompson}}]{imbriale/etal:2011}
{Imbriale}, W.~A., {Gundersen}, J., \& {Thompson}, K.~L. 2011, IEEE
  Transactions on Antennas and Propagation, 59, 1972

\bibitem[{{Irwin} \& {Hilton}(2005)}]{irwin/hilton:2005}
{Irwin}, K. \& {Hilton}, G. 2005, {Cryogenic Particle Detection:
  Transition-Edge Sensors chapter} (Springer)

\bibitem[{{Jarosik} et~al.(2003)}]{jarosik/etal:2003}
{Jarosik}, N., et~al. 2003, \apjs, 145, 413

\bibitem[{{Jarosik} et~al.(2007)}]{jarosik/etal:2007}
---. 2007, \apjs, 170, 263

\bibitem[{{Johnson}(2011)}]{johnson:private}
{Johnson}, B. 2011, private communication

\bibitem[{{Johnson} et~al.(2007)}]{johnson/etal:2007}
{Johnson}, B.~R., et~al. 2007, \apj, 665, 42

\bibitem[{Jones(2005)}]{jones:2005}
Jones, W.~C. 2005, Ph.D. thesis, "California Institute of Technology"

\bibitem[{{Kamionkowski} et~al.(1997){Kamionkowski}, {Kosowsky}, \&
  {Stebbins}}]{kamionkowski/etal:1997}
{Kamionkowski}, M., {Kosowsky}, A., \& {Stebbins}, A. 1997, \prd, 55, 7368

\bibitem[{{Keating} et~al.(2001){Keating}, {O'Dell}, {de Oliveira-Costa},
  {Klawikowski}, {Stebor}, {Piccirillo}, {Tegmark}, \&
  {Timbie}}]{keating/etal:2001}
{Keating}, B.~G., {O'Dell}, C.~W., {de Oliveira-Costa}, A., {Klawikowski}, S.,
  {Stebor}, N., {Piccirillo}, L., {Tegmark}, M., \& {Timbie}, P.~T. 2001,
  \apjl, 560, L1

\bibitem[{{Keating} et~al.(1998){Keating}, {Timbie}, {Polnarev}, \&
  {Steinberger}}]{keating/etal:1998}
{Keating}, B.~G., {Timbie}, P.~T., {Polnarev}, A., \& {Steinberger}, J. 1998,
  \apj, 495, 580

\bibitem[{{Keisler} et~al.(2011)}]{keisler/etal:2011}
{Keisler}, R., et~al. 2011, \apj, 743, 28

\bibitem[{{Klypin} et~al.(1992){Klypin}, {Strukov}, \&
  {Skulachev}}]{klypin/etal:1992}
{Klypin}, A.~A., {Strukov}, I.~A., \& {Skulachev}, D.~P. 1992, \mnras, 258, 71

\bibitem[{{Knoke} et~al.(1984){Knoke}, {Partridge}, {Ratner}, \&
  {Shapiro}}]{knoke/etal:1984}
{Knoke}, J.~E., {Partridge}, R.~B., {Ratner}, M.~I., \& {Shapiro}, I.~I. 1984,
  \apj, 284, 479

\bibitem[{{Kogut} et~al.(1996){Kogut}, {Banday}, {Bennett}, {G{\'o}rski},
  {Hinshaw}, \& {Reach}}]{kogut/etal:1996}
{Kogut}, A., {Banday}, A.~J., {Bennett}, C.~L., {G{\'o}rski}, K.~M., {Hinshaw},
  G., \& {Reach}, W.~T. 1996, \apj, 460, 1

\bibitem[{{Kogut} et~al.(1992)}]{kogut/etal:1992}
{Kogut}, A., et~al. 1992, \apj, 401, 1

\bibitem[{{Kogut} et~al.(2011)}]{kogut/etal:2011}
---. 2011, \jcap, 7, 25

\bibitem[{{Komatsu} et~al.(2011)}]{komatsu/etal:2011}
{Komatsu}, E., et~al. 2011, \apjs, 192, 18

\bibitem[{Korsch(1991)}]{korsch:1991}
Korsch, D. 1991, Reflective optics (Academic Press)

\bibitem[{{Kovac} et~al.(2002){Kovac}, {Leitch}, {Pryke}, {Carlstrom},
  {Halverson}, \& {Holzapfel}}]{kovac/etal:2002}
{Kovac}, J.~M., {Leitch}, E.~M., {Pryke}, C., {Carlstrom}, J.~E., {Halverson},
  N.~W., \& {Holzapfel}, W.~L. 2002, Nature, 420, 772

\bibitem[{{Kuo} et~al.(2004)}]{kuo/etal:2004}
{Kuo}, C.~L., et~al. 2004, \apj, 600, 32

\bibitem[{{Kuo} et~al.(2008)}]{kuo/etal:2008}
{Kuo}, C.~L., et~al. 2008, in Society of Photo-Optical Instrumentation
  Engineers (SPIE) Conference Series, Vol. 7020

\bibitem[{{Lamarre}(1986)}]{lamarre:1986}
{Lamarre}, J.~M. 1986, Applied Optics, 25, 870

\bibitem[{{Lasenby} \& {Davies}(1983)}]{lasenby/davis:1983}
{Lasenby}, A.~N. \& {Davies}, R.~D. 1983, \mnras, 203, 1137

\bibitem[{{Lay} \& {Halverson}(2000)}]{lay/halverson:2000}
{Lay}, O.~P. \& {Halverson}, N.~W. 2000, \apj, 543, 787

\bibitem[{{Ledden} et~al.(1980){Ledden}, {Broderick}, {Brown}, \&
  {Condon}}]{ledden/etal:1980}
{Ledden}, J.~E., {Broderick}, J.~J., {Brown}, R.~L., \& {Condon}, J.~J. 1980,
  \aj, 85, 780

\bibitem[{{Leitch} et~al.(1997){Leitch}, {Readhead}, {Pearson}, \&
  {Myers}}]{leitch/etal:1997}
{Leitch}, E.~M., {Readhead}, A.~C.~S., {Pearson}, T.~J., \& {Myers}, S.~T.
  1997, \apjl, 486, L23

\bibitem[{{Leitch} et~al.(2000){Leitch}, {Readhead}, {Pearson}, {Myers},
  {Gulkis}, \& {Lawrence}}]{leitch/etal:2000}
{Leitch}, E.~M., {Readhead}, A.~C.~S., {Pearson}, T.~J., {Myers}, S.~T.,
  {Gulkis}, S., \& {Lawrence}, C.~R. 2000, \apj, 532, 37

\bibitem[{{Levy} et~al.(2008)}]{levy/etal:2008}
{Levy}, A.~R., et~al. 2008, \apjs, 177, 419

\bibitem[{{Lin}(2009)}]{lin:2009}
{Lin}, J. 2009, private communication for a study of a ``feed farm'' satellite
  concept available from
  http://phy-page-g5.princeton.edu/\~page/cmbpol\_feedfarm.pdf

\bibitem[{Love et~al.(1978)Love, Antennas, \& Society}]{love:1978}
Love, A., Antennas, I., \& Society, P. 1978, Reflector antennas, IEEE Press
  selected reprint series (IEEE Press)

\bibitem[{{Lubin} et~al.(1983){Lubin}, {Epstein}, \& {Smoot}}]{lubin/etal:1983}
{Lubin}, P.~M., {Epstein}, G.~L., \& {Smoot}, G.~F. 1983, Physical Review
  Letters, 50, 616

\bibitem[{{Maffei} et~al.(2010)}]{maffei/etal:2010}
{Maffei}, B., et~al. 2010, \aap, 520, A12

\bibitem[{{Mandolesi} et~al.(1986){Mandolesi}, {Calzolari}, {Cortiglioni},
  {Delpino}, \& {Sironi}}]{mandolesi/etal:1986}
{Mandolesi}, N., {Calzolari}, P., {Cortiglioni}, S., {Delpino}, F., \&
  {Sironi}, G. 1986, \nat, 319, 751

\bibitem[{{Mather}(1982)}]{mather:1982}
{Mather}, J.~C. 1982, Applied Optics, 21, 1125

\bibitem[{{McMahon} et~al.(2012)}]{mcmahon/etal:2012}
{McMahon}, J., et~al. 2012, ArXiv e-prints

\bibitem[{{McMahon} et~al.(2009)}]{mcmahon/etal:2009}
{McMahon}, J.~J., et~al. 2009, in American Institute of Physics Conference
  Series, ed. {B.~Young, B.~Cabrera, \& A.~Miller}, Vol. 1185,  511--514

\bibitem[{{Meinhold} \& {Lubin}(1991)}]{meinhold/lubin:1991}
{Meinhold}, P. \& {Lubin}, P. 1991, \apjl, 370, L11

\bibitem[{{Meinhold} et~al.(1993){Meinhold}, {Chingcuanco}, {Gundersen},
  {Schuster}, {Seiffert}, {Lubin}, {Morris}, \& {Villela}}]{meinhold/etal:1993}
{Meinhold}, P.~R., {Chingcuanco}, A.~O., {Gundersen}, J.~O., {Schuster}, J.~A.,
  {Seiffert}, M.~D., {Lubin}, P.~M., {Morris}, D., \& {Villela}, T. 1993, \apj,
  406, 12

\bibitem[{{Meinhold} et~al.(2005)}]{meinhold/etal:2005}
{Meinhold}, P.~R., et~al. 2005, \apjs, 158, 101

\bibitem[{{Mennella} et~al.(2011)}]{planck-LFI-inst:2011}
{Mennella}, A., et~al. 2011, ArXiv e-prints

\bibitem[{{Meyer} et~al.(1991){Meyer}, {Cheng}, \& {Page}}]{meyer/etal:1991}
{Meyer}, S.~S., {Cheng}, E.~S., \& {Page}, L.~A. 1991, \apjl, 371, L7

\bibitem[{{Miller} et~al.(1999)}]{miller/etal:1999}
{Miller}, A.~D., et~al. 1999, \apjl, 524, L1

\bibitem[{{Mizuguchi} et~al.(1978){Mizuguchi}, {Akagawa}, \&
  {Yokoi}}]{mizuguchi/ak/yok:1978}
{Mizuguchi}, Y., {Akagawa}, M., \& {Yokoi}, H. 1978, Electronics Communications
  of Japan, 61, 58

\bibitem[{{Mizuguchi} \& {Yokoi}(1974)}]{mizuguchi/yokoi:1974}
{Mizuguchi}, Y. \& {Yokoi}, H. 1974, Int. Conv. of IECE, Japan, 801

\bibitem[{{Mizugutch} et~al.(1976){Mizugutch}, {Akagawa}, \&
  {Yokoi}}]{mizuguchi/ak/yokoi:1976}
{Mizugutch}, Y., {Akagawa}, M., \& {Yokoi}, H. 1976, Digest of 1976 AP-S
  International Symposium on Antennas and Propagation

\bibitem[{{Mizugutch} \& {Yokoi}(1975)}]{mizuguchi/yokoi:1975}
{Mizugutch}, Y. \& {Yokoi}, H. 1975, Trans of IECE of Japan, 58-3

\bibitem[{{Muehlner}(1977)}]{muehlner:1977}
{Muehlner}, D. 1977, in Astrophysics and Space Science Library, Vol.~63,
  Infrared and submillimeter astronomy, ed. {G.~G.~Fazio},  143--152

\bibitem[{{Nanos}(1979)}]{nanos:1979}
{Nanos}, Jr., G.~P. 1979, \apj, 232, 341

\bibitem[{{Niemack} et~al.(2008)}]{niemack/etal:2008}
{Niemack}, M.~D., et~al. 2008, Journal of Low Temperature Physics, 151, 690

\bibitem[{{Niemack} et~al.(2010)}]{niemack/etal:2010}
{Niemack}, M.~D., et~al. 2010, in Society of Photo-Optical Instrumentation
  Engineers (SPIE) Conference Series, Vol. 7741

\bibitem[{{O'Brient} et~al.(2008)}]{obrient/etal:2008}
{O'Brient}, R., et~al. 2008, Journal of Low Temperature Physics, 151, 459

\bibitem[{{O'Dea} et~al.(2007){O'Dea}, {Challinor}, \&
  {Johnson}}]{odea/etal:2007}
{O'Dea}, D., {Challinor}, A., \& {Johnson}, B.~R. 2007, \mnras, 376, 1767

\bibitem[{{Olver} et~al.(1994){Olver}, {Clarricoats}, {Kishk}, \&
  {Shafai}}]{olver/etal:1994}
{Olver}, A., {Clarricoats}, P., {Kishk}, A., \& {Shafai}, L. 1994, Microwave
  Horns and Feeds (IEEE Press and IEE)

\bibitem[{{Olver}(1991)}]{olver:1991}
{Olver}, A.~D. 1991, in {Antennas and Propagation, 1991. ICAP 91., Seventh
  International Conference on (IEE)}, Vol.~1,  99 -- 108

\bibitem[{{O'Sullivan} et~al.(1995)}]{osullivan/etal:1995}
{O'Sullivan}, C., et~al. 1995, \mnras, 274, 861

\bibitem[{{O'Sullivan} et~al.(2008)}]{osullivan/etal:2008}
---. 2008, Infrared Physics and Technology, 51, 277

\bibitem[{{Padin} et~al.(2001)}]{padin/etal:2001}
{Padin}, S., et~al. 2001, \apjl, 549, L1

\bibitem[{{Padin} et~al.(2008)}]{padin/etal:2008}
---. 2008, Applied Optics, 47, 4418

\bibitem[{{Page} et~al.(2003)}]{page/etal:2003}
{Page}, L., et~al. 2003, \apj, 585, 566

\bibitem[{{Page} et~al.(1994){Page}, {Cheng}, {Golubovic}, {Meyer}, \&
  {Gundersen}}]{page/etal:1994}
{Page}, L.~A., {Cheng}, E.~S., {Golubovic}, B., {Meyer}, S.~S., \& {Gundersen},
  J. 1994, \ao, 33, 11

\bibitem[{{Page} et~al.(1990){Page}, {Cheng}, \& {Meyer}}]{page/etal:1990}
{Page}, L.~A., {Cheng}, E.~S., \& {Meyer}, S.~S. 1990, \apjl, 355, L1

\bibitem[{{Pardo} et~al.(2001){Pardo}, {Cernicharo}, \&
  {Serabyn}}]{pardo/etal:2001}
{Pardo}, J.~R., {Cernicharo}, J., \& {Serabyn}, E. 2001, IEEE Transactions on
  Antennas and Propagation, 49, 1683

\bibitem[{{Pariiskii} \& {Pyatunina}(1971)}]{pariiskii/pyatunina:1971}
{Pariiskii}, Y.~N. \& {Pyatunina}, T.~B. 1971, \sovast, 14, 1067

\bibitem[{{Parijskij}(1973)}]{pariiskii:1973}
{Parijskij}, Y.~N. 1973, \apjl, 180, L47

\bibitem[{{Partridge}(1980)}]{partridge:1980}
{Partridge}, R.~B. 1980, \apj, 235, 681

\bibitem[{{Partridge}(1995)}]{partridge:1995}
---. 1995, {3K: The Cosmic Microwave Background Radiation} (Cambridge, UK:
  Cambridge University Press)

\bibitem[{{Pascale} et~al.(2008)}]{pascale/etal:2008}
{Pascale}, E., et~al. 2008, \apj, 681, 400

\bibitem[{{Pearson} et~al.(2003)}]{pearson/etal:2003}
{Pearson}, T.~J., et~al. 2003, \apj, 591, 556

\bibitem[{{Peebles}(1994)}]{peebles:1994}
{Peebles}, P.~J.~E. 1994, \apjl, 432, L1

\bibitem[{{Peebles} et~al.(2009){Peebles}, {Page}, \& {Partridge}}]{ftbb:2009}
{Peebles}, P. J.~E., {Page}, L., \& {Partridge}, B. 2009, {Finding the Big
  Bang} (Cambridge, UK: Cambridge University Press)

\bibitem[{{Peterson} et~al.(2000)}]{peterson/etal:2000}
{Peterson}, J.~B., et~al. 2000, \apjl, 532, L83

\bibitem[{{Piccirillo}(1991)}]{piccirillo:1991}
{Piccirillo}, L. 1991, Review of Scientific Instruments, 62, 1293

\bibitem[{{Piccirillo} \& {Calisse}(1993)}]{piccirillo/calisse:1993}
{Piccirillo}, L. \& {Calisse}, P. 1993, \apj, 411, 529

\bibitem[{{Piccirillo} et~al.(1997)}]{piccirillo/etal:1997}
{Piccirillo}, L., et~al. 1997, \apjl, 475, L77

\bibitem[{{Piccirillo} et~al.(2008)}]{piccirillo/etal:2008}
{Piccirillo}, L., et~al. 2008, in Society of Photo-Optical Instrumentation
  Engineers (SPIE) Conference Series, Vol. 7020

\bibitem[{{Planck HFI Core Team} et~al.(2011)}]{planck-HFI-inst:2011}
{Planck HFI Core Team}, et~al. 2011, ArXiv e-prints

\bibitem[{Pospieszalski(1992)}]{pospieszalski:1992}
Pospieszalski, M.~W. 1992, IEEE MTT-S Digest, 1369

\bibitem[{{Pryke}(2011)}]{pryke:private}
{Pryke}, C. 2011, private communication

\bibitem[{{QUIET Collaboration} et~al.(2011)}]{bischoff/etal:2011}
{QUIET Collaboration}, et~al. 2011, \apj, 741, 111

\bibitem[{{Rabii} et~al.(2006)}]{rabii/etal:2006}
{Rabii}, B., et~al. 2006, Review of Scientific Instruments, 77, 071101

\bibitem[{{Rahmat-Samii} et~al.(1995){Rahmat-Samii}, {Imbriale}, \&
  {Galindo-Isreal}}]{YRS:DADRA}
{Rahmat-Samii}, Y., {Imbriale}, W., \& {Galindo-Isreal}, V. 1995, DADRA, {YRS
  Associates}, rahmat@ee.ucla.edu

\bibitem[{{Readhead} et~al.(1989){Readhead}, {Lawrence}, {Myers}, {Sargent},
  {Hardebeck}, \& {Moffet}}]{readhead/etal:1989}
{Readhead}, A.~C.~S., {Lawrence}, C.~R., {Myers}, S.~T., {Sargent}, W.~L.~W.,
  {Hardebeck}, H.~E., \& {Moffet}, A.~T. 1989, \apj, 346, 566

\bibitem[{{Reichborn-Kjennerud} et~al.(2010)}]{reichborn/etal:2010}
{Reichborn-Kjennerud}, B., et~al. 2010, in Society of Photo-Optical
  Instrumentation Engineers (SPIE) Conference Series, Vol. 7741

\bibitem[{{Richards}(1994)}]{richards:1994}
{Richards}, P.~L. 1994, Journal of Applied Physics, 76, 1

\bibitem[{{Rudnick}(1978)}]{rudnick:1978}
{Rudnick}, L. 1978, \apj, 223, 37

\bibitem[{{Schaffer} et~al.(2011)}]{schaffer/etal:2011}
{Schaffer}, K.~K., et~al. 2011, \apj, 743, 90

\bibitem[{{Schlaerth} et~al.(2010)}]{schlaerth/etal:2010}
{Schlaerth}, J.~A., et~al. 2010, in Society of Photo-Optical Instrumentation
  Engineers (SPIE) Conference Series, Vol. 7741

\bibitem[{{Scott} et~al.(1996)}]{scott/etal:1996}
{Scott}, P.~F., et~al. 1996, \apjl, 461, L1

\bibitem[{{Seielstad} et~al.(1981){Seielstad}, {Masson}, \&
  {Berge}}]{seielstad/etal:1981}
{Seielstad}, G.~A., {Masson}, C.~R., \& {Berge}, G.~L. 1981, \apj, 244, 717

\bibitem[{{Seo} et~al.(2008)}]{seo/etal:2008}
{Seo}, E.~S., et~al. 2008, Advances in Space Research, 42, 1656

\bibitem[{{Sheehy} et~al.(2010)}]{sheehy/etal:2010}
{Sheehy}, C.~D., et~al. 2010, in Society of Photo-Optical Instrumentation
  Engineers (SPIE) Conference Series, Vol. 7741

\bibitem[{{Shimon} et~al.(2008){Shimon}, {Keating}, {Ponthieu}, \&
  {Hivon}}]{shimon/etal:2008}
{Shimon}, M., {Keating}, B., {Ponthieu}, N., \& {Hivon}, E. 2008, Phys. Rev. D,
  77, 083003

\bibitem[{{Shirokoff} et~al.(2011)}]{shirokoff/etal:2011}
{Shirokoff}, E., et~al. 2011, \apj, 736, 61

\bibitem[{{Singal} et~al.(2011)}]{singal/etal:2011}
{Singal}, J., et~al. 2011, \apj, 730, 138

\bibitem[{{Smoot} et~al.(1990)}]{smoot/etal:1990}
{Smoot}, G., et~al. 1990, \apj, 360, 685

\bibitem[{{Smoot} et~al.(1977){Smoot}, {Gorenstein}, \&
  {Muller}}]{smoot/etal:1977}
{Smoot}, G.~F., {Gorenstein}, M.~V., \& {Muller}, R.~A. 1977, Physical Review
  Letters, 39, 898

\bibitem[{{Smoot} et~al.(1991)}]{smoot/etal:1991}
{Smoot}, G.~F., et~al. 1991, \apjl, 371, L1

\bibitem[{{Staniszewski} et~al.(2009)}]{staniszewski/etal:2009}
{Staniszewski}, Z., et~al. 2009, \apj, 701, 32

\bibitem[{{Stankevich}(1974)}]{stankevich:1974}
{Stankevich}, K.~S. 1974, \sovast, 18, 126

\bibitem[{{Staren} et~al.(2000)}]{staren/etal:2000}
{Staren}, J., et~al. 2000, \apj, 539, 52

\bibitem[{{Strukov} \& {Skulachev}(1984)}]{strukov/skulachev:1984}
{Strukov}, I.~A. \& {Skulachev}, D.~P. 1984, Soviet Astronomy Letters, 10, 1

\bibitem[{{Strukov} \& {Skulachev}(1986)}]{strukov/skulachev:1986}
---. 1986, Itogi Nauki i Tekhniki Seriia Astronomiia, 31, 37

\bibitem[{{Strukov} et~al.(1987){Strukov}, {Skulachev}, {Boyarskiy}, \&
  {Tkachev}}]{strukov/etal:1987}
{Strukov}, I.~A., {Skulachev}, D.~P., {Boyarskiy}, M.~N., \& {Tkachev}, A.~N.
  1987, JPRS Report Science Technology USSR Space, 3, 59

\bibitem[{{Strukov} et~al.(1988){Strukov}, {Skulachev}, \&
  {Klypin}}]{strukov/etal:1988}
{Strukov}, I.~A., {Skulachev}, D.~P., \& {Klypin}, A.~A. 1988, in IAU
  Symposium, Vol. 130, Large Scale Structures of the Universe, ed. {J.~Audouze,
  M.-C.~Pelletan, A.~Szalay, Y.~B.~Zel'Dovich, \& P.~J.~E.~Peebles },  27--+

\bibitem[{{Su} et~al.(2011){Su}, {Yadav}, {Shimon}, \&
  {Keating}}]{meng/etal:2011}
{Su}, M., {Yadav}, A.~P.~S., {Shimon}, M., \& {Keating}, B.~G. 2011, Phys. Rev.
  D, 83, 103007

\bibitem[{{Sunyaev} \& {Zeldovich}(1980)}]{sunyaev/zeldovich:1980}
{Sunyaev}, R.~A. \& {Zeldovich}, I.~B. 1980, \araa, 18, 537

\bibitem[{{Swetz} et~al.(2011)}]{swetz/etal:2011}
{Swetz}, D.~S., et~al. 2011, \apjs, 194, 41

\bibitem[{{Takahashi} et~al.(2010)}]{takahashi/etal:2010}
{Takahashi}, Y.~D., et~al. 2010, \apj, 711, 1141

\bibitem[{{Tatarskii}(1961)}]{tatarskii:1961}
{Tatarskii}, V.~I. 1961, {Wave propagation in a turbulent medium} (McGraw-Hill)

\bibitem[{{Tauber} et~al.(2010{\natexlab{a}})}]{tauber/etal:2010}
{Tauber}, J.~A., et~al. 2010{\natexlab{a}}, \aap, 520, A2+

\bibitem[{{Tauber} et~al.(2010{\natexlab{b}})}]{tauber/etal:2010b}
---. 2010{\natexlab{b}}, \aap, 520, A1+

\bibitem[{{Tegmark} \& {Zaldarriaga}(2009)}]{tegmark/zaldarriaga:2009}
{Tegmark}, M. \& {Zaldarriaga}, M. 2009, \prd, 79, 083530

\bibitem[{{Tegmark} \& {Zaldarriaga}(2010)}]{tegmark/zaldarriaga:2010}
---. 2010, \prd, 82, 103501

\bibitem[{{The COrE Collaboration} et~al.(2011)}]{core}
{The COrE Collaboration}, et~al. 2011, ArXiv e-prints

\bibitem[{{The Qubic Collaboration} et~al.(2011)}]{battistelli/etal:2011}
{The Qubic Collaboration}, et~al. 2011, Astroparticle Physics, 34, 705

\bibitem[{{Timbie} \& {Wilkinson}(1988)}]{timbie/wilkinson:1988}
{Timbie}, P.~T. \& {Wilkinson}, D.~T. 1988, Review of Scientific Instruments,
  59, 914

\bibitem[{{Timbie} \& {Wilkinson}(1990)}]{timbie/wilkinson:1990}
---. 1990, \apj, 353, 140

\bibitem[{{Timbie} et~al.(2006)}]{timbie/etal:2006}
{Timbie}, P.~T., et~al. 2006, \nar, 50, 999

\bibitem[{{Toral} et~al.(1989){Toral}, {Ratliff}, {Lecha}, {Maruschak}, \&
  {Bennett}}]{toral/etal:1989}
{Toral}, M.~A., {Ratliff}, R.~B., {Lecha}, M.~C., {Maruschak}, J.~G., \&
  {Bennett}, C.~L. 1989, IEEE Transactions on Antennas and Propagation, 37, 171

\bibitem[{{Tran} et~al.(2008){Tran}, {Lee}, {Hanany}, {Milligan}, \&
  {Renbarger}}]{tran/etal:2008}
{Tran}, H., {Lee}, A., {Hanany}, S., {Milligan}, M., \& {Renbarger}, T. 2008,
  \ao, 47, 103

\bibitem[{{Tran} \& {Page}(2009)}]{tran/page:2009}
{Tran}, H. \& {Page}, L. 2009, Journal of Physics Conference Series, 155,
  012007

\bibitem[{{Tran} et~al.(2010)}]{tran/etal:2010}
{Tran}, H., et~al. 2010, in Society of Photo-Optical Instrumentation Engineers
  (SPIE) Conference Series, Vol. 7731

\bibitem[{{Tucker} et~al.(1993){Tucker}, {Griffin}, {Nguyen}, \&
  {Peterson}}]{tucker/etal:1993}
{Tucker}, G.~S., {Griffin}, G.~S., {Nguyen}, H.~T., \& {Peterson}, J.~B. 1993,
  \apjl, 419, L45+

\bibitem[{{Tucker} et~al.(1997){Tucker}, {Gush}, {Halpern}, {Shinkoda}, \&
  {Towlson}}]{tucker/etal:1997}
{Tucker}, G.~S., {Gush}, H.~P., {Halpern}, M., {Shinkoda}, I., \& {Towlson}, W.
  1997, \apjl, 475, L73

\bibitem[{{Uson} \& {Wilkinson}(1982)}]{uson/wilkinson:1982}
{Uson}, J.~M. \& {Wilkinson}, D.~T. 1982, Physical Review Letters, 49, 1463

\bibitem[{{Watson} et~al.(1992){Watson}, {Gutierrez de La Cruz}, {Davies},
  {Lasenby}, {Rebolo}, {Beckman}, \& {Hancock}}]{watson/etal:1992}
{Watson}, R.~A., {Gutierrez de La Cruz}, C.~M., {Davies}, R.~D., {Lasenby},
  A.~N., {Rebolo}, R., {Beckman}, J.~E., \& {Hancock}, S. 1992, Nature, 357,
  660

\bibitem[{{Weiss}(1980)}]{weiss:1980}
{Weiss}, R. 1980, \araa, 18, 489

\bibitem[{{Welton} \& {Winston}(1978)}]{welford/winston:1978}
{Welton}, W. \& {Winston}, R. 1978, The Optics of Nonimaging Concentrators
  (Academic Press.)

\bibitem[{{Wilkinson} \& {Partridge}(1967)}]{wilkinson/partridge:1967}
{Wilkinson}, D.~T. \& {Partridge}, R.~B. 1967, \nat, 215, 719

\bibitem[{{Wilson} \& {Penzias}(1967)}]{wilson/penzias:1967}
{Wilson}, R.~W. \& {Penzias}, A.~A. 1967, Science, 156, 1100

\bibitem[{{Wollack} et~al.(1993){Wollack}, {Jarosik}, {Netterfield}, {Page}, \&
  {Wilkinson}}]{wollack/etal:1993}
{Wollack}, E.~J., {Jarosik}, N.~C., {Netterfield}, C.~B., {Page}, L.~A., \&
  {Wilkinson}, D. 1993, \apjl, 419, L49

\bibitem[{{Yadav} et~al.(2010){Yadav}, {Su}, \&
  {Zaldarriaga}}]{yadav/su/zal:2010}
{Yadav}, A.~P.~S., {Su}, M., \& {Zaldarriaga}, M. 2010, \prd, 81, 063512

\bibitem[{{Yoon} et~al.(2006)}]{yoon/etal:2006}
{Yoon}, K.~W., et~al. 2006, in Society of Photo-Optical Instrumentation
  Engineers (SPIE) Conference Series, Vol. 6275

\bibitem[{{Yoon} et~al.(2009)}]{yoon/etal:2009}
{Yoon}, K.~W., et~al. 2009, in American Institute of Physics Conference Series,
  ed. {B.~Young, B.~Cabrera, \& A.~Miller}, Vol. 1185,  515--518

\bibitem[{{Zaldarriaga}(2006)}]{zaldarriaga:private}
{Zaldarriaga}, M. 2006, private communication

\bibitem[{{Zaldarriaga} \& {Seljak}(1997)}]{zal+seljak:1997}
{Zaldarriaga}, M. \& {Seljak}, U. 1997, \prd, 55, 1830

\bibitem[{{Zmuidzinas}(2003)}]{zmuidzinas:2003}
{Zmuidzinas}, J. 2003, Applied Optics, 42, 4989

\end{thebibliography}

\end{document}